\documentclass[a4paper,11pt]{article}

\usepackage[all]{xypic}
\usepackage{a4wide,latexsym,amssymb,amsmath,lscape,setspace,amsthm}
\usepackage[normalem]{ulem}
\numberwithin{equation}{section}
\usepackage[pdftex]{graphicx}
\usepackage{tikz}        
\usepackage{color}
\usepackage{mathtools}
\usepackage{mathptmx}
\usepackage{slashed,epsfig}
\usepackage{amsfonts}
\usepackage{cite}
\usepackage{hyperref}
\hypersetup{
    colorlinks=true, 
    linktoc=all,     
    linkcolor=blue,  
}
\newcommand{\blu}[1]{\textcolor{blue}{#1}}

\newcommand{\qee}{\mbox{\hspace{0.2mm}}\hfill$\triangle$}
\newtheorem{thm}{Theorem}[section]

\newtheorem{definizione}{Definition}[section]

\theoremstyle{remark}
\newtheorem{rem}[thm]{Remark}
\newtheorem{ex}[thm]{Example}
\newenvironment{remark}{\begin{rem}\rm}{\qee\end{rem}}

\newcommand{\C}{{\mathbb C}}

\newcommand{\cO}{{\mathcal O}}

\newcommand{\ft}[2]{{\textstyle\frac{#1}{#2}}}
 \newcommand{\Z}{{\mathbb Z}}
\newcommand{\Q}{{\mathbb Q}}

\begin{document}
\begin{titlepage}
\begin{center}
\begin{flushright}
ARC-17-6
\end{flushright}
\vskip20pt
{\Large \bf Crepant Resolutions of  $\mathbb{C}^3/\mathbb{Z}_4$ and the
Generalized Kronheimer \\[8pt] Construction
 (in view of the  Gauge/Gravity Correspondence )}\\[6mm]
{{\sc Ugo Bruzzo${}^{\; a,g}$,  Anna Fino$^{\; b,g}$,
Pietro Fr\'e${}^{\; c,f,g}$, \\
Pietro Antonio Grassi$^{\;d,f,g}$  and Dimitri Markushevich${}^{\;e,g}$}
\\[5pt] \small \sl
${}^a$ SISSA (Scuola Internazionale Superiore di Studi Avanzati),    \\
via Bonomea 265, 34136 Trieste, Italy; \\ INFN (Istituto Nazionale di Fisica Nucleare),  Sezione di
Trieste; \\ IGAP (Institute for Geometry and Physics), Trieste \\
\emph{e-mail:} {   \tt bruzzo@sissa.it}\\[4pt]
${}^b$ Dipartimento di Matematica G. Peano,
Universit\`a di Torino, \\ Via Carlo Alberto 10, 10123 Torino,
Italy \\
\emph{e-mail:} \quad {\small {\tt annamaria.fino@unito.it}}\\[4pt]
${}^c$  Dipartimento di Fisica, Universit\`a
di Torino, via P. Giuria 1, 10125 Torino, Italy\\
\emph{e-mail:}  {\tt pfre@unito.it}\\[4pt]
$^{d}$   Dipartimento di Scienze e Innovazione Tecnologica, Universit\`a del Piemonte Orientale, \\
 viale T. Michel 11, 15121 Alessandria, Italy \\
\emph{e-mail:}   {\tt pietro.grassi@uniupo.it}\\[4pt]
${}^e$  Department de Math\'ematique, Universit\'e de Lille, B\^atiment M2, \\ Cit\'e Scientifique, 59655 Villeneuve-d'Ascq, France \\
\emph{e-mail:}    {\tt dimitri.markouchevitch@univ-lille.fr}\\[4pt]
$^{f}$  INFN (Istituto Nazionale di Fisica Nucleare),  Sezione di Torino \\[4pt]
$^g$  Arnold-Regge Center for Algebra, Geometry and Theoretical Physics, \\  via P. Giuria 1,  10125 Torino,
Italy}

\bigskip
February 18, 2019;  revised \today


\bigskip
\begin{abstract}
As a continuation of a general program started in two previous
publications, in the present paper we study the K\"ahler quotient
resolution of the orbifold $\mathbb{C}^3/\mathbb{Z}_4$, comparing
with the results of a toric description of the same.
In this way we
determine the algebraic structure of the exceptional divisor, whose
compact component is the second Hirzebruch surface $\mathbb F_2$. We determine the
explicit K\"ahler geometry of the smooth resolved manifold $Y$, which is the total space
of the canonical bundle of  $\mathbb F_2$. We study in detail the chamber structure
of the space of stability parameters (corresponding in gauge theory to the Fayet-Iliopoulos
parameters) that are involved in the construction of the desingularizations either by generalized
Kronheimer quotient, or as   algebro-geometric quotients. The walls of the chambers correspond
to two degenerations; one is a partial desingularization of the quotient, which is the total space
of the canonical bundle of the weighted projective space $\mathbb P[1,1,2]$, while the other
is the product of the ALE space $A_1$ by a line, and is related to the full resolution is a subtler way.
These geometrical results will be used to look for exact supergravity brane
solutions and dual superconformal gauge theories.

\end{abstract}
\end{center}
\end{titlepage}

\setcounter{tocdepth}{2}
\tableofcontents \noindent {}

\bigskip
\section{Introduction}
\label{introito} The present paper is the next step  of
an investigation program initiated in
\cite{pappo1,Bruzzo:2017fwj}, whose final aim is to fully
elucidate the relation among the physical building blocks of
$D=3/D=4$  supersymmetric gauge theories, the mathematical
constructions of the generalized McKay correspondence, and   the
generalized Kronheimer-like resolution of $\mathbb{C}^n/\Gamma$
singularities.

From the physics side, the context of these investigations is, in a
broad sense, the \textit{gauge/gravity correspondence}
\cite{Maldacena:1997re,Kallosh:1998ph,Ferrara:1998jm,Ferrara:1998ej,sergiotorino,witkleb,Fabbri:1999hw,Fabbri:1999ay,
Aharony:2008ug,Gaiotto:2007qi,Gaiotto:2009tk}. In particular there
are two main paradigms:
\begin{description}
  \item[a)] The case of M2-branes solutions of $D=11$ supergravity
  where the eight-dimensional space transverse to the brane world
  volume is taken to be, to begin with, of the form $\mathbb{C}\times
  \frac{\mathbb{C}^3}{\Gamma}$, having denoted by $\Gamma \subset
  \mathrm{SU(3)}$ a finite subgroup.
  \item[b)] The case of D3-brane solutions of type IIB supergravity
  where the 6-dimensional transverse space to the brane is taken to
  be, to begin with, just the singular orbifold $\frac{\mathbb{C}^3}{\Gamma}$
  mentioned in the previous lines.
\end{description}
In both cases the idea is the following. Provided one takes into
account also the \text{twisted states}, String Theory can
confortably live also on singular orbifolds, on the contrary
Supergravity does  require smooth manifolds. Hence we are
interested in the crepant\footnote{We recall that a morphism of varieties
$X \to Y$, which in particular can be a resolution of singularities,   is crepant when the canonical sheaf of $X$ is the pullback of
the canonical sheaf of $Y$.} resolution of the orbifold
singularity:
\begin{equation}\label{resolvo}
    Y \longrightarrow \mathbb{C}^n/\Gamma
\end{equation}
and we look for exact solutions of $D=11$ or $D=10$ type IIB
supergravities, respectively of the M2-brane and of the D3-brane
type, where the transverse space is, respectively, $\mathbb{C}\times
Y$  or $Y$. At the same time we are interested in the construction
of the $D=3$, respectively, $D=4$ matter coupled supersymmetric
gauge theories that become superconformal at a suitable infrared
fixed point and are, supposedly, the dual partners of such brane
solutions.
\paragraph{Matter content of the gauge theory.}
As it was strongly emphasized  in \cite{pappo1,Bruzzo:2017fwj}, the
mathematical lore on the generalization of the  McKay correspondence
\cite{mckay} and of the Kronheimer construction of ALE manifolds
\cite{kro1,kro2} enter at this point in an essential way. Indeed
the basic starting point of such constructions, \textit{i.e.} the
space\footnote{For the definition of the space \eqref{Sgamma} see
\cite{Bruzzo:2017fwj}. For its realization in the specific model
studied in the present paper see  eq.~\eqref{carnevalediPaulo}.}
\begin{equation}\label{Sgamma}
    \mathcal{S}_\Gamma \, \equiv \,
    \mathrm{Hom}_\Gamma\left( R,\mathcal{Q}\otimes R\right),
\end{equation}
 where $\mathcal{Q}$ is the given embedding of $\Gamma$ into $\mathrm{SU(n)}$ and $R$ is the regular representation of $\Gamma$.
The structure of the representation ring of $\Gamma$ is described by the McKay quiver matrix,\footnote{See both \cite{Bruzzo:2017fwj} for the general
definition and below, (eq.~\eqref{carteodollo}) for the present
$\mathbb{Z}_4$ case.} and the latter determines
the gauge group
$\mathcal{F}_\Gamma$ and the whole spectrum of  Wess-Zumino  hypermultiplets of the associated gauge theory.\footnote{ Singularities
$\mathbb{C}^n/\Gamma$ can have a crepant resolution and fall in the
class analyzed in \cite{Bruzzo:2017fwj} only if $\Gamma$ acts
through its representation $Q$ as a subgroup of $\mathrm{SU}(n)$. It
is important to recall that the $\mathbb{Z}_k$ quotient of
$\mathbb{C}^4$ utilized in the ABJM construction
\cite{Aharony:2008ug} does not satisfy this condition and indeed admits
only discrepant resolutions.}
\paragraph{The McKay quivers.}
The quiver matrix admits a representation by means of a quiver
diagram and quiver gauge theories have been over  the last twenty years
the target of a large
physical literature  (see for
instance \cite{Bianchi:2000de,Bianchi:2009bg,Bianchi:2007wy} and all
references therein). Quiver diagrams interpreted as a recipe to
construct a supersymmetric gauge theory are more general than McKay
quivers and up to our knowledge the inverse problem of defining
necessary and sufficient conditions  which,  in the vast class of
quivers, select those that are of McKay type is unsolved. Each McKay
quiver is associated with a discrete group $\Gamma\subset
\mathrm{SU(n)}$ and its nodes are in one-to-one correspondence with
the irreducible representations of $\Gamma$. Its lines codify the
data to construct the above mentioned space
$\mathrm{Hom}_\Gamma\left( R,\mathcal{Q}\otimes R\right)$ which
hosts the matter multiplets of the corresponding gauge theory. In a
general quiver the lines provide the same information about matter
multiplets, yet there is no a priori guarantee that these latter
fill a space $\mathrm{Hom}_\Gamma\left( R,\mathcal{Q}\otimes
R\right)$ for any suitable $\Gamma$. Said differently, any quiver
diagram provides information for the construction of a K\"ahler or
HyperK\"ahler quotient. McKay quivers provide information how to
resolve a $\mathbb{C}^n/\Gamma$ by means of a K\"ahler or
HyperK\"ahler quotient defined according with a generalized
Kronheimer construction; yet not all (Hyper)K\"ahler quotients are
devised to solve quotient singularities.  A notable counterexample
is provided by the case where the (Hyper)K\"ahler quotient provides
the resolution of a conifold singularity. The relation between gauge
theories of the McKay type and of conifold type is addressed in a
forthcoming publication \cite{conmasbia}.
\paragraph{The first example: ALE manifolds.}
Historically the first case that was fully developed both
mathematically and physically is that of the Kleinian singularities
$\mathbb{C}^2/\Gamma$, where the finite groups $\Gamma \subset
\mathrm{SU(2)}$ admit the celebrated ADE classification (for a
comprehensive recent review see chapter 8 of \cite{advancio}).
Relying on the properties of the relevant McKay quiver matrices
which, due to the ADE classification, are identified with the
extended Cartan matrices of the ADE Lie algebras, and on the
recently introduced hyperk\"ahler quotient constructions\cite{HKLR},
Kronheimer \cite{kro1,kro2} succeeded in providing an algorithmic
construction of all four-dimensional ALE gravitational instantons,
the A subclass of which had been previously exhibited by Gibbons and
Hawking as multicenter metrics \cite{Gibbons:1979zt,Gibbons:1979xm},
generalizing the first and simplest example of the Eguchi-Hanson
one--center metric \cite{eguccio}. ALE-manifolds and more general
gravitational instantons were extensively studied and utilized in
various capacities in string theory and in supergravity
\cite{Billo:1992zv,Billo:1992uw,Billo:1992ei,Bianchi:1994gi,Bianchi:1995ad,Bianchi:1996zj}.
The first paper where the Kronheimer construction was applied to
2D-conformal field theories in stringy setups dates back to 1994
\cite{mango}, while the first example of an exact D3-brane solution
of type IIB supergravity  with $\mathrm{ALE}\times \mathbb{C}$
transverse space was produced in 2000 \cite{Bertolini:2002pr}.

The ALE resolutions of Kleinian singularities are hyperk\"ahler
manifolds\ and the Kronheimer construction is based on  the
hyperk\"ahler quotient. All this goes hand in hand with
$\mathcal{N}=2$ supersymmetry in $D=4$ and $\mathcal{N}=4$ (broken
to $\mathcal{N}=3$ by Chern-Simons couplings
\cite{ringoni,Fre1999xp}) in $D=3$.
\paragraph{The generalization to three dimensions.}
In the second half of the nineties of the last century, a few
mathematicians from the algebraic geometry community addressed the
generalization of the Kronheimer construction to the case of
$\mathbb{C}^3/\Gamma$ singularities
\cite{crawthesis,itoriddo,roanno,marcovaldo,SardoInfirri:1994is,SardoInfirri:1996ga,SardoInfirri:1996gb}.
In this case the resolved variety is simply a K\"ahler manifold and
the Kronheimer construction reduces to a K\"ahler quotient. All this
goes hand in hand with $\mathcal{N}=1$ supersymmetry in $D=4$ or
$\mathcal{N}=2$ in $D=3$. As we discussed at length in
\cite{Bruzzo:2017fwj} the additional necessary item in the
generalized Kronheimer construction, besides the real moment maps,
is a universal holomorphic equation which substitutes the
holomorphic part of the tri-holomorphic moment maps and amounts to a
universal cubic superpotential. Instead the generalized McKay quiver
diagrams have the same group theoretical definition as in the
$\mathbb{C}^2$ case although they no longer represent extended
Cartan matrices. The choice of the appropriate gauge group
$\mathcal{F}_\Gamma$ follows from the quiver in exactly the same way
as in the $n=2$ case. Henceforth the finite group $\Gamma$ and its
homomorphism $\mathcal{Q}$ into $\mathrm{SU(3)}$ provide, once more
as in the original Kronheimer case,  all the data to construct the
corresponding unique supersymmetric gauge theory on the brane
world-volume.
\paragraph{The Ito--Reid theorem and the
tautological bundles} The two most important results of the
mentioned mathematical activity are:\footnote{In view of the
previous discussion about McKay quivers as a subclass of the set of all
possible quivers, the below listed mathematical results apply only
to case of supersymmetric gauge theories derived from a McKay quiver
and hence associated with a $\mathbb{C}^3/\Gamma$ singularity. They
do not apply to general quiver gauge theories.}
\begin{description}
  \item[1)] the Ito-Reid theorem
\cite{itoriddo} that relates the generators of the cohomology groups
$H^{(q,q)}(Y)$ of the resolved variety to the  conjugacy classes of
$\Gamma$ that in the $\mathcal{Q}$ homomorphic image have
\textit{age} $q$.
  \item[2)] the definition \cite{degeratu} of the
tautological vector bundles  $E_i \longrightarrow Y$ associated with
each of the irreducible nontrivial  {representations} $\mathcal{D}_i$ of
$\Gamma$, with
$$ \text{rank} \, E_i \, = \, n_i \, \equiv \, \text{dim} \, \mathcal{D}_i $$
\end{description}

As we emphasized in \cite{Bruzzo:2017fwj}, the first Chern forms
$\omega_i^{(1,1)}$ of the tautological bundles $E_i$  {usually provide a
redundant set} of generators for the $H^{(1,1)}(Y)$
group, whose dimension is fixed by the Ito-Reid theorem. One would
like to have a constructive approach to single out both the
components of the exceptional divisor $D_E$ and a good
basis of homological compact two-cycles $\mathcal{C}_I$
($I=1,\dots,\ell$) so as to be able to calculate the periods of the
$\omega_i^{(1,1)}$ on them:
\begin{equation}\label{seniorito}
    \Pi^i_I \, \equiv \, \int_{\mathcal{C}_i} \omega_I^{(1,1)} \quad
    ; \quad I=1,\dots, r\, = \, |\Gamma|-1 \quad ; \quad i\, = \,
    1,\dots , \ell \, = \, \# \, \mbox{ of senior c.c.}
\end{equation}
The rationale of the above counting is the following. Because of
Poincar\'e duality between $H^{(2,2)}(Y)$ and $H^{(1,1)}_c(Y)$ with
compact support, the number of senior classes is equal to the number
of $\omega_i^{(1,1)}$ with compact support.
\paragraph{Relevance of the cocycles and of the cycles for the
gauge/gravity dual pairs.} The above geometrical information is of
vital importance for the physical interpretation of the orbifold
resolution within the correspondence between supergravity brane
solutions and gauge theories on the world volume. The levels $\zeta$
of the moment maps are in the gauge-theory the Fayet-Iliopoulos
parameters. On the supergravity side these latter emerge as fluxes
of $p$-forms partially or fully wrapped on homology cycles of the
space transverse to the brane\footnote{See in particular
\cite{Bertolini:2002pr} for an exemplification of this mechanism in
the case of a D3-brane with transverse space $\mathbb{C}\times
\mathrm{ALE}$.}.
\par
More specifically, given the blow--down morphism
\begin{equation}\label{bolligiu}
    Y \, \longrightarrow \, \mathbb{C}^3/\Gamma
\end{equation}
in order to construct a \textit{bona--fide} M2--brane or D3--brane
solution of the relevant supergravity that is dual to the considered
world--volume gauge--theory we need a \textit{Ricci flat} K\"ahler
metric on the resolved manifold $Y$. This latter, which is clearly
identified, both topologically and algebraically, by the Kronheimer
construction, is not endowed by the corresponding K\"ahler quotient
with a Ricci-flat metric. The derivation of a Ricci-flat K\"ahler on
$Y$ is a difficult mathematical problem discussed in a forthcoming
publication \cite{conmasbia}, yet one thing is clear a priori. The
parameters of such a Ricci-flat K\"ahler metric must be into
correspondence with the Fayet-Iliopoulos parameters appearing in the
Kronheimer construction and should parameterize the volumes of the
homology cycles of $Y$. When one or more of these cycles shrink to
zero  a  singularity develops. When all cycles shrink to zero we
come back to singular orbifold $\mathbb{C}^3/\Gamma$ both on the
supergravity and on the gauge theory side.
\paragraph{What we do in this paper.} In this article we study a
specific nontrivial example of orbifold, whose analysis was only
sketched in \cite{Bruzzo:2017fwj}. It corresponds to a specific
embedding:
$$ \mathcal{Q} \quad : \quad \mathbb{Z}_4 \, \longrightarrow \, \mathrm{SU(3)}$$
In the case of cyclic groups the McKay correspondence
and the costruction of the desingularizations of the singular quotient can be realized using
toric geometry \cite{itoriddo}. It is then fairly straightforward to identify the possible resolutions
and study in full detail their geometry. In particular, one identifies the toric divisors
and curves, including the compact divisors (a copy of the second Hirzebruch surface
$\mathbb F_2$ for the full resolution, and the weighted projective space $\mathbb P[1,1,2]$
for one of the partial desingularizations).
 Actually utilizing the equations of the divisors provided by toric
geometry we are able, by means of the restriction of the moment map
equations to these special loci, to compute the periods of the
$(1,1)$-forms $\omega^{(1,1)}_{1,2,3}$ on the basis of compact
cycles in the general case.

The construction of the desingularizations, be it made as a generalized Kronheimer
quotient \cite{degeratu,Bruzzo:2017fwj}, or as an algebro-geometric GIT quotient \cite{itoriddo,CrawIshii},
depends on a set of parameters, living in a linear space, called the {\em stability parameter space.}
This is partitioned in chambers, with the property that the quotient does not change when one moves
within the interior of a chamber, while nontrivial topological changes may occur while crossing a wall
between two chambers. In Sections \ref{camerataccademica} to \ref{summatheologica} we study in great
detail this chamber structure, by means of explicit calculations of the periods of the Chern classes
of the tautological bundles on the the cycles that generate the 2-homology of the resolutions.

This geometrical study provides the basis for the construction of an
explicit dual pair either in M-theory or in type IIB theory. This we
plan to do in a new publication \cite{conmasbia}. Further comments
on the next  steps in our program are left for the conclusions.

\section{The $\mathbb{C}^3/\mathbb{Z}_4$ model, its McKay quiver and
the associated Kronheimer construction} \label{maccaius} The action
of the group $\mathbb{Z}_4$ on $\mathbb{C}^3$ is defined by
introducing the three-dimensional unitary representation
$\mathcal{Q}(A)$ of its abstract generator $A$ that satisfies the
defining relation $A^4 \, = \, \mathbf{e}$. We set:
\begin{equation}\label{generatoreAZ4}
   \mathcal{Q}(A) \, = \, \left(
\begin{array}{ccc}
 i & 0 & 0 \\
 0 & i & 0 \\
 0 & 0 & -1 \\
\end{array}
\right) \quad ; \quad \mathcal{Q}(A)^4 \, = \, \left(
\begin{array}{ccc}
 1 & 0 & 0 \\
 0 & 1 & 0 \\
 0 & 0 & 1 \\
\end{array}
\right)
\end{equation}
Since $\mathbb{Z}_4$ is abelian and cyclic, each of its four
elements corresponds to an entire conjugacy class of which we can
easily calculate the age-vector and the ages according with the
conventions established in \cite{Bruzzo:2017fwj}. We obtain:
\begin{equation}\label{panefresco}
    \begin{array}{|c||c|c|c|c|}
    \hline
    \text{Conj.
    Class}&\text{Matrix}&\text{age-vector}&\text{age}&\text{name}\\
    \hline
       \mathrm{Id} & \left(
\begin{array}{ccc}
 1 & 0 & 0 \\
 0 & 1 & 0 \\
 0 & 0 & 1 \\
\end{array}
\right) & \ft 14 \, (0,0,0) & 0 & \text{null class} \\
\hline
 \mathcal{Q}(A) & \left(
\begin{array}{ccc}
 i & 0 & 0 \\
 0 & i & 0 \\
 0 & 0 & -1 \\
\end{array}
\right) & \ft 14 \, (1,1,2) & 1 & \text{junior class} \\
\hline \mathcal{Q}(A)^2 & \left(
\begin{array}{ccc}
 -1 & 0 & 0 \\
 0 & -1 & 0 \\
 0 & 0 & 1 \\
\end{array}
\right) & \ft 14 \, (2,2,0) & 1 & \text{junior class} \\
\hline
 \mathcal{Q}(A)^3  & \left(
\begin{array}{ccc}
 -i & 0 & 0 \\
 0 & -i & 0 \\
 0 & 0 & -1 \\
\end{array}
\right) & \ft 14 \, (3,3,2) & 2 & \text{senior class}\\
       \hline
     \end{array}
\end{equation}
Therefore, according with the theorem of Ito and Reid
\cite{itoriddo}, as reviewed in \cite{Bruzzo:2017fwj}, in the
crepant resolution:
\begin{equation}\label{pirollo}
   Y \, \equiv \, \mathcal{M}_\zeta \, \longrightarrow \, \frac{\mathbb{C}^3}{\mathbb{Z}_4}
\end{equation}
the Hodge numbers of the smooth resolved variety are as follows:
\begin{equation}\label{ganimellus}
    h^{(0,0)}\left(\mathcal{M}_\zeta\right) \, = \,1 \quad ; \quad h^{(1,1)}\left(\mathcal{M}_\zeta\right) \, =
    \,2 \quad ; \quad h^{(2,2)}\left(\mathcal{M}_\zeta\right) \, =
    \,1
\end{equation}
Furthermore the existence of a senior class implies that one of the
two generators of $H^{(1,1)}\left(\mathcal{M}_\zeta\right)$
 {can be chosen to have} compact support while the other  {will have} non-compact support. As we
later discuss studying the resolution with the help of toric
geometry, this distinction goes hand in hand with the structure of
the exceptional divisor that has two components, one compact and one
non-compact.

The character table of the $\mathbb{Z}_4$ group is easily calculated
and it foresees four one-dimensional representations that we
respectively name  {$\mathcal{D}_i$}, $(i\, = \, 0,1,2,3)$. The table is
given below.
\begin{equation}\label{caratteruccio}
\begin{array}{c||cccc}
\begin{array}{ccc}
  \text{Irrep}& \setminus & \text{C.C.} \\
\end{array} & \mathbf{e} &A & A^2 & A^3\\
\hline \hline
 {\mathcal{D}_0} &1 & 1 & 1 & 1 \\
 {\mathcal{D}_1}& 1 & i & -1 & -i \\
 {\mathcal{D}_2} & 1 & -1 & 1 & -1 \\
 {\mathcal{D}_3} & 1 & -i & -1 & i \\
\end{array}
\end{equation}
\subsection{The McKay quiver diagram and its representation}
\begin{figure}
\centering
\includegraphics[height=7cm]{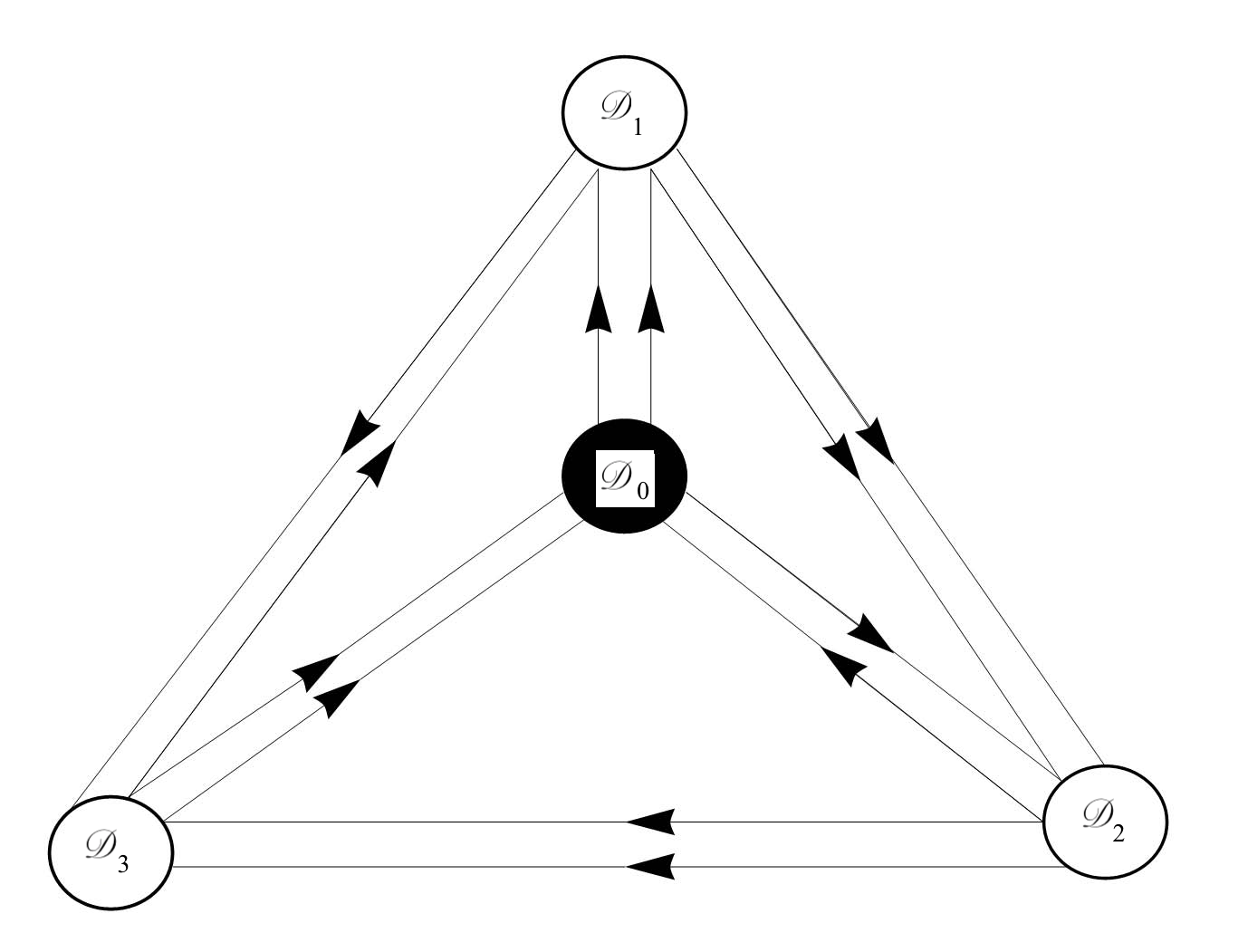}
\caption{ \label{mckayquivz4} The McKay quiver of the $\mathbb{Z}_4$
group embedded into $\mathrm{SU(3)}$ according {to}
eq.~\eqref{panefresco}.}
\end{figure}

The information encoded in
eqs. \eqref{panefresco}, \eqref{caratteruccio} is sufficient to
calculate the McKay quiver matrix defined by:
\begin{equation}\label{carteodollo}
    \mathcal{Q} \otimes \mathcal{D}_I\, = \, \bigoplus_{J=0}^3 \,\mathcal{A}_{IJ} \,\mathcal{D}_J
\end{equation}
where $\mathcal{D}_I$ denotes the $4$ irreducible representation of
the group
 {defined in eq.} \eqref{caratteruccio}, while $\mathcal{Q}$ is the
representation \eqref{panefresco} describing the embedding
$\mathbb{Z}_4 \hookrightarrow \mathrm{SU(3)}$. Explicitly we obtain
\begin{equation}\label{quiverroz4}
   \mathcal{A}_{IJ} \, = \, \left(
\begin{array}{cccc}
 0 & 2 & 1 & 0 \\
 0 & 0 & 2 & 1 \\
 1 & 0 & 0 & 2 \\
 2 & 1 & 0 & 0 \\
\end{array}
\right)
\end{equation}
A graphical representation of the quiver matrix \eqref{quiverroz4}
is provided in fig.~\ref{mckayquivz4}.

Every node of the diagram corresponds to an irreducible
representation $\mathcal{D}_I$ and, in the Kronheimer construction, to a gauge
group factor $\mathrm{U}\left(\dim \mathcal{D}_I \right )$.

As one sees in every node enter three lines and go out three lines.
Both the incoming and the outgoing lines are subdivided in a double
line in some direction (or from some direction) and a single line to
some direction, or from some direction.

This information is sufficient to derive the number of Wess-Zumino
multiplets in the corresponding supersymmetric gauge theory and
assign their representations with respect to the three gauge groups
(actually four minus one for the barycentric motion) as we will more
extensively discuss in the forthcoming paper \cite{conmasbia}.
\par
From the mathematical viewpoint each line of the diagram corresponds
to an independent parameter appearing in the explicit construction
of the space $\mathcal{S}_{\mathbb{Z}_4}$. This latter is
constructed as follows. Let $R$ denote the regular representation of
$\Gamma$. We consider the space of triplets of $4\times 4$ complex
matrices:
\begin{eqnarray}
    p \in \mathcal{P}_{\mathbb{Z}_4} \, \equiv \, \mbox{Hom}\left(R,\mathcal{Q}\otimes R\right) \, \Rightarrow\,
    p\,=\, \left(\begin{array}{c}
                   A \\
                   B \\
                   C
                 \end{array}
     \right) \label{homqg}
\end{eqnarray}
The action of the discrete group $\mathbb{Z}_4$ on the space
$\mathcal{P}_\Gamma$ is defined in full analogy with the Kronheimer
case by:
\begin{equation}\label{gammazione}
    \forall \gamma \in \mathbb{Z}_4: \quad \gamma\cdot p \,\equiv\, \mathcal{Q}(\gamma)\,\left(\begin{array}{c}
                  R(\gamma)\, A \, R(\gamma^{-1})\\
                   R(\gamma)\, B \, R(\gamma^{-1}) \\
                  R(\gamma)\, C \, R(\gamma^{-1})
                 \end{array}
     \right)
\end{equation}
where  $R(\gamma)$ denotes its $4 \times 4$-matrix image in the
regular representation.

The subspace $\mathcal{S}_{\mathbb{Z}_4}$ is obtained by setting:
\begin{equation}
\mathcal{S}_{\mathbb{Z}_4} \, \equiv \,
\mbox{Hom}\left(R,\mathcal{Q}\otimes R\right)^{\mathbb{Z}_4}\, = \,
\left\{p\in\mathcal{P}_{\mathbb{Z}_4} / \forall \gamma\in
\mathbb{Z}_4 , \gamma\cdot p = p\right\}\,\,
\label{carnevalediPaulo}
\end{equation}
As we know from the general theory exposed in \cite{Bruzzo:2017fwj}
the space $\mathcal{S}_{\mathbb{Z}_4}$ must have complex dimension
$3\times |\mathbb{Z}_4| \, = \, 12$ which is indeed the number of
lines in the quiver diagram of fig. \ref{mckayquivz4}. In the basis
where the regular representation has been diagonalized with the help
of the character Table \eqref{caratteruccio} the general form of the
triplet of matrices composing
$\mbox{Hom}_{\mathbb{Z}_4}\left(R,\mathcal{Q}\otimes R\right)$ and
therefore providing the representation of the quiver diagram of fig.
\ref{mckayquivz4} is the following one:
\begin{equation}\label{pastrugno}
\begin{array}{ccccccc}
       A & = & \left(
\begin{array}{cccc}
 0 & 0 & 0 & \Phi^{(1)}_{0,3} \\
 \Phi^{(1)}_{1,0} & 0 & 0 & 0 \\
 0 & \Phi^{(1)}_{2,1} & 0 & 0 \\
 0 & 0 & \Phi^{(1)}_{3,2} & 0 \\
\end{array}
\right) & ; & B & = & \left(\begin{array}{cccc}
 0 & 0 & 0 & \Phi^{(2)}_{0,3} \\
 \Phi^{(2)}_{1,0} & 0 & 0 & 0 \\
 0 & \Phi^{(2)}_{2,1} & 0 & 0 \\
 0 & 0 & \Phi^{(2)}_{3,2} & 0 \\
\end{array}
\right) \\
\null&\null& \null & \null & \null&\null& \null \\
       C & = & \left(
\begin{array}{cccc}
 0 & 0 & \Phi^{(3)}_{0,2} & 0 \\
 0 & 0 & 0 & \Phi^{(3)}_{1,3} \\
 \Phi^{(3)}_{2,0} & 0 & 0 & 0 \\
 0 & \Phi^{(3)}_{3,1} & 0 & 0 \\
\end{array}
\right) & \null & \null & \null & \null
     \end{array}
\end{equation}
The twelve complex parameters $\Phi^{(J)}_{p,q}$ with $J=1,2,3$,
$p,q=0,1,2,3$, promoted to be functions of the space-time
coordinates $\xi^\mu$:
$$\Phi^{(J)}_{p,q}(\xi)$$
are the complex scalar fields filling the flat K\"ahler manifold of
the Wess-Zumino multiplets in the microscopic lagrangian of the
corresponding gauge theory.
\subsection{The locus $\mathbb{V}_6 \subset \mathcal{S}_{\mathbb{Z}_4}$}
As it was explained in the context of the general framework in
\cite{Bruzzo:2017fwj}, the  $3|\Gamma|$-dimensional flat K\"ahler
manifold $\mathcal{S}_\Gamma$ contains always a  {subvariety}
$\mathbb{V}_{|\Gamma|+2} \subset \mathcal{S}_\Gamma$ of dimension
$|\Gamma|+2$ which is singled out by the following set of quadratic
equations:
\begin{equation}\label{ceramicus}
    \left[A\, , \, B\right] \, = \, \left[B\, , \, C\right]\, = \, \left[C\, , \,
    A\right]\, = \, 0
\end{equation}
From the physical point of view, the holomorphic equation
\eqref{ceramicus} occurs as the vanishing of the superpotential
derivatives $\partial_i\mathcal{W}(\Phi) \, = \, 0$ while looking
for the scalar potential extrema, namely for the classical vacua of
the gauge theory. From the mathematical viewpoint the locus
$\mathbb{V}_{|\Gamma|+2}$ is the one we start from in order to
calculate the K\"ahler quotient $\mathcal{M}_\zeta$ which
provides the crepant resolution of the singularity. At the end of
the day $\mathcal{M}_\zeta$ is just the manifold of
classical vacua of the gauge theory.

As discussed in \cite{Bruzzo:2017fwj}, the vanishing locus
\eqref{ceramicus}  {consists of several irreducible components} of different
dimensions,  {and $\mathbb{V}_{|\Gamma|+2}=\mathbb{V}_6$ is the only component of dimension 6. It can be
represented in the form $\mathcal{G}_\Gamma\cdot L_\Gamma$, where
$\mathcal{G}_\Gamma$ is the holomorphic quiver group,
defined in the next Section, and $L_\Gamma$ is the three dimensional locus
that we will shortly characterize. An open part of this principal component $\mathbb{V}_6$ can be given by the following explicit equations:}
\begin{eqnarray}\label{bagnomarietta}
  && \Phi^{(2)}_{1,0}\, = \,  \frac{\Phi^{(1)}_{1,0} \Phi^{(2)}_{0,3}}{\Phi^{(1)}_{0,3}},\quad \Phi^{(2)}_{2,1}
   \, = \,  \frac{\Phi^{(1)}_{2,1}
   \Phi^{(2)}_{0,3}}{\Phi^{(1)}_{0,3}},\quad \Phi^{(2)}_{3,2}
   \, = \,  \frac{\Phi^{(1)}_{3,2} \Phi^{(2)}_{0,3}}{\Phi^{(1)}_{0,3}},\nonumber\\
   && \Phi^{(3)}_{1,3}\, = \,
   \frac{\Phi^{(1)}_{1,0} \Phi^{(3)}_{0,2}}{\Phi^{(1)}_{3,2}},\quad
   \Phi^{(3)}_{2,0}\, = \,  \frac{\Phi^{(1)}_{1,0} \Phi^{(1)}_{2,1}
   \Phi^{(3)}_{0,2}}{\Phi^{(1)}_{0,3} \Phi^{(1)}_{3,2}},\quad \Phi^{(3)}_{3,1}\, = \,
   \frac{\Phi^{(1)}_{2,1} \Phi^{(3)}_{0,2}}{\Phi^{(1)}_{0,3}}\end{eqnarray}
\subsection{The holomorphic quiver group $\mathcal{G}_{\mathbb{Z}_4}$ and the gauge group $\mathcal{F}_{\mathbb{Z}_4}$ }
Following the general scheme outlined in \cite{Bruzzo:2017fwj}, we
see that the locus $\mathcal{S}_{\mathbb{Z}_4}$ is mapped into
itself by the action of the \textit{complex quiver group}:
\begin{equation}\label{cromostatico}
    \mathcal{G}_{\mathbb{Z}_4} \, = \, \mathbb{C}^\star \times \mathbb{C}^\star
    \times \mathbb{C}^\star \, \simeq \,\pmb{\Lambda} \, \equiv \, \left(
                                           \begin{array}{c|c|c|c}
                                             \mathbb{C}^\star & 0 & 0 & 0 \\
                                             \hline
                                             0 & \mathbb{C}^\star & 0 & 0 \\
                                             \hline
                                             0 & 0 & \mathbb{C}^\star & 0 \\
                                             \hline
                                             0 & 0 & 0 & \mathbb{C}^\star\\
                                           \end{array}
                                         \right) \quad ; \quad
                                         \mbox{det}\, \pmb{\Lambda}\, =
                                         \, 1
\end{equation}
The gauge group of the final gauge theory is the maximal compact
subgroup of $\mathcal{G}_{\mathbb{Z}_4}$, namely:
\begin{equation}\label{cromodinamico}
    \mathcal{F}_{\mathbb{Z}_4} \, = \, \mathrm{U(1)} \times \mathrm{U(1)}
    \times \mathrm{U(1)} \, \simeq \,\pmb{\Xi} \, \equiv \, \left(
                                           \begin{array}{c|c|c|c}
                                             \mathrm{U(1)} & 0 & 0 & 0 \\
                                             \hline
                                             0 & \mathrm{U(1)} & 0 & 0 \\
                                             \hline
                                             0 & 0 & \mathrm{U(1)} & 0 \\
                                             \hline
                                             0 & 0 & 0 & \mathrm{U(1)}\\
                                           \end{array}
                                         \right) \quad ; \quad
                                         \mbox{det}\, \pmb{\Xi}\, =
                                         \, 1
\end{equation}
The explicit form of the matrices $\pmb{\Lambda}$ and $\pmb{\Xi}$ is
that appropriate to the basis where the regular representation is
diagonalized. In that basis the charge assignments (representation
assignments) of the scalar fields are read off from the
transformation rule:
\begin{equation}\label{caricopendente}
    A(\Phi^\prime) \, = \, \pmb{\Xi}^{-1} \, A(\Phi)\, \pmb{\Xi},
    \quad B(\Phi^\prime) \, = \, \pmb{\Xi}^{-1} \, B(\Phi) \,\pmb{\Xi},
    \quad C(\Phi^\prime) \, = \, \pmb{\Xi}^{-1} \, C(\Phi) \,\pmb{\Xi}
\end{equation}
Then we consider the locus $L_{\mathbb{Z}_4}$ made by those triplets
of matrices $A,B,C$ that belong to $\mathcal{S}_\Gamma$ and are
diagonal in the natural basis of the regular representation. In the
diagonal basis of the regular representation the same matrices
$A,B,C$ have the following form:
\begin{equation}\label{baldovinus}
\begin{array}{ccccccc}
       A_0 & = & \left(
\begin{array}{cccc}
 0 & 0 & 0 & Z^1 \\
 Z^1 & 0 & 0 & 0 \\
 0 & Z^1 & 0 & 0 \\
 0 & 0 & Z^1 & 0 \\
\end{array}
\right) & ; & B_0 & = & \left(\begin{array}{cccc}
 0 & 0 & 0 & Z^2 \\
 Z^2 & 0 & 0 & 0 \\
 0 & Z^2 & 0 & 0 \\
 0 & 0 & Z^2 & 0 \\
\end{array}
\right) \\
\null&\null& \null & \null & \null&\null& \null \\
       C_0 & = & \left(
\begin{array}{cccc}
 0 & 0 & Z^3 & 0 \\
 0 & 0 & 0 & Z^3\\
 Z^3 & 0 & 0 & 0 \\
 0 & Z^3 & 0 & 0 \\
\end{array}
\right) & \null & \null & \null & \null
     \end{array}
\end{equation}
The fields $Z^{1,2,3}$ provide a set of three coordinates spanning
the three-dimensional locus $L_{\mathbb{Z}_4}$. The complete locus
$\mathbb{V}_6$ coincides with the orbit of $L_{\mathbb{Z}_4}$ under
the free action of $\mathcal{G}_{\mathbb{Z}_4}$:
\begin{equation}\label{gestaltus}
    \mathbb{V}_6 \, = \,
{\mathcal{G}_{\mathbb{Z}_4}\cdot L_{\mathbb{Z}_4}}
    \, = \, \left(\begin{array}{c}
                    \pmb{\Lambda}^{-1} \, A_0 \,\,\pmb{\Lambda} \\
                    \pmb{\Lambda}^{-1} \, B_0 \,\,\pmb{\Lambda} \\
                    \pmb{\Lambda}^{-1} \, C_0 \,\,\pmb{\Lambda}
                  \end{array}
     \right)
\end{equation}
\subsection{The moment map equations}
Implementing once more the general procedure outlined in
\cite{Bruzzo:2017fwj} we arrive at the moment map equations and at
the final crepant resolution of the singularity in the following
way.

We refer the reader to \cite{Bruzzo:2017fwj} for the general
definition of the moment map
$$\mu \quad : \quad \mathcal{S}_\Gamma \, \longrightarrow \, \mathbb{F}_\Gamma^\star$$
where $\mathbb{F}_\Gamma^\star$ denotes the dual (as vector spaces)
of the Lie algebra $\mathbb{F}_\Gamma$ of the gauge group. We recall
that the preimage of the level zero moment map is the $\mathcal{F}_\Gamma$
orbit of the locus $L_\Gamma$:
\begin{equation}\label{quadriglia}
  \mu^{-1}\left(0\right) \, = \,
 {\mathcal{F}_\Gamma\cdot L_\Gamma.}
\end{equation}
Note that $L_\Gamma$   is actually  $\mathbb C^3$,
so that the {image of $L_\Gamma$ in} the K\"ahler quotient of level zero
 coincides with the original singular variety
$\mathbb{C}^3/\Gamma$ (cf.~Lemma 3.1 in \cite{kro1}).

Next, as in \cite{Bruzzo:2017fwj}, we  consider the following
decomposition of the Lie quiver group algebra:
\begin{eqnarray}
  \mathbb{G}_{\mathbb{Z}_4} &=& \mathbb{F}_{\mathbb{Z}_4} \oplus
  \mathbb{K}_{\mathbb{Z}_4}\\
  \left[\mathbb{F}_{\mathbb{Z}_4} \, , \,
  \mathbb{F}_{\mathbb{Z}_4}\right] &\subset & \mathbb{F}_{\mathbb{Z}_4}
  \quad ; \quad
\left[\mathbb{F}_{\mathbb{Z}_4} \, , \,
\mathbb{K}_{\mathbb{Z}_4}\right] \,\subset \,
\mathbb{K}_{\mathbb{Z}_4} \quad ; \quad
\left[\mathbb{K}_{\mathbb{Z}_4} \, , \,
\mathbb{K}_{\mathbb{Z}_4}\right] \,\subset \,
\mathbb{F}_{\mathbb{Z}_4} \label{salameiolecco}
\end{eqnarray}
where $\mathbb{F}_{\mathbb{Z}_4}$ is the maximal compact subalgebra.

A special feature  of all the quiver  {g}roups and Lie  {a}lgebras is that
$\mathbb{F}_\Gamma$ and $\mathbb{K}_\Gamma$ have the same real
dimension $|\Gamma|-1$ and one can choose a basis of  {H}ermitian
generators $T^I$ such that:
\begin{equation}\label{sacherdivuli}
    \begin{array}{ccccccc}
       \forall \pmb{\Phi} \in \mathbb{F}_\Gamma & : & \pmb{\Phi} & = &
       {\rm i} \times \sum_{I=1}^{|\Gamma|-1} c_I T^I & ; &
       c_I \in \mathbb{R} \\
       \forall \pmb{K} \in \mathbb{K}_\Gamma & : & \pmb{K} & = &
       \sum_{I=1}^{|\Gamma|-1} b_I T^I & ; &
       b_I \in \mathbb{R} \\
     \end{array}
\end{equation}
Correspondingly a generic element $g\in \mathcal{G}_{\mathbb{Z}_4}$
can be split as follows:
\begin{equation}\label{consolatio}
   \forall g \in \mathcal G_{\mathbb{Z}_4} \quad : \quad g=\mathcal{U} \,
   \mathcal{H} \quad  ; \quad \mathcal{U} \in
   \mathcal{F}_{\mathbb{Z}_4} \quad ; \quad  \mathcal{H} \in
   \exp\left[ \mathbb{K}_{\mathbb{Z}_4}\right]
\end{equation}
Using the above property we arrive at the following parametrization
of the space $\mathbb{V}_6$
\begin{equation}\label{krumiro}
    \mathbb{V}_6 \, = \,
     {\mathcal{F}_{\mathbb{Z}_4}\cdot}
    \left(\exp\left[
    \mathbb{K}_{\mathbb{Z}_4}\right]\cdot L_{\mathbb{Z}_4}\right)
\end{equation}
where, by definition, we have set:
\begin{eqnarray}
  p\in \exp\left[
    \mathbb{K}_{\mathbb{Z}_4}\right]\cdot L_{\mathbb{Z}_4}  &\Rightarrow &
    p=\left\{\exp\left[-\pmb{K}\right]\, A_0
    \exp\left[\pmb{K}\right], \, \exp\left[-\pmb{K}\right]\, B_0\,
    \exp\left[\pmb{K}\right],\, \exp\left[-\pmb{K}\right]\, C_0
    \exp\left[\pmb{K}\right]\right\} \nonumber\\
 \left\{ A_0, \, B_0,\,  C_0\right\} &\in & L_{\mathbb{Z}_4} \nonumber\\
 { \pmb{K}} & { \in}&  { \mathbb{K}_{\mathbb{Z}_4}} \label{centodiquestigiorni}
\end{eqnarray}
In our case the three generators $T^I$ of the real subspace
 {$\mathbb{K}_{\mathbb{Z}_4}$} have been chosen as follows:
\begin{equation}\label{TIgener}
    \begin{array}{ccccccc}
       T^1 & = & \left(
\begin{array}{cccc}
 1 & 0 & 0 & 0 \\
 0 & -1 & 0 & 0 \\
 0 & 0 & 0 & 0 \\
 0 & 0 & 0 & 0 \\
\end{array}
\right) & ; & T^2 & = & \left(
\begin{array}{cccc}
 0 & 0 & 0 & 0 \\
 0 & 1 & 0 & 0 \\
 0 & 0 & -1 & 0 \\
 0 & 0 & 0 & 0 \\
\end{array}
\right) \\
       \null & \null & \null & \null & \null & \null & \null \\
       T^3 & = & \left(
\begin{array}{cccc}
 0 & 0 & 0 & 0 \\
 0 & 0 & 0 & 0 \\
 0 & 0 & 1 & 0 \\
 0 & 0 & 0 & -1 \\
\end{array}
\right) & \null & \null & \null & \null
     \end{array}
\end{equation}
So that the relevant real group element takes the following form:
\begin{equation}\label{sirenus}
    \exp[\pmb{K}] \, = \, \left(
\begin{array}{cccc}
 \mathfrak{H}_1 & 0 & 0 & 0 \\
 0 & \frac{\mathfrak{H}_2}{\mathfrak{H}_1}
   & 0 & 0 \\
 0 & 0 &
   \frac{\mathfrak{H}_3}{\mathfrak{H}_2} &
   0 \\
 0 & 0 & 0 & \frac{1}{\mathfrak{H}_3} \\
\end{array}
\right)
\end{equation}
where $\mathfrak{H}_I$ are, by definition, real. Relying on this, in
the K\"ahler quotient we can invert the order of the operations.
First we quotient the action of the compact gauge group
$\mathcal{F}_{\mathbb{Z}_4}$ and then we implement the moment map
constraints. We have:
\begin{equation}\label{cascapistola}
 \mathbb{V}_6/\!\!/_{\mathcal{F}_{\mathbb{Z}_4}}\, =
  \, {\left(\exp\left [\mathbb{K}_{\mathbb{Z}_4}\right]\cdot L_{\mathbb{Z}_4}\right)/\mathbb{Z}_4,}
\end{equation}
 {where $\mathbb{Z}_4$ acts on $L_{\mathbb{Z}_4}$ via the action induced by that of the stabiliser of $L_{\mathbb{Z}_4}$ in $\mathcal{F}_{\mathbb{Z}_4}$.}
The explicit form of  {the triple} of matrices mentioned in
eq.~\eqref{centodiquestigiorni} is easily calculated on the basis of
eqs.~\eqref{baldovinus} and \eqref{sirenus}
\begin{equation}\label{caglioccio}
    p\, = \,\left(\begin{array}{c}
                    A \\
                    B \\
                    C
                  \end{array}
     \right) \, \equiv \, \left(\begin{array}{c}
                  \exp[-\pmb{K}] \, A_0 \, \exp[\pmb{K}]\\
                  \exp[-\pmb{K}] \, B_0 \, \exp[\pmb{K}]\\
                  \exp[-\pmb{K}] \, C_0 \, \exp[\pmb{K}]
                  \end{array}
     \right)
\end{equation}
The moment map  {is given by:}
\begin{eqnarray}\label{momentidimappa}
\mu\left(p\right)& = & \left\{ \mathfrak{P}_1,\mathfrak{P}_2,\mathfrak{P}_3 \right\} \nonumber\\
    \mathfrak{P}_I & = &\mathrm{Tr}\left[ T_I \, \left(\left[A\,
    , \, A^\dagger\right]\, +\, \left[B\,
    , \, B^\dagger\right]\, +\, \left[C\,
    , \, C^\dagger\right]\right)\right]
\end{eqnarray}
Imposing the moment map constraint we find:
\begin{equation}\label{carampana}
\mu^{-1}\left( \zeta\right)/\!\!/_{\mathcal{F}_{\mathbb{Z}_4}}\, =
\, \left\{ p\, \in \exp\left [\mathbb{K}_{\mathbb{Z}_4}\right]\cdot
L_{\mathbb{Z}_4}\,
\parallel \, \mathfrak{P}_I (p) \,= \, \zeta_I \right\} {/\mathbb{Z}_4.}
\end{equation}
Eq.\,\eqref{carampana} provides an explicit algorithm to calculate
the K\"ahler potential of the final resolved manifold if we are able
to solve the constraints for $\mathfrak{H}_I$ in terms of the triple
of complex coordinates $Z^i$ ($i = 1,2,3$).  Indeed we recall that the
K\"ahler potential $\mathcal{K}_{\mathcal{M}_\zeta}$ of the resolved variety
$\mathcal M_\zeta = \mathcal N_\zeta/\mathcal{F}_{\mathbb{Z}_4}$, where $ \mathcal N_\zeta = \mu^{-1}(\zeta)\subset \mathcal S_\Gamma$,
is given by the
celebrated formula (see \cite[eq.~(3.58)]{HKLR} and also \cite{Billo:1993rd})
\begin{equation}\label{celeberro}
  \mathcal{K}_{\mathcal{M}_\zeta}\, =   \mathcal{K}_{\mathcal{N}_\zeta} \,
  + \zeta_I  \, \mathfrak{C}^{IJ} \,
 \log
  { \mathfrak{H}_J^{2\alpha_{\zeta_J}}} .  \end{equation}
 Here $ \mathcal{K}_{\mathcal{N}_\zeta}$ is the restriction to $\mathcal{N}_\zeta$ of the  K\"ahler potential of the flat K\"ahler metric on $\mathcal S_\Gamma$,
 which is $\mathcal{F}_{\mathbb{Z}_4}$-invariant and therefore can be regarded as a function on $\mathcal{M}_\zeta$.
 The positive rational constants $\alpha_\zeta$ are to be chosen so that the functions $  { \mathfrak{H}_J^{2\alpha_{\zeta_J}}}$
are  hermitian fiber metrics on the three
tautological bundles; these constants are completely determined by the geometry, but it will be easier to fix them later on,
using the fact that Chern characters of the tautological line bundles are a basis of the cohomology ring of $M_\zeta$ \cite{CrawIshii}.
Moreover,
\begin{equation}\label{romualdo}
\mathfrak{C}^{IJ} \, = \,
\mbox{Tr}\left(T^I \, T^J\right ) \, = \, \left(
\begin{array}{ccc}
2 & -1 & 0 \\
-1 & 2 & -1 \\
0 & -1 & 2 \\
\end{array}
\right)
\end{equation}
is the matrix of scalar products of the gauge group generators.

The final outcome of this calculation was already presented in
Section 9 of \cite{Bruzzo:2017fwj}. As it was done there, it is
convenient to consider the following linear combinations:
\begin{equation}\label{innominata}
    \left(
\begin{array}{ccc}
 1 & 0 & -1 \\
 1 & -1 & 1 \\
 0 & 1 & 0 \\
\end{array}
\right)\, \left(\begin{array}{c}
                  \mathfrak{P}_1 -\zeta_1\\
                  \mathfrak{P}_2 -\zeta_2\\
                  \mathfrak{P}_3-\zeta_3
                \end{array}
\right) \, = \, 0
\end{equation}
In this way we obtain:
\begin{eqnarray}
\left(
\begin{array}{c}
-\frac{\left(X_1^2-X_3^2\right) \left(X_1 X_3
 \left(\Delta _1^2+\Delta _2^2\right)+\left(1+X_2^2\right)
 \Delta _3^2\right)}{X_1 X_2 X_3} \\
 \frac{\left(X_2+X_2^3-X_1 X_3 \left(X_1^2+X_3^2\right)\right)
 \left(\Delta _1^2+\Delta _2^2\right)}{X_1 X_2 X_3} \\
 -\frac{\left(-1+X_2^2\right) \left(X_2 \left(\Delta _1^2+\Delta _2^2\right)
 +\left(X_1^2+X_3^2\right)
 \Delta _3^2\right)}{X_1 X_2 X_3} \\
\end{array}
\right) & = & \left(
\begin{array}{c}
 \zeta _1-\zeta _3 \\
 \zeta _1-\zeta _2+\zeta _3 \\
 \zeta _2 \\
\end{array}
\right) \label{sakerdivoli}
\end{eqnarray}
where $\Delta_i \, = \, \mid Z^{ i}\mid$ are the moduli of the three
complex coordinates $Z^i$, and $X_J= \mathfrak{H}_J^2$ for $J=1,2,3$.

Applying the general framework developed in \cite{Bruzzo:2017fwj} we
have
\begin{equation}
\mathcal{H}\text{ }\equiv \text{ }\left(
\begin{array}{|c|c|c|}
\hline
 \mathfrak{H}_1  & 0 & 0  \\
\hline
 0 & \mathfrak{H}_2 & 0 \\
\hline
 0 & 0 &  \mathfrak{H}_{3}  \\
\hline
\end{array}
\right)\label{tautobundmetro}
\end{equation}
and the positive definite hermitian matrix $\mathcal{H}^{2\alpha_{\zeta_J}}$  is the
fiber metric on the direct sum:
\begin{equation}\label{direttosummo}
    \mathcal{R}\,=\,\bigoplus_{I=1}^{r} \, \mathcal{R}_I
\end{equation}
of the $r=3$ tautological bundles that, by construction, are
holomorphic vector bundles with rank equal to the dimensions of the
three nontrivial irreducible representations of $\Gamma$, which in
this case is always one:
\begin{equation}\label{tautibundiEach}
    \mathcal{R}_I \, \stackrel{\pi}{\longrightarrow}\,
    \mathcal{M}_\zeta\quad\quad ;
    \quad \quad\forall p \in \mathcal{M}_\zeta\quad :\quad
    \pi^{-1}(p) \approx \mathbb{C}^{n_I}
\end{equation}
The compatible connection\footnote{Following standard mathematical
nomenclature, we call compatible connection on a holomorphic vector
bundle,  one whose $(0,1)$ part is the Cauchy-Riemann operator of
the bundle.} on the holomorphic vector bundle $\mathcal{R}=\bigoplus_I\mathcal R_I$ is given
by $\vartheta = \bigoplus_I\vartheta_I$,  where
\begin{eqnarray}\label{comancio}
    \vartheta_I = \alpha_{\zeta_I} \,\partial \log {X}_I = \mathcal{H}^{-2{\alpha_{\zeta_I}}} \,\partial\mathcal{H}^{2\alpha_{\zeta_I}}\end{eqnarray}
which is a 1-form with values in $\C$, the Lie algebra of the
structural group $\C^\ast $ of the $I$-th tautological vector
bundle. The natural connection of the $\mathcal{F}_{\mathbb{Z}_4}$
principal bundle, mentioned in eq.\,\eqref{cromodinamico} is just,
according
 {to the} universal scheme, the imaginary part of the
connection $\vartheta$.

In order to solve the system of equations \eqref{sakerdivoli}  it is
convenient to change variables and write:
\begin{eqnarray}\label{lobus}
   && \Sigma \,= \, \Delta_1^2 + \Delta_2^2 \quad ; \quad U \, = \,
   \Delta_3^2
\end{eqnarray}
In this way we obtain:
\begin{equation}
   \left\{ \begin{array}{lcl}
 -U X_2^2 X_1^2-U X_1^2+U X_2^2 X_3^2+U X_3^2-\zeta _1 X_2 X_3 X_1+\zeta _3 X_2 X_3 X_1-\Sigma  X_3 X_1^3
 +\Sigma  X_3^3 X_1 & = &0 \\
 -\zeta _1 X_2 X_3 X_1+\zeta _2 X_2 X_3 X_1-\zeta _3 X_2 X_3 X_1-\Sigma  X_3 X_1^3-\Sigma  X_3^3 X_1+\Sigma  X_2^3
 +\Sigma  X_2 & = &0\\
 -U X_1^2 X_2^2-U X_3^2 X_2^2+U X_1^2+U X_3^2-\zeta _2 X_1 X_3 X_2-\Sigma  X_2^3+\Sigma  X_2& = &0 \\
\end{array}\right. \label{sistemico}
\end{equation}
This is the fundamental algebraic system encoding all information
about the singularity resolution.
\section{Properties of the moment map algebraic system and chamber
structure}\label{GenMap} The resolubility of the system
\eqref{sistemico}, viewed as a set of algebraic equations of higher
order has some very peculiar properties that actually encode the
topology and analytic structure of the resolved manifold $Y$ and of
its possible degenerations. The most relevant property of
\eqref{sistemico} is that for generic values $U>0,\Sigma
>0$ it has always one and only one root where all $X_i > 0$ are real
positive. This is of course expected from the general theory, as we know there
are a well-defined quotient, and a K\"aher metric on the quotient,  for every generic choice of the
{\em level parameters} $\zeta$ (these  are called {\em Fayet-Iliopoulos parameters} in gauge theory,
while in algebraic geometry they correspond to the so-called {\em stability parameters}); but it
has also been verified   numerically at an arbitrary large
collection of random points $\zeta \in \mathbb{R}^3$ and for
an arbitrary large collection of points $\{U,\Sigma\} \in
\mathbb{R}^2_+$. \paragraph{The special surface $\mathcal{S}_2
$.} Although it is not a wall there is in the $\zeta$ space a
planar surface defined by the following conditions
\begin{equation}\label{essedue}
 \mathcal{S}_2 \, \equiv \,  \left\{\zeta_1\, =\, \zeta_3\, =\, a
  , \quad\zeta_2\, =\, b \neq 2a \right\}
\end{equation}
where the algebraic system \eqref{sistemico} acquires a more
manageable form without loosing generality. The  strong
simplification is encoded in the following condition:
\begin{equation}\label{simplicius}
    X_1 \, = \, X_2
\end{equation}
Thanks to eq.\eqref{simplicius}, on the plane $\mathcal{S}_2$ the
moment map system reduces to a system of two rather than three
equations. To understand eq.\eqref{simplicius} we need to recall
some results already obtained in \cite{Bruzzo:2017fwj}. Considering
the system \eqref{sistemico}, it was there observed that one of
the three  functions $X_i$ can always be algebraically solved
in terms of the other two. Indeed we can write:
\begin{eqnarray}\label{caragamba}
    X_3 &=&
   X_1 \sqrt[2]{\frac{\zeta
   _2+\zeta _3 \left(X_2^2-1\right)}{\zeta
   _2+\zeta _1 \left(X_2^2-1\right)}}
\end{eqnarray}
This relation is the algebraic counterpart, in the moment map
equations of the topological result that the homology and cohomology
of the resolved variety $Y$ has dimension $2$. If, inspired by
eq.\eqref{caragamba} we consider the case where all parameters
$\zeta_i$ are different from zero  but two of them, namely $\zeta_1$
and $\zeta_3$ are equal among themselves, \text{i.e.} we localize
our calculations on the surface $\mathcal{S}_2$, then
eq.\eqref{simplicius} follows, yielding a reduced system of moment
map equations:
\begin{equation}\label{grumildus}
    \left(
\begin{array}{c}
 -2 a X_2 X_1^2+b X_2 X_1^2-2 \Sigma  X_1^4+\Sigma
   X_2^3+\Sigma  X_2 \\
 -b X_1^2 X_2-2 U X_1^2 X_2^2+2 U X_1^2-\Sigma  X_2^3+\Sigma
   X_2 \\
\end{array}
\right) \, = \, \left(
                  \begin{array}{c}
                    0 \\
                    0 \\
                  \end{array}
                \right)
\end{equation}
\subsection{Generalities on the chamber structure}
\label{classiwall} In general, the space of    parameters
$\zeta$ (which are closely related to the
{\rm stability parameters} of the GIT quotient construction)\footnote{Geometric
invariant theory, usually shortened into GIT, is the standard way of taking quotients
in algebraic geometry \cite{mumford-GIT,newstead-GIT}. The relation between
the GIT approach and the K\"ahler quotient \`a la Kronheimer in the problem at hand
is explored in \cite{degeratu}.}
has a chamber structure. Let $C$ be a
chamber in that space, and $\mathcal{W}$   a wall of $C$; denote   by
$\mathcal M_C$ the resolution corresponding to a generic $\zeta$ in
$C$ (they are all isomorphic), and by $\mathcal M_{\mathcal{W}}$ the resolution
corresponding to a generic $\zeta$ in ${\mathcal{W}}$. There is a well-defined
morphism $\gamma\,_{\mathcal{W}}\colon \mathcal M_C \to \mathcal M_{\mathcal{W}}$ (actually
one should take the normalization of the second space, but we skip
such details). In general, the morphism $\gamma\,_{\mathcal{W}}$ contracts
curves or divisors;  in \cite{CrawIshii} the walls are classified
according to the nature of the contractions performed by
$\gamma\,_{\mathcal{W}}$. One says that ${\mathcal{W}}$ is of
\begin{enumerate}\item  type 0 if $\gamma\,_{\mathcal{W}}$ is an isomorphism;
\item type I if  $\gamma\,_{\mathcal{W}}$ contracts a curve to a point;
\item type III if  $\gamma\,_{\mathcal{W}}$ contracts a divisor to a curve.
\end{enumerate}
Walls of type II, that should contract a divisor to a point, do not
actually exist, as shown in \cite{CrawIshii}.

The chamber structure pertaining to our example is analyzed and
reconstructed in detail in Section \ref{camerataccademica}. A guide
to the localization of walls and chambers is provided by the
existence of some lines in $\zeta$ space where the system
\eqref{sistemico} becomes solvable by radicals or reduces to a
single algebraic equation. These lines  {either turn out} to be edges
of the convex chambers occuring at   intersections of walls, or
 just belong to  walls. We begin by analyzing such solvable
lines.

\subsection{The solvable lines  located in  $\zeta$ space} In
the $\zeta$ moduli space there are few subcases where the
solution of the algebraic system \eqref{sistemico} can be reduced to
finding the roots of a single algebraic equation whose order is
equal or less than $6$. As anticipated these solvable cases are
located on walls of the chamber and in most case occur at the
intersection of two walls.
\begin{description}
\item[A)] \textbf{Case Cardano  I, $\zeta_1=0, \, \zeta_2=\zeta_3 =s$}. With this choice the general
solution of the system \eqref{sistemico} is provided by  setting the
ansatz displayed  below and by solving the quartic equation for $X$
contained in the next line:
\begin{eqnarray}
    && X_1\, = \, 1, \quad X_2\, = \,X, \quad X_3\, = \, X \nonumber \\
    &&s X^2-U X^4+U-\Sigma  X^3+\Sigma  X \, = \, 0 \label{equatura}
\end{eqnarray}
Obviously we need to choose a branch of the solution such that $X$
is real and positive. As we discuss later on, this is always
possible for all values of $U$ and $\Sigma$ and the required branch
is unique.

To this effect a simple, but very crucial observation is the
following. The arithmetic square root $\sqrt{|s|}$ of the level
parameter $s$ can be used as length scale of the considered space by
rescaling the coordinates as follows: $Z^i \, \to \, \sqrt{|s|}
\,\tilde{Z}^i$ so that equation \eqref{equatura} can be rewritten as
follows
 \begin{equation}\label{baldop}
   -\tilde{U} X^4+U-\tilde{\Sigma}  X^3+  \mathfrak{s}
   \,X^2+\tilde{\Sigma } X \, = \,0
 \end{equation}
where $\mathfrak{s}$ denotes the sign of the moment map level. This
implies that we have only three cases to be studied, namely:
\begin{equation}\label{xenofonte}
   \mathfrak{s} \, = \, \left \{ \begin{array}{c}
                                   1 \\
                                   0\\
                                   -1
                                 \end{array}\right.
\end{equation}
The second case corresponds to the original singular orbifold while
the first and the third yield one instance of what we name the
Cardano manifold. We will see that it corresponds to one of the
possible degenerations of the full resolution $Y$. In the following
we disregard the tildas and we simply write:
\begin{equation}\label{pirettus}
   -{U} X^4+U-{\Sigma}  X^3 \pm
   \,X^2+{\Sigma } X \, = \,0
 \end{equation}
  \item[B)] \textbf{Case Cardano II $\zeta_3=0, \, \zeta_1=\zeta_2 =s$}. With this choice the general
  solution of the system \eqref{sistemico} is provided by  setting the
  ansatz displayed  below and by solving the quartic equation for
  $X$ contained in the next line:
  \begin{eqnarray}\label{raschiotto}
    && X_1\, = \, X,\quad X_2\, = \,X,  \quad X_3\, = \, 1
    \nonumber\\
    &&-s X^2-U X^4+U-\Sigma  X^3+\Sigma  X\, = \, 0
    \label{konigsberg}
  \end{eqnarray}
As one sees eq.~\eqref{konigsberg} can be reduced to the form
\eqref{pirettus} by means of a  rescaling similar to that utilized
in the previous case. All previous conclusions apply to this case
upon the exchange of $X_1$ and $X_3$.
\item[C)] \textbf{Case Eguchi-Hanson  $\zeta_2=0, \, \zeta_1=\zeta_3 =s $}.
With this choice the general
  solution of the system \eqref{sistemico} is provided by  setting the
  ansatz displayed  below and by solving the quartic equation for
  $X$ contained in the next line:
  \begin{eqnarray}\label{raschiotto}
    && X_1\, = \, X,\quad X_2\, = \,1\,  \quad X_3\, = \, X
    \nonumber\\
    &&-2 s X^2+2 \Sigma -2 \Sigma  X^4\, = \, 0 \label{ridiculite}
  \end{eqnarray}
The unique real positive branch of the solution to
eq.~\eqref{ridiculite} is given below:
\begin{equation}\label{segretusquid}
  X\to \frac{\sqrt{\frac{\sqrt{s^2+4 \Sigma ^2}}{\Sigma }-\frac{s}{\Sigma
   }}}{\sqrt{2}}
\end{equation}
We will see in a later Section that eq.\eqref{segretusquid} leads to
a complex three-dimensional manifold that is the tensor product
$\mathrm{EH} \times \mathbb{C}$, having denoted by $\mathrm{EH}$ the
Eguchi-Hanson hyperk\"ahler manifold.
\item[D)] \textbf{Case Kamp\'{e}  $\zeta_2=2s,
\, \zeta_1=\zeta_3 =s$}. With this choice the general solution of
the system \eqref{sistemico} is provided by  setting the ansatz
displayed below and by solving the sextic equation for $X$ contained
in the next line:
\begin{eqnarray}\label{raschiotto}
    && X_1\, = \, \frac{\sqrt[4]{Z^3+Z}}{\sqrt[4]{2}},
    \quad X_2\, = \,Z,  \quad X_3\, = \, \frac{\sqrt[4]{Z^3+Z}}{\sqrt[4]{2}}
    \nonumber\\
    && 2 \left(Z^2+1\right) \left(s Z-U Z^2+U\right)^2-\Sigma ^2 Z
\left(Z^2-1\right)^2\, = \,0
\end{eqnarray}
As in other cases the root of the sextic equation must be chosen
real and positive. Furthermore the absolute value of the parameter
$s$ can be disposed off by means of a rescaling.
\end{description}
\subsection{The K\"ahler potential of the quotient manifolds}
\label{kallusquidam} Before discussing the chamber structure guided
by the discovery of the above mentioned solvable edges $A,B,C,D$ it
is useful to complete the determination of the K\"ahler manifolds
singled out by such edges. This requires considering the explicit
form of the K\"ahler potential for the quotient manifolds. Following
the general rules of the K\"ahler quotient resolution \`a la
{Kronheimer}, as developed in \cite{Bruzzo:2017fwj}, the restriction
of the K\"ahler potential of the linear space $\mathcal{S}_\Gamma =
{\mbox{Hom}_\Gamma\left(R,\mathcal{Q}\otimes R\right)}$ to the
algebraic locus $\mathcal{D}(L_\Gamma )$ and then to the level
surface $\mathcal{N}_\zeta$ is, for the case under
consideration, the following one:
\begin{equation}\label{KelloN}
   \mathcal{K}_0 \mid_{\mathcal{N}_\zeta} \, = \,\frac{U \left(X_2^2+1\right)
   \left(X_1^2+X_3^2\right)+\Sigma
   \left(X_2^3+X_2+X_1\, X_3
   \left(X_1^2+X_3^2\right)\right)}{X_1
   X_2 X_3}
\end{equation}
The complete K\"ahler potential of the resolved variety is given by:
\begin{equation}\label{caramboletta}
    \mathcal{K}_{\mathcal{M}_\zeta} \, = \, \mathcal{K}_0 \mid_{\mathcal{N}_\zeta}
    \, + \alpha_\zeta   \, \zeta_I  \, \mathfrak{C}^{IJ} \,
    \log\left[X_J\right]
\end{equation}

The main point we need to stress is that, depending on the
choices of the moduli $\zeta_I$ (up to rescalings) we can obtain
substantially different manifolds, both topologically and
metrically.

The generic case which  captures the entire algebraic structure of
the resolved variety, to be discussed in later sections by means of
toric geometry, is provided by
\begin{equation}\label{comancho}
    \zeta_1 \neq 0 \quad, \quad \zeta_2 \neq 0 \quad, \quad \zeta_3 \neq 0
\end{equation}
We name the corresponding K\"ahler manifold $Y$. The explicit
calculation of the K\"ahler geometry of the manifold $Y$ is
discussed  in the later Section \ref{Ysezia}, relying on the
particular case $\zeta_1=\zeta_3=\ft 12, \,\, \zeta_2=2$.

For the solvable edges of \textit{moduli space} which we have
classified in the previous Section we have instead the following
results
\begin{description}
  \item[A)] \textbf{Cardano case $\mathcal{M}_{0,1,1}$}. We name
  \textit{Cardano manifold} the one emerging from the choice
  $\zeta_1=0,\, \zeta_2=1, \, \zeta_3 \, = \, 1$ where the
  solution of the moment map equations is reduced to the solution
  of the quartic algebraic equation \eqref{pirettus}. Choosing the
  sign plus in that equation and performing the substitution $X_1 =
  1,\, X_2= X, \, X_3=X$ the K\"ahler potential of the
  Cardano manifold $\mathcal{M}_{0,1,1}$ takes the form:
\begin{equation}\label{KpotCardan}
    \mathcal{K}_{\mathcal{M}_{0,1,1}} \, = \, 2\, {\alpha_\zeta}\, \log{X} \, + \,
    \frac{\left(X^2+1\right) \left(U \left(X^2+1\right)+2 \Sigma  X\right)}{X^2}
\end{equation}
where, depending on the $\Sigma,U$ region, $X$ is the positive real
root of the quartic equation
\begin{equation}\label{baldoppo}
   -U X^4+U-\Sigma  X^3+ X^2+\Sigma  X \, = \,0
\end{equation}
We already argued that this exists and is unique in all regions of
the $\Sigma,U$ plane.
\item[B)] \textbf{Cardano case $\mathcal{M}_{1,1,0}$}. This turns
out to be an identical copy of the previous Cardano manifold. It
emerges from the choice $\zeta_1=1,\, \zeta_2=1, \, \zeta_3 \, = \,
0$ for which the solution of the moment map equations are also
reduced to the solution of the quartic algebraic equation
\eqref{pirettus}. Performing the substitution $X_1 = X,\, X_2= X,
\, X_3=1$ the K\"ahler potential of the Cardano manifold
$\mathcal{M}_{1,1,0}$ takes the form:
\begin{equation}\label{croccus}
\mathcal{K}_{\mathcal{M}_{1,1,0}} \, = \, 2 \, {\alpha_\zeta} \,\log
(X)+\frac{\left(X^2+1\right) \left(U X^2+U+2 \Sigma X\right)}{X^2}
\end{equation}
which is identical with eq. \eqref{KpotCardan} and $X$ is once again
the positive real root of the quartic equation \eqref{baldoppo}.

Let us name $B_i(\Sigma,U)$ the four roots of eq.\eqref{baldoppo}
enumerated in the order chosen by MATHEMATICA to implement Cardano's
formula.  For all the points $\Sigma>0,U>0$  the fourth branch
$B_4(\Sigma,U)$ is the unique  real  positive one.  This property is
visualized in fig. \ref{brancettus}.
\begin{figure}\label{brancettus}
\begin{center}
\includegraphics[height=8cm]{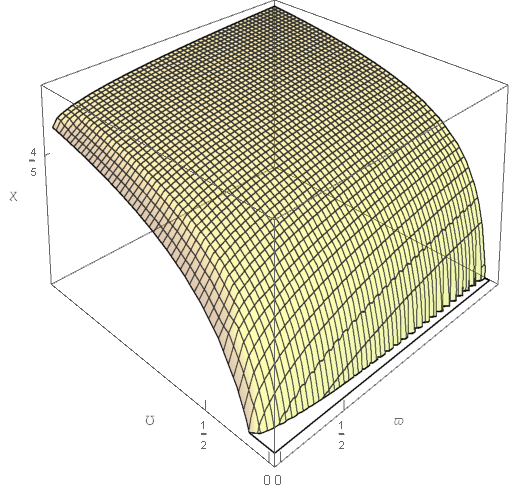}
\caption{\label{brancettus}{\small Plot of the 4th branch of the
solution to the quartic equation \eqref{baldop}. This branch is the
unique real positive one. The surface is plotted with respect to the
new variables $\varpi$ and $\mho$ defined in eq. \eqref{mhovarpi}} }
\end{center}
\end{figure}
Hence let us consider the K\"ahler potential of
eq.~\eqref{KpotCardan} where  $X\to \mathfrak{X}(\varpi,\mho)$ is
the positive real root of the quartic equation \eqref{baldoppo}.

We write the positive real  solution to the quartic equation
\eqref{baldoppo} in terms of two new variables defined below:
\begin{equation}\label{mhovarpi}
    U \, = \, \sqrt{\mho} \quad ; \quad \Sigma\, = \, \sqrt{\varpi}
    \, \sqrt[4]{\mho}
\end{equation}
As the reader will appreciate later on, these variables are
specially prepared to perform the limit to the compact exceptional
divisor and are justified by the toric analysis of Section
\ref{toricresolution}.

The explicit form of $\mathfrak{X}(\varpi,\mho)$ (plotted in
fig.\ref{brancettus}) is the following one:
\begin{eqnarray}
\mathfrak{X}&=&\frac{1}{12 \mho^{1/4}}\, \left(\sqrt{6}
\sqrt{-\frac{3 \sqrt{3} \mathcal{C}}{\sqrt{\frac{\frac{4 \sqrt[3]{2}
\mathcal{A}}{\sqrt[3]{\mathcal{D}}}+2\
   2^{2/3} \sqrt[3]{\mathcal{D}}+3 \varpi -8}{\varpi }}}-\frac{2 \sqrt[3]{2}
   \mathcal{A}}{\sqrt[3]{\mathcal{D}}}-2^{2/3} \sqrt[3]{\mathcal{D}}+3 \varpi -8}\right.\nonumber\\
   &&\left.+\sqrt{3} \sqrt{\frac{4 \sqrt[3]{2}
   \mathcal{A}}{\sqrt[3]{\mathcal{D}}}+2\ 2^{2/3} \sqrt[3]{\mathcal{D}}+3 \varpi -8}-3 \sqrt{\varpi
   }\right)
\label{caponatasiciliana}
\end{eqnarray}
The symbols $\mathcal{A},\mathcal{B},\mathcal{C},\mathcal{D}$
utilized in \eqref{caponatasiciliana} are just shorthands for
certain combinations of the parameters $(\varpi,\mho)$ that we
mention below.
 \begin{eqnarray}
   \mathcal{A} &=& 3 \varpi  \sqrt{\mho }-12 \mho +1\nonumber \\
   \mathcal{B} &=& 9 \varpi  \sqrt{\mho }+72 \mho +2 \nonumber\\
   \mathcal{C} &=& \varpi -8 \sqrt{\mho }-4 \nonumber\\
   \mathcal{D}&=& \sqrt{\mathcal{B}^2-4
   \mathcal{A}^3}+\mathcal{B}\label{princiapisellus}
 \end{eqnarray}
\item[C)] \textbf{Eguchi Hanson case $\mathcal{M}_{s,0,s}$}.
When we choose $\zeta_1=s,\, \zeta_2=0, \, \zeta_3 \, = \,  s$ the
moment map system reduces to eqs.  \eqref{ridiculite}. Performing
the substitution $X_1 = X,\, X_2= 1, \, X_3=X$ and using
eq.~\eqref{segretusquid} the K\"ahler potential of the manifold
$\mathcal{M}_{s,0,s}$ takes the form:
  \begin{equation}\label{canebirbo}
  \mathcal{K}_{\mathcal{M}_{s,0,s}}\, =\underbrace{4\, \alpha\, s \, \log [X] \,+2 \Sigma  X^2+\frac{2 \Sigma
  }{X^2}}_{\mathcal{K}_2}\, +\,4 U
  \end{equation}
What we immediately observe from eq.~\eqref{canebirbo} is that the
K\"ahler potential is of the form:
\begin{equation}\label{cancellino}
    \mathcal{K}_{\mathcal{M}_{s,0,s}}\, =
    \,\underbrace{\mathcal{K}_2\left(Z_1,Z_2,\bar{Z}_1,\bar{Z}_2\right)}_{\text{K\"ahler potential of a two-fold}}\, + \,
    4\times|Z_3|^2
\end{equation}
Hence the manifold $\mathcal{M}_{s,0,s}$ is a direct product:
\begin{equation}\label{sapientulo}
    \mathcal{M}_{s,0,s} \, = \, \mathcal{M}_2 \, \times \, \mathbb{C}
\end{equation}
It is not difficult to realize that the manifold $\mathcal{M}_2$ is
just the Eguchi-Hanson space $EH \, \equiv \, ALE_{\mathbb{Z}_2}$.
To this effect it suffices to set $s= \frac{\ell}{2}$ and rescale
the coordinates $Z^{1,2} \, \to \, \frac{\check{Z}^{1,2}}{2}$. This
implies $\Sigma = \ft 14 |\check{\mathbf{Z}}|^2$ and the K\"ahler
potential $\mathcal{K}_2\left(Z_1,Z_2,\bar{Z}_1,\bar{Z}_2\right)$
turns out to be
\begin{equation}\label{certamino}
    \mathcal{K}_2 \, = \, \sqrt{\ell^2 +|\check{\mathbf{Z}}|^4} \, +
    \, {\alpha_\zeta} \,\ell \, \log
    \left[\frac{-\ell + \sqrt{\ell^2 +|\check{\mathbf{Z}}|^4}}{|\check{\mathbf{Z}}|^2}\right]
    \,
\end{equation}
which is essentially equivalent to the form of the Eguchi-Hanson
K\"ahler potential given  in eq. (7.22) of \cite{Bruzzo:2017fwj}.
This might be already conclusive, yet for later purposes it is
convenient to consider the further development of the result
\eqref{certamino} since the known and fully computable case of the
Eguchi-Hanson space allows us to calibrate  the general formula for
the K\"ahler potential \eqref{celeberro} fixing the value of the so
far unknown parameter $\alpha_\zeta$ that, in this case,
turns out to be $\alpha_\zeta \, = \, 1$. To this effect
recalling the topological structure of the Eguchi Hanson space that is the
total space of the line bundle $\mathcal{O}_{\mathbb{P}^1}(-2)$ we
perform the change of variables:
\begin{equation}\label{ciucius}
    \check{Z}_1 \, = \, u \, \sqrt{v} \quad ; \quad \check{Z}_2 \, = \,\sqrt{v}
\end{equation}
where $u$ is the complex coordinate of the compact base
$\mathbb{P}^1$, while $v$ is the complex coordinate spanning the
non-compact fiber. Upon such a change the K\"ahler potential
\eqref{certamino} becomes:
\begin{equation}\label{cabacchio}
  \mathcal{K}_2 \, = \,  \sqrt{|v|^2   (|u|^2  +1)^2+\ell ^2}\,+\,  {\alpha_\zeta} \,\ell  \log \left(\frac{\sqrt{|v|^2
     (|u|^2  +1)^2+\ell ^2}-\ell }{(|u|^2  +1) \sqrt{|v|^2
    }}\right)
\end{equation}

A further important information can be extracted from the present
case. Setting $v=0$ we perform the reduction to the exceptional
divisor of this partial resolution which is just the base manifold
$\mathbb{P}^1$ of Eguchi-Hanson space. The reduction of the K\"ahler
2-form to this divisor is very simple and it is the following one:
\begin{equation}\label{lampsuco}
\mathbb{K} \mid_{\mathbb{P}^1} \, = \,   \frac{\rho  \sqrt{\ell ^2}
d\rho\wedge d\theta}{\pi \left(\rho
   ^2+1\right)^2}
\end{equation}
where we have set $u=\rho \, \exp[i\,\theta]$. It follows that the
period integral of the K\"ahler 2-form on the unique homology cycle
$C_1$ of the partial resolution $\mathrm{EH}\times
\mathbb{C}$ which is the above mentioned $\mathbb{P}^1$ is:
\begin{equation}\label{califragilisti}
    \int_{C_1} \,\mathbb{K} \, = \, 2\pi \, \int_0^\infty
    \frac{\rho  \sqrt{\ell ^2}
d\rho}{\pi \left(\rho
   ^2+1\right)^2} \, = \, \sqrt{\ell ^2}
\end{equation}
Equation \eqref{califragilisti} sends us two important messages:
\begin{itemize}
  \item Whether the level parameter $s=\ell/2$ is positive or
  negative does not matter.
  \item The absolute value $|s|$ encodes the size of the homology
  cycle in the exceptional divisor. When it vanishes the homology
  cycle shrinks to a point and we have a further degeneration.
\end{itemize}
\item[D)] \textbf{Sextic case or the \textit{Kamp\'{e} manifold}\footnote{Since
it is generally stated that the roots of a general sextic equation
can be written in terms of Kamp\'{e} de  {F\'{e}riet} functions,
although explicit generic formulae are difficult to be found, we
have decided to call  $\mathcal{M}_{s,2s,s}$ the \textit{Kamp\'{e}
manifold} with the same logic that led us to name
$\mathcal{M}_{0,s,s}$ the Cardano manifold.}
 $\mathcal{M}_{s,2s,s}$}.
When we choose $\zeta_1=s,\, \zeta_2=2s, \, \zeta_3 \, = \, s$ the
moment map system reduces to eqs.  \eqref{raschiotto}. With the same
positions used there,
the K\"ahler potential of the $\mathcal{M}_{s,2s,s}$-manifold turns
out to be the following one:
\begin{equation}\label{rodriguez}
   \mathcal{K}_{\mathcal{M}_{1,2,1}} \, = \,  2\, s\,
   \log (Z)\, +\, \frac{\sqrt{2}
   \sqrt{Z^3+Z} \left(\sqrt{2} U \sqrt{Z^3+Z}+2 \Sigma
   Z\right)}{Z^2}
\end{equation}
where the function $Z(\Sigma,U)$ of the complex coordinates is the
positive real root, depending on the $\Sigma,U$ region of the sextic
equation:
\begin{equation}\label{sesticina}
     2\left(Z^2+1\right) \left(s Z-U Z^2+U\right)^2-\Sigma ^2 Z
\left(Z^2-1\right)^2\, = \,0
\end{equation}
\end{description}
In the next Section we discuss the geometry of the crepant
resolution of the singularity $\frac{\mathbb{C}^3}{\mathbb{Z}_4}$
utilizing toric geometry. Then we return to the formulae for the
K\"ahler potential displayed in the present Section in order to see
how the K\"ahler geometry of the entire space and in particular of
the various components of the exceptional divisor is realized in the
various corners of the moduli-space. This will allow us to discuss
the Chamber Structure of this particular instance of K\"ahler
quotient resolution ${\grave{a}}$ la Kronheimer.

As we are going to see, all the four cases analyzed in the present
Section, the two Cardano cases, the Eguchi-Hanson case and the
Kamp\'e case correspond to partial resolutions of the orbifold
singularity and indeed they are located on walls or even on edges
where some homology cycles shrink to zero.  The K\"ahler geometry of
the full resolution will be analyzed in Section \ref{Ysezia}.


\section{Toric geometry description of the crepant resolution}
\label{toricresolution}
As announced above in the present Section we study the full and partial resolutions of
the singularity $\mathbb{C}^3/\mathbb{Z}_4$ in terms of
toric geometry. Both resolutions turn to be the total space of the canonical line bundle over an algebraic surface, the second Hirzebruch surface $\mathbb F_2$ and the
weighted projective plane $\mathbb P[1,1,2]$.
The main output of this study is provided by two
informations:
\begin{enumerate}
  \item The identification as algebraic varieties of the irreducible components of the
  exceptional divisor $\mathcal{D}_E$ introduced by the resolution.
  \item The explicit form of the atlas of coordinate patches that describe the
  resolved manifold and the coordinate transformation from the
  original $Z_i$ to the new $u,v,w$ (appropriate to each patch) that constitute
  the blowup of the singularities.
\end{enumerate}
The second information of the above list and in particular the
equation of the exceptional divisor in each patch is the main tool
that allows to connect the K\"ahler quotient description outlined in
the previous Section with the algebraic description. In particular
by this token we arrive at the determination of the K\"ahler metric
of the exceptional divisor components induced by the Kronheimer
construction.

  \subsection{An initial cautionary remark}
  Let $\Gamma$ be a finite subgroup of $\operatorname{SL}(3,\mathbb{C})$, and let
 let $X_0 = (\mathbb C^3)^{ss}/\!\!/_0\ \Gamma$.\footnote{{For an affine scheme, the GIT quotient is constructed
 as the spectrum of the $\Gamma$-invariant subring of the coordinate ring of the affine scheme.
 In the non-affine case the GIT quotient is obtained by glueing local affine quotients.  It is a {\em categorical} quotient of the open subscheme of $\theta$-semistable points and is denoted by the symbol}  $X^{ss} /\!\!/\!_\theta\,\Gamma$, after choosing a stability parameter $\theta$.
  See e.g.~\cite{mumford-GIT,newstead-GIT}.}
 By general theory (see e.g.~\cite{Hauzer-Langer}) for every generic value of the stability parameter $\theta$ there is a morphism
 $\mathbb C^3/\!\!/_\theta\ \Gamma \to X_0 $. Sardo Infirri \cite[Thm.~4,4 and Rmk.~4.5]{SardoInfirri:1996ga} and Craw-Ishii \cite[Prop.~2.2]{CrawIshii}
notice that there always is a closed embedding $\mathbb C^3/\Gamma \to X_0$ (this makes
$\mathbb C^3/\Gamma $ into an irreducible component of $X_0$), and that this is an isomorphism if and only if $\Gamma$ acts freely away from 0. This happens for   $\mathbb C^2/\Z_n$ both for the standard $\mbox{SU}(2)$ and $\mbox{U}(2)$ actions, and   for the  $\mathbb C^3/\Z_3$ case treated in
\cite{Bruzzo:2017fwj}, but not for the present $\mathbb C^3/\Z_4$ case, where each point of the $z$ axis has a   $\Z_2$ isotropy subgroup. On the other hand, one can shows the existence
of at least one stability chamber such that for generic  values of $\theta$ in that chamber,
$ \mathbb C^3 /\!\!/_\theta\ \Gamma$ is a resolution of singularities of $\mathbb C^3/\Gamma$ \cite{CrawIshii}.
So in that case one has a commutative diagram
$$ \xymatrix{ \mathbb C^3   /\!\!/_\theta\  \Gamma \ar[d]\ar[dr] \\
\mathbb C^3/\Gamma\  \ar@{^{(}->}[r] & X_0
}$$
Actually we shall see in Sections \ref{camerataccademica} and \ref{Periodstaut} that in the present $\mathbb C^3/\Z_4$ case {\em all} stability chambers
correspond to the full resolution of singularities.

  \subsection{The singular variety $Y_0=\mathbb C^3/\mathbb Z_4$}

We shall denote by
$(x,y,z)$ the coordinates of $\mathbb{C}^3$, by $\{\mathbf e_i\}$
the standard basis of $\mathbb R ^3$ and by $\{ \epsilon^i\}$ the
dual basis. The action of $\mathbb Z_4$ is given by
$$(x,y,z) \mapsto (\omega x, \omega y, \omega^2 z )$$
with $\omega^4=1$. It is easy to find a basis for the space of
invariant Laurent polynomials:
\begin{equation}\label{formulauno}
    \mathcal{I}_1 \, = \,x \, y^{-1}, \quad \mathcal{I}_2 \, =
    \,y^2\,z^{-1}, \quad \mathcal{I}_3 \, = \, z^2
\end{equation}
so that the three vectors
$$ \mathbf u^1 = \epsilon^1 -  \epsilon^2, \qquad
\mathbf u^2 =2 \epsilon^2 -  \epsilon^3, \qquad
\mathbf u^3 = 2\epsilon^3
$$
generate the lattice $M$ of invariants, which is a sublattice of the
standard (dual) lattice $M_0$. The lattice $N$ dual to $M$ is a
superlattice of the standard lattice $N_0$, and is generated by the
vectors
\begin{equation}\label{doppievu}
\mathbf w_1 = \mathbf e_1, \qquad
\mathbf w_2 =\tfrac12 \mathbf e_1+\tfrac12 \mathbf e_2, \qquad
\mathbf w_3 =\tfrac14 \mathbf e_1+\tfrac14 \mathbf e_2+\tfrac12 \mathbf e_3.
\end{equation}
The generators of the rays giving the cone associated with the variety $Y_0$
are obtained by inverting these relations, i.e.,
$$\mathbf v_1 = \mathbf w_1, \qquad
\mathbf v_2 = 2  \mathbf w_2 - \mathbf w_1, \qquad
\mathbf v_3 = - \mathbf w_2+2  \mathbf w_3.
$$
From now on, unless differently stated,  coordinate expressions will
always refer to this basis $\{\mathbf w_i\}$
of $N$. So the rays of the fan $\Sigma_0$ of $Y_0$
 are
 $$ \mathbf v_1 = (1,0,0),\qquad \mathbf v_2=(-1,2,0),\qquad \mathbf v_3=(0,-1,2)$$
 They do not form a basis of $N$, according to the fact that
 $Y_0$ is singular (note indeed that that $N/\sum_i\Z\mathbf v_i=\Z_4$).

  \subsection{The full resolution $Y$ of  $Y_0=\mathbb C^3/\mathbb Z_4$}
  In this Section we study the full (smooth) resolution $Y$ of the   singular quotient
$Y_0$, describing its torus-invariant divisors and curves and the natural coordinate systems
on its affine patches.

\subsubsection{The fan}
The  fan of $Y$ is obtained by adding to the fan
$\Sigma_0$ the rays generated by the lattice points lying
on the triangle with vertices $\{\mathbf v_i\}$. These
are
$$\mathbf w_2 = (0,1,0),\qquad \mathbf w_3 = (0,0,1).$$
The  torus invariant divisors corresponding to the two new rays
of the  fan are the components of the exceptional divisor. Since $\mathbf w_3$ is in the interior of the triangle,
the corresponding component of the exceptional divisor is compact, while
the component corresponding to $\mathbf w_2$, which lies on the border,  is noncompact. Note indeed that,
according to the equations \eqref{doppievu}, $\mathbf w_2$ and $\mathbf w_3$
correspond to the junior classes $\frac12(1,1,0)$ (noncompact) and $\frac14(1,1,2)$ (compact)
associated with the given representation of $\mathbb Z_4$.

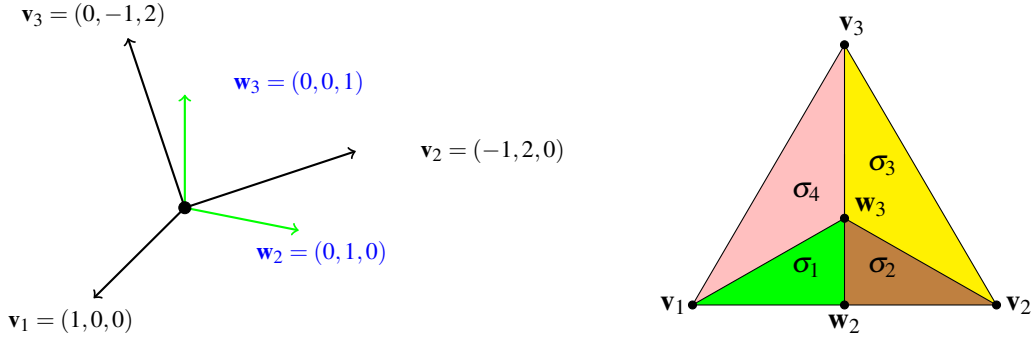
\begin{figure}
\begin{center}
\begin{tikzpicture}[scale=1.50]
\draw [thick,->] (0,0) -- (-0.5,1.5) ;
\draw [thick,->] (0,0) -- (1.5,0.5) ;
\draw [thick,->] (0,0) -- (-0.8,-0.8) ;
\draw [thick,->,green] (0,0) -- (1,-0.2);
\node at (1.2,-0.4){\blu{\footnotesize $\mathbf w_2= (0,1,0)$}};
\node at (-1,-1) {\footnotesize$\mathbf v_1=(1,0,0)$};
\node at (2.7,0.5) {\footnotesize$\mathbf v_2=(-1,2,0)$};
\node at (-0.8,1.7) {\footnotesize$\mathbf v_3=(0,-1,2)$};
\draw [thick,->,green] (0,0) -- (0,1);
\node at (1,1.1){\blu{\footnotesize $\mathbf w_3= (0,0,1)$}};

  \draw [fill] (0,0) circle (1.5pt) ;

\end{tikzpicture}\hskip1cm
\begin{tikzpicture}[scale=0.50]
  \path [fill=pink] (0,0) to (4,2.3) to  (4,6.9) to (0,0) ;
    \path [fill=yellow] (8,0) to (4,2.3) to  (4,6.9) to (8,0) ;
      \path [fill=green] (0,0) to (4,2.3)  to  (4,0) to (0,0);
      \path [fill=brown] (4,0) to (4,2.3)  to  (8,0) to (4,0);
  \draw [fill] (0,0) circle (3pt);
    \draw [fill]  (8,0) circle (3pt);
      \draw [fill] (4,6.9) circle (3pt);
        \draw [fill] (4,2.3) circle (3pt);
\draw (0,0) -- (8,0); \draw (0,0) -- (4,6.9); \draw (8,0) --
(4,6.9); \draw (0,0) -- (4,2.3); \draw (8,0) -- (4,2.3); \draw
(4,6.9) -- (4,2.3); \node at (-0.5,0) {$\mathbf v_1$}; \node at (8.6,0)
{$\mathbf v_2$}; \node at (4.2,7.4)
{$\mathbf v_3$}; \node at
(4.7,2.6) {$\mathbf w_3$};
\node at (3,3) {$\sigma_4$}; \node at (5,3.7) {$\sigma_3$};
\node at
(4,-0.5) {$\mathbf w_2$}; \draw [fill]  (4,0) circle (3pt); \draw
(4,0) -- (4,4);
\node at (5,1) {$\sigma_2$}; \node at (3,1) {$\sigma_1$};
\end{tikzpicture}

 \caption{\label{SigmaY}  \small The fan $\Sigma_Y$ of the resolution $Y$ and the associated planar graph}

\end{center}
\end{figure}

\begin{figure}
\centering
\includegraphics[height=8cm]{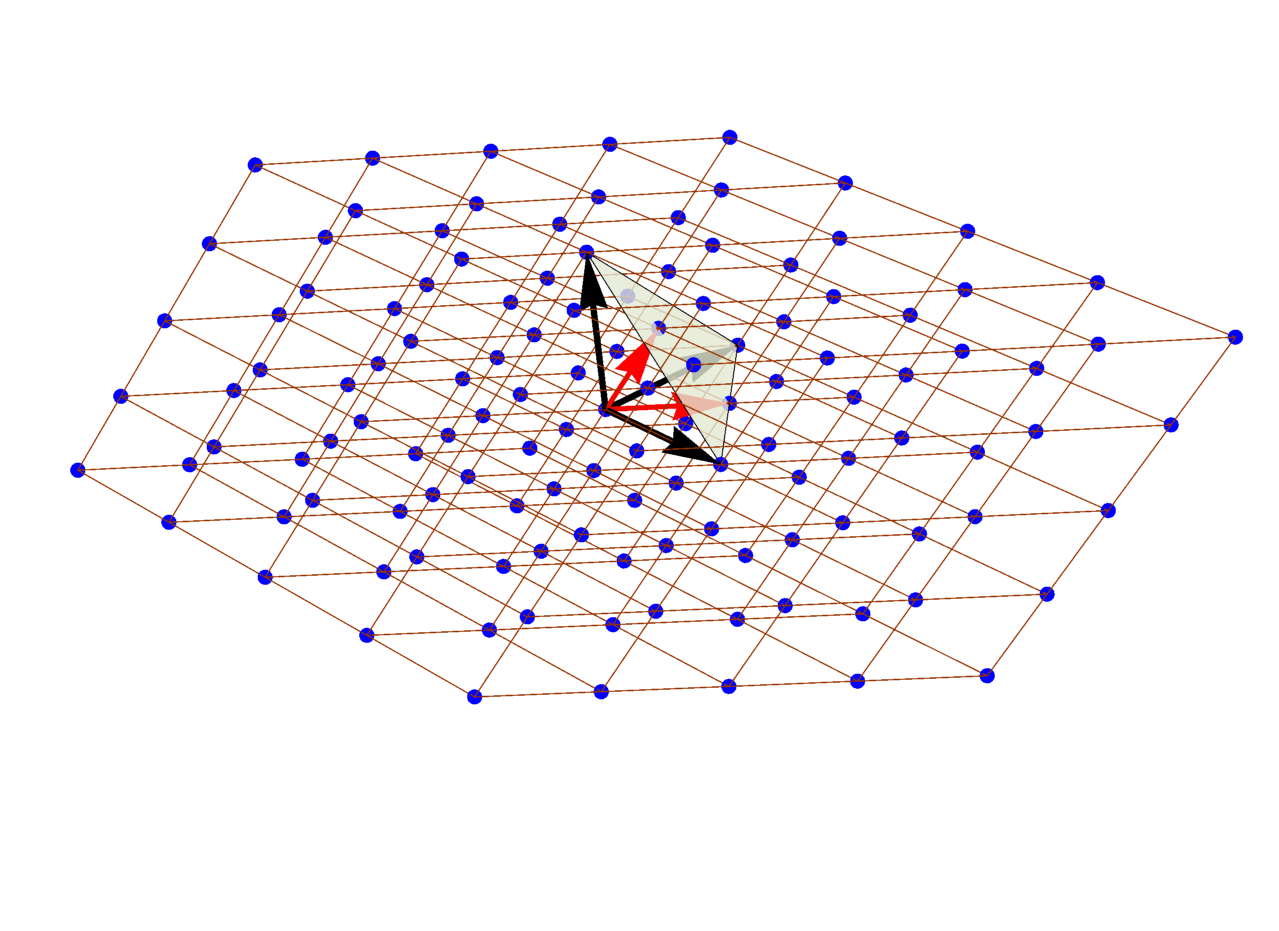}
\vskip -2cm \caption{ \label{z4trian} The figure displays a finite
portion of the lattice $N$ dual {to} the lattice $M$ of
$\mathbb{Z}_4$-invariants and the generators of the cone
$\sigma$ describing the singular quotient  $\mathbb{C}^3/\mathbb Z_4$ marked as fat dark arrows. The
extremal points of the three generators single out a triangle,
which  intersects  the lattice $N$ in two
additional points, namely the extremal points of the vector
$\mathbf w_3$  and of the vector $\mathbf w_2$, marked as lighter
arrows in the figure. These vectors have to be added to the
fan and divide  the original cone into four maximal cones,
corresponding to as many open charts of the resolved smooth toric
variety.}
\end{figure}

We shall denote by $Y$ this resolution of singularities. Figure \ref{SigmaY} shows the fan of $Y$ and the associated
planar graph. The planar graph is obtained by projecting the generators of the rays onto the triangle formed by the three original vertices; this is shown in a 3-dimensional perspective in Figure
\ref{z4trian}.

One can explicitly check that all cones of $\Sigma_Y$ are smooth, so that $Y$ is indeed smooth.
Note that all cones of $\Sigma_Y$ are contained in the cones of $\Sigma_0$, which corresponds to the existence of
a morphism $Y \to  Y_0$.

 If we first make the blowup corresponding to the ray $\mathbf w_3$,
i.e., to the junior class $\frac14(1,1,2)$, according to the general theory the exceptional divisor is
a copy of the weighted projective plane $\mathbb P[1,1,2]$. When we make the second blowup, i.e. we blow up the $z$ axis,
we also blowup $\mathbb P[1,1,2]$ at its singular point, so that the compact component of
the exceptional divisor of the resolution of $Y_0$ is a copy of the second Hirzebruch surface
$\mathbb F_2$. Moreover, the noncompact component of the exceptional divisor is isomorphic to $\mathbb P^1\times\mathbb C$
(which, by the way, turns out to be the weighted projective space $\mathbb P[1,1,0]$). This will be shown in more detail in the next sections   (in particular, the compact exceptional divisor will be characterized as the Hirzebruch surface $\mathbb F_2$ by computing its fan).

By general theory \cite{itoriddo} (see also \cite{Bruzzo:2017fwj}) we know
$$h_2(Y,\Q) = 2, \qquad h^2(Y,\Q)=2, \qquad h^2_c(Y,\Q)=1, \qquad h^4(Y,\Q)=1.$$

 \subsubsection{Divisors}  
\begin{figure}
\begin{center}
\begin{tikzpicture}
\draw [thick,->] (0,0) -- (1,0) ;  \node at (1.2,0) {\small $E_3$};
\draw [thick,->] (0,0) -- (-1,2) ;  \node at (-1.1,2.2) {\small $E_1$};
\draw [thick,->] (0,0) -- (0,-1) ;  \node at (0,-1.4) {\small $E_4$};
\draw [thick,->] (0,0) -- (0,1);  \node at (0,1.2) {\small $E_2$};
  \draw [fill] (0,0) circle (1.5pt) ;
  \node at (0,-2) {$D_c$};
  \node at (0.5,0.5) {$\tau_2$} ;
    \node at (0.5,-0.5) {$\tau_3$} ;
        \node at (-0.7,-0.2) {$\tau_4$} ;
            \node at (-0.4,1.5) {$\tau_1$} ;
\end{tikzpicture}
\hskip5mm
\begin{tikzpicture}
\draw [thick,->] (0,0) -- (1,0) ;
\draw [thick,->] (0,0) -- (-1,0) ;
\draw [thick,->] (0,0) -- (0,1);
  \draw [fill] (0,0) circle (1.5pt) ;
  \node at (0,-1.3) {$D_{nc}$};
\end{tikzpicture}
\hskip5mm
\begin{tikzpicture}
\draw [thick,->] (0,0) -- (1,0) ;
\draw [thick,->] (0,0) -- (-1,2) ;
\draw [thick,->] (0,0) -- (0,1);
  \draw [fill] (0,0) circle (1.5pt) ;
  \node at (0,-1.3) {$D_{EH}$, $D'_{EH}$};
\end{tikzpicture}
\hskip5mm
\begin{tikzpicture}
\draw [thick,->] (0,0) -- (1,0) ;
\draw [thick,->] (0,0) -- (-1,4) ;
\draw [thick,->] (0,0) -- (0,1);
  \draw [fill] (0,0) circle (1.5pt) ;
  \node at (0,-1.3) {$D_4$};
\end{tikzpicture}
\caption{\label{DivY}  \small The fans of the 5 toric divisors of $Y$. In the fan of $D_c$ we have labelled the rays with the names of the corresponding divisors; the $\tau_i$'s are the maximal cones. }

\end{center}
\end{figure}
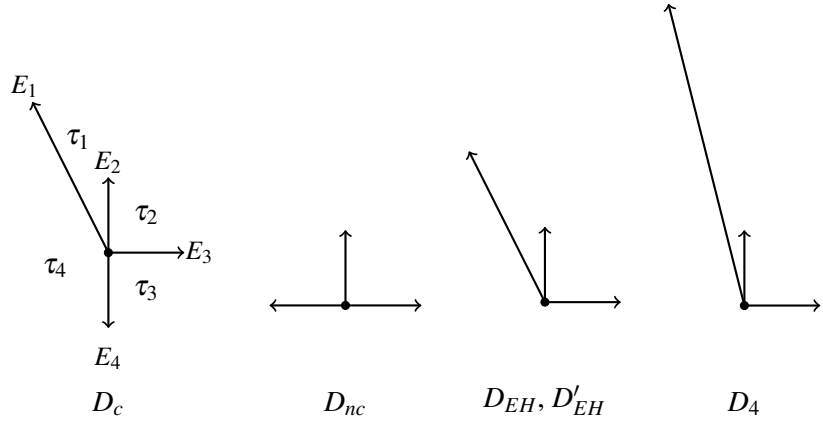

We analyze the toric divisors of $Y$; they are summarized in Table \ref{tableDivY}. Each of these is associated with a ray of the fan $\Sigma_Y$. The divisors corresponding
to $\mathbf w_3$, $\mathbf w_2$, $\mathbf v_1$, $\mathbf v_3$, $\mathbf v_2$ will be denoted $D_c$, $D_{nc}$, $D_{EH}$,
$D_{4}$, $D'_{EH}$ respectively. Since $Y$ is smooth all of them are Cartier. Table \ref{divisors} shows the fans of these divisors and what variety they are as intrinsic varieties. The fans are depicted in Figure \ref{DivY}.\footnote{In the cases when the ray associated to the divisor is not a coordinate axis we made a change of basis.} The fan of $D_c$ is
generated by the rays $\mathbf v_2$, $\mathbf w_2$, $\mathbf v_1$, $\mathbf v_3$, which shows that $D_c$ is the second Hirzebruch surface $\mathbb F_2$.  The corresponding curves in $D_c$ have been denoted $E_1$, $E_2$, $E_3$, $E_4$ respectively. From the self-intersections of these curves (in $D_c$)
$$E_1^2=0, \qquad   E_2^2 = -2,\qquad  E_3^2 = 0, \qquad  E_4^2 = 2  $$
(see \cite[Example 6.4.6]{CoxLS}) we see that $E_2$ is the section of
$\mathbb F_2 \to \mathbb P^1$ which squares to $-2$, i.e., the exceptional divisor of the blowup
$\mathbb F_2 \to \mathbb P[1,1,2]$, while $E_4$ is the section that squares to 2, and $E_1$, $E_3$ are the toric fibers of $\mathbb F_2 \to \mathbb P^1$.

\begin{table}[ht]
\renewcommand{\arraystretch}{1.50}
\caption{Toric Divisors in $Y$. The last column shows the components of the divisor class on the basis given by $(D_{nc}, D_c)$.   The variety $\mbox{ALE}_{A_1}$  is the Eguchi-Hanson space,
the crepant resolution of the singular space $\mathbb C^2/\Z_2$.
\label{tableDivY}}
\vskip10pt
\centering 
\begin{tabular}{|c |c| c| c| c |} 
 \hline 
Ray & Divisor   & Fan  & Variety & Components \\ [1ex] 
\hline 
$\mathbf w_3$ & $D_c$  & \small $(1,0),\ (-1,2),\ (0,-1),\ (0,1)$ & $\mathbb F_2$ & $ (0,1)$\\   \hline 
$\mathbf w_2$ & $D_{nc}$ & \small  $(1,0),\ (-1,0),\ (0,1)$ & $\mathbb P^1\times\mathbb C$ & $(1,0)$ \\  \hline
$\mathbf v_1$ & $D_{EH}$  & $ (1,0),\ (-1,2),\ (0,1) $&  $\mbox{ALE}_{A_1}$ & $(-\frac12,-\frac14)$  \\  \hline
$\mathbf v_3$ & $D_{4}$ &  $ (1,0),\ (-1,4),\ (0,1) $ & \small tot  $(\mathcal O(-4) \to \mathbb P^1$) & $(0,-\frac12)$\\  \hline
$\mathbf v_2$ & $D'_{EH}$ & $ (1,0),\ (-1,2),\ (0,1) $ &  $\mbox{ALE}_{A_1}$  & $(-\frac12,-\frac14)$ \\     [1ex] 
\hline 
\end{tabular}
\label{divisors} 
\end{table}

Among the 5 divisors only 2 are independent in cohomology.  We consider  the exact sequence \cite[Thm.~4.2.1]{CoxLS}
\begin{equation}\label{Pic}
0 \to M \xrightarrow{A} \operatorname{Div}_{\mathbb T}(Y) \xrightarrow{B}   \operatorname{Pic}(Y) \to  0
\end{equation}
where $M$ is the dual lattice, $ \operatorname{Div}_{\mathbb T}(Y) $ is the group of torus-invariant divisors,
and $\operatorname{Pic}(Y) $ is the Picard group\footnote{The Picard group $\operatorname{Pic}(X)$ of a complex variety $X$ is the group of isomorphism classes of holomorphic line bundles on $X$. Using \v Cech cohomology it can be represented as the cohomology group $H^1(X,\cO_X^\ast)$, where $
\cO_X^\ast$ is the sheaf of nowhere vanishing holomorphic functions on $X$.} of the full resolution $Y$. The morphism $B$ simply takes the
class of a divisor in the Picard group, while for every $m\in M$, $A(m)$ is the divisor associated with the rational
function defined by $m$.
Moreover we know that the classes of the divisors $D_{nc}$ and $D_{c}$ generate $\operatorname{Pic}(Y) $ over $\Q$ \cite{itoriddo}.
With that choice of basis in $\operatorname{Pic}(Y) \otimes\Q$, with the basis given by the 5 divisors in $\operatorname{Div}_{\mathbb T}(Y)  \otimes\Q$, and the basis in $M\otimes\Q$ given by the duals of the $\{\mathbf w_i\}$, the morphisms $A$ and $B$ are represented over the rationals  by the matrices
$$ A =\begin{pmatrix} 1 & 0 & 0 \\ -1 & 2 & 0 \\ 0 & -1 & 2 \\ 0 & 1 & 0 \\ 0 & 0 & 1 \end{pmatrix},\qquad
B =\begin{pmatrix}  -\tfrac12 &  -\tfrac12 & 0 & 1 & 0 \\
 -\tfrac14 &  -\tfrac14 &  -\tfrac12 & 0 & 1
\end{pmatrix}
$$
We deduce that the relations among the classes of the 5 toric divisors in the Picard group are\footnote{The notation $[D]$ means the class in the Picard group of the line bundle $\cO_X(D)$.}
\begin{gather} [D_{EH}]  =  [D'_{EH}]=  -\tfrac12 [D_{nc}] - \tfrac14 [D_c] \label{Div1} \\
{[}D_{4} ]   =  -\tfrac12 [D_c] \label{eqDiv2}
\end{gather}
Since the canonical divisor can be written as minus the sum of the torus-invariant divisors,
one has
$$ [ K_Y] = - [D_{nc}] - [D_c] - [D_{EH}] - [D'_{EH}] - [D_4] = 0 $$
consistently with the fact that the resolution $Y \to Y_0$ is crepant. (Note that $\operatorname{Pic}(Y)$ is free
over $\Z$ \cite[Prop.~4.2.5]{CoxLS}, so that the equality $ [ K_Y] =0$ in $\operatorname{Pic}(Y) \otimes\Q$ also implies that $ [ K_Y] =0$  in  $\operatorname{Pic}(Y)$).

The matrix $B$ can also be chosen as
$$ B = \begin{pmatrix}  1 & 1 & 0 & -2 & 0 \\
0 & 0 &  1 & 1 & -2 \end{pmatrix} $$
which corresponds to taking the classes of $D_{EH}$ and $D_4$ as basis of the Picard group.
In this way the matrix $B$ is integral, which means that $D_{EH}$ and $D_4$ generate
$\operatorname{Pic}(Y)$ over the integers.

\subsubsection{Toric curves and intersections}  \label{curvesY}
The planar graph in Figure \ref{SigmaY} shows that $Y$ has 4 compact toric curves, corresponding to the inner edges of the graph.
The intersection between a smooth irreducible curve $C$ and a Cartier divisor $D$ is defined as
$$ C\cdot D = \deg (f^\ast \cO_Y(D))$$
where $f\colon C \to Y$ is the embedding, and $\cO_Y(D)$ is the line bundle associated to the divisor\footnote{This also allows one to compute the intersection between a curve $C$ and a Weil divisor $D$. Indeed simplicial toric varieties are $\Q$-factorial, i.e., every Weil divisor has a multiple that is Cartier. So if $mD$ is Cartier, one defines
$$ C \cdot D = \tfrac1m \,C\cdot(mD).$$ This may be a rational number. \label{nota}}
(we shall use this definition in Section \ref{Periodstaut} to compute the periods of the tautological line bundles).
Inspection of the fan allows one to detect when the intersection is transversal (in which case the intersection mumber is 1), empty (intersection number 0), or the curve is inside the divisor.
The intersections are shown in Table \ref{tableintersec}.

\begin{table}[ht]
\caption{Intersections among the   toric curves and divisors in the full resolution $Y$. For the curves that are inside $D_c$ the last two columns also show the identifications with the curves corresponding to the rays in the fan of $D_c$ of Figure \ref{DivY}, and what they are inside the second Hirzebruch surface. Note that $C_1$ is the intersection between the two components of the exceptional divisor.
The basis of the fibration $D_c \to \mathbb P^1$ may be identified with $C_1$, while $C_2$, $C_4$
are the fibers over two toric points, which correspond to the cones $\sigma_1$ and $\sigma_2$.
\label{tableintersec}} 
\vskip10pt
\centering 
\begin{tabular}{|c|c |c| c| c| c| c| c| c| } 
 \hline 
 Edge/face & Curve & $D_c$ & $D_{nc}$   & $D_{EH}$   & $D'_{EH}$ & $D_4$ & Inside $D_c$ &   \\ \hline
 $(\mathbf w_2\mathbf w_3)$ & $C_1$ & 0 & - 2 & 1 & 1 & 0 & $ E_2$ & $-2$-section \\ \hline
 $(\mathbf v_1\mathbf w_3)$ & $C_2$ & -2 & 1 & 0 & 0 & 1 & $E_3$ & fiber \\ \hline
  $(\mathbf v_2\mathbf w_3)$ & $C_4$ & -2 & 1 & 0 & 0 & 1 & $E_1$ & fiber \\ \hline
   $(\mathbf v_3\mathbf w_3)$ & $C_5$ & -4 & 0 & 1 & 1 & 2 & $E_4$ & 2-section\\ \hline
    $(\mathbf v_1\mathbf w_2)$ & $C_3$ & 1 & 0 &  0  & 0 &  0 & &  \\
 \hline 
\end{tabular}
\label{intersections} 
\end{table}

The intersection numbers of $D_{EH}$ and $D_4$ with $C_1$, $C_2$ show that
the latter are a basis of $H_2(Y,\Z)$ dual to $\{[D_4],[D_{EH}]\}$.

\subsubsection{Coordinate systems and curves}\label{coorcurves}
The four 3-dimensional cones in the fan of $Y$ correspond to four affine open varieties, and since all cones are smooth (basic), they are copies of $\mathbb C^2$. The variables attached to the rays generating a cone provide a coordinate system on the corresponding affine set. A face between two 3-dimensional cones corresponds to the intersection between the two open sets. Note that all charts have a common intersection,
as they all contain the 3-dimensional torus corresponding to the origin of the fan.
Table \ref{coordinates} shows the association among cones, rays, coordinates and coordinate expressions of toric  curves.
Below we provide a list of  {coordinate systems}, with all transition functions between them, and the expressions of the toric curves in the coordinate systems of the charts they belong to.
Table \ref{changes} displays the coordinate transformations among the four coordinate systems.
We have denoted
$C_i$, $i=1\dots 5$ as before, and moreover $C_6$, $C_7$, $C_8$ are the noncompact toric curves in the charts 2, 3 and 4 (analogously, $C_3$ was the noncompact curve in the chart 1). The column ``Dual gen.'' displays the generators of the dual cone.

In each chart, the coordinates $(u,v,w)$ are related to the invariants  \eqref{formulauno} as follows:
\begin{equation}\label{settorio}
    \begin{array}{lclcrcccccl}
\left\{u,v,w\right\}_1 &=&\{&\frac{x}{y} &,& \frac{y^2}{z} &,&
   z^2&\} \\
\left\{u,v,w\right\}_2 &=&\{&\frac{y}{x} &,& \frac{x^2}{z} &,&
   z^2&\} \\
\left\{u,v,w\right\}_3&=&\{& \frac{y}{x} &,& \frac{z}{x^2} &,&
   x^4 &\}\\
\left\{u,v,w\right\}_4 &=&\{&\frac{x}{y} &,& y^4 &,&
   \frac{z}{y^2}&\} \\
\end{array}
\end{equation}
Eqs. ~\eqref{settorio} can be easily inverted and one obtains:
\begin{equation}\label{subberulle}
\begin{array}{clclcl}
\mbox{Chart } X_{\sigma_1}& x\to u \sqrt{v} \sqrt[4]{w} &,&
   y\to \sqrt{v} \sqrt[4]{w} &,&
   z\to \sqrt{w} \\
\mbox{Chart } X_{\sigma_2}& x\to \sqrt{v} \sqrt[4]{w} &,&
   y\to u \sqrt{v} \sqrt[4]{w}
   &,& z\to \sqrt{w} \\
\mbox{Chart } X_{\sigma_3}& x\to \sqrt[4]{w} &,& y\to u
   \sqrt[4]{w} &,& z\to v
   \sqrt{w} \\
\mbox{Chart } X_{\sigma_4}& x\to u \sqrt[4]{v} &,& y\to
   \sqrt[4]{v} &,& z\to \sqrt{v}
   w \\
\end{array}
\end{equation}
The irrational coordinate transformations \eqref{subberulle} derived
from the toric construction are the essential tool to relate the
results of the K\"ahler quotient construction with the geometry of
the exceptional divisor as identified by the toric resolution of the
singularity.

The coordinates $x,y,z$ in the above equation are to be identified
with the $Z^{1,2,3}$ that parameterize the locus $L_{\mathbb{Z}_4}$
composed by the matrices $A_0,B_0,C_0$ of eq.~\eqref{baldovinus}. As
we know  this locus is lifted to the resolved variety $Y$ by the
action of the quiver group element $\exp [\pmb{\Phi}]$, whose
corresponding Lie algebra element $\pmb{\Phi}$ satisfies the moment
map equations \eqref{sakerdivoli}.

\newcommand{\eps}[1]{\epsilon_{#1}}

\begin{table}[ht] \small
\caption{For each cone the table assigns a name to the coordinates associated to the rays. The third column lists the generators of the dual cones. The $\epsilon$'s here are the dual basis to the $\mathbf w$. We also write the equations of the toric curves in these coordinates.}
\vskip10pt
\centering 
\begin{tabular}{|c|c |c| c| c| } 
 \hline 
 Cone & Rays & Dual gen. & Coordinates & Curves \\ \hline
 $\sigma_1$ &$ \mathbf v_1  \mathbf w_2  \mathbf  w_3 $ & $\epsilon_1,\epsilon_2,\epsilon_3 $ &
 $u_1,v_1,w_1$ & $C_1: v_1=w_1=0, \ C_2: u_1=w_1=0,\ C_3 = u_1 = v_1 =0 $\\ \hline
 $\sigma_2$ & $ \mathbf v_2  \mathbf w_2  \mathbf  w_3 $  &$ -\eps1,\eps2+2\eps1,\eps3$ &
 $u_2,v_2,w_2$ & $ C_1: v_2=w_2=0, \  C_4: u_2=v_2=0, \ C_6 : u_2 = w_2=0 $\\
 \hline
 $\sigma_3$ &  $ \mathbf v_2  \mathbf v_3  \mathbf  w_3 $  &$-\eps1,-2\eps1-\eps2,\eps3+2\eps2+4\eps1$ &
 $u_3,v_3,w_3$ & $C_4: u_3=w_3=0, \ C_5: v_3=w_3=0, C_7: u_3=v_3=0 $ \\
 \hline
 $\sigma_4$ & $ \mathbf v_1  \mathbf v_3  \mathbf  w_3 $ &$\eps1,2\eps2+\eps3,-\eps2$  &
 $u_4,v_4,w_4$ & $C_2: u_4=w_4=0,\ C_5: v_4=w_4=0, C_8: u_4=v_4=0$ \\
 \hline
\end{tabular}
\label{coordinates} 
\end{table}

\begin{table}[ht] \footnotesize
\hskip-2mm\parbox{\textwidth}{
\caption{Coordinate changes among the charts described in Table \ref{coordinates}}
\vskip10pt
\centering 
\begin{tabular}{|c|c |c| c| c| } 
 \hline 
 & $\sigma_1$  & $\sigma_2$ & $\sigma_3$ & $\sigma_4$ \\ \hline
$\sigma_1$ & id  & $ u_2=\frac1{u_1}, v_2 = u_1^2v_1, w_2=w_1 $  & $ u_3=\frac1{u_1}, v_3 = \frac{1}{u_1^2v_1}, w_3 = u_1v_1^2w_1 $ &
$u_4=u_1, \ v_4 = v_1^2w_1,\ w_4=\frac1{v_1}$ \\ \hline
$\sigma_2$ & $u_1 = \frac1{u_2},v_1=u_2^2v_2,w_1=w_2 $ & id & $u_3=u_2,v_3=\frac1{v_2},w_3=v_2^2w_2$ &
$u_4=\frac1{u_2},v_4=u_2^4v_2^2w_2,w_4=\frac1{u_2^2v_2}$\\ \hline
$\sigma_3$ &$ u_1=\frac1{u_3},v_1=\frac{u_3^2}{v_3},w_1=v_3^2w_3 $ & $u_2=u_3,v_2=\frac1{v_3},w_2={v_3^2}{w_3}$& id &
$u_4=\frac1{u_3},v_4=u_3^4w_3,w_4=\frac{v_3}{u_3^2}$\\ \hline
$\sigma_4$ & $ u_1=u_4,v_1=\frac1{w_4},w_1=v_4w_4^2$& $u_2=\frac1{u_4},v_2=\frac{u_2^4}{w_4},w_2=v_4w_4^2$&$u_3=\frac1{u_4},v_3=
\frac{w_4}{u_4^2},w_3=u_4^4v_4$& id \\ \hline
\end{tabular}
\label{changes} 
}
\end{table}

\subsubsection{$Y$ as a line bundle on $\mathbb F_2$}\label{Ylinebundle}
The full resolution $Y$ is the total space
of the canonical bundle of the second Hirzebruch surface $\mathbb F_2$; this is quite clear
from the blowup procedure (cf. Section 2.3 in \cite{itoriddo})
and was
explicitly noted in \cite{bouchard}.
Following \cite[\S 7.3]{CoxLS} we give a toric description of
this fact. The canonical bundle of $\mathbb F_2$ is
 the line bundle $\cO_{\mathbb F_2}(-2H)$, where $H$ is the section of $\mathbb F_2\to\mathbb P^1$ squaring to 2.
We regard $H$ as the toric divisor $E_4$, see Figure \ref{DivY}. To each of the cones $\tau_i$ of the fan
of $\mathbb F_2$
one associates a 3-dimensional cone $\tilde\sigma_i$, obtaining
\begin{eqnarray*}
\tilde\sigma_1 &=& \mbox{Cone}((0,0,1),(0,1,0),(1,0,0)) \\
\tilde\sigma_2 &=& \mbox{Cone}((0,0,1),(-1,2,0),(0,1,0)) \\
\tilde\sigma_3 &=& \mbox{Cone}((0,0,1),(0,-1,2),(-1,2,0)) \\
\tilde\sigma_4 &=& \mbox{Cone}((0,0,1),(1,0,0),(0,-1,2))
\end{eqnarray*}
This is the fan of $Y$. So,
$ \mbox{tot}(\cO_{\mathbb F_2}(-2H)) \simeq Y$,
i.e., $Y$ is the total space of the canonical bundle of $\mathbb F_2$.
Thus, the canonical bundle of $Y$ is trivial.\footnote{Let $L$ be the total space
of $\cO_{\mathbb F_2}(-2E)$, with projection $\pi\colon L\to \mathbb F_2$. Then we have an exact sequence
$$ 0 \to \pi^\ast\Omega^1_{\mathbb F_2} \to \Omega^1_L \to \Omega^1_{L/\mathbb F_2} \to 0.$$
The bundle of relative differentials $ \Omega^1_{L/\mathbb F_2}$ is  isomorphic
to $\pi^\ast L^\ast$. As a result,
$$ K_L = \det (\Omega^1_L ) \simeq \pi^\ast K_{\mathbb F_2}\otimes \pi^\ast L^\ast
\simeq   \pi^\ast\cO_{\mathbb F_2}(-2H) \otimes  \pi^\ast\cO_{\mathbb F_2}(2H) \simeq \cO_L.$$}
This will be the key to the computations of the K\"ahler potential of $Y$, of its K\"ahler metric,
and of the K\"ahler 2-form integrals on the homology cycles that
we shall perform in Section \ref{Ysezia}.

\subsection{The partial resolution $Y_3$} \label{Y3sezia}
The full resolution $Y$ is obtained by adding two rays to the fan of $Y_0=\mathbb C^3/\Z_4$. If we add just one we obtain a partial resolution. Here we examine the partial resolution that will occur in correspondence of some walls of the stability parameters space.

\subsubsection{The fan}
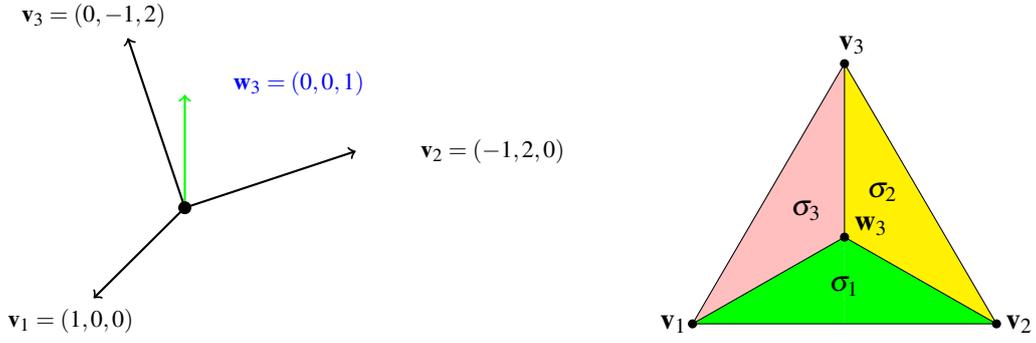
\begin{figure}
\begin{center}
\begin{tikzpicture}[scale=1.50]
\draw [thick,->] (0,0) -- (-0.5,1.5) ;
\draw [thick,->] (0,0) -- (1.5,0.5) ;
\draw [thick,->] (0,0) -- (-0.8,-0.8) ;
\draw [thick,->,green] (0,0) -- (0,1);
\node at (1,1.1){\blu{\footnotesize $\mathbf w_3= (0,0,1)$}};
\node at (-1,-1) {\footnotesize$\mathbf v_1=(1,0,0)$};
\node at (2.7,0.5) {\footnotesize$\mathbf v_2=(-1,2,0)$};
\node at (-0.8,1.7) {\footnotesize$\mathbf v_3=(0,-1,2)$};

  \draw [fill] (0,0) circle (1.5pt) ;

\end{tikzpicture}\hskip1cm
\begin{tikzpicture}[scale=0.50]
  \path [fill=pink] (0,0) to (4,2.3) to  (4,6.9) to (0,0) ;
    \path [fill=yellow] (8,0) to (4,2.3) to  (4,6.9) to (8,0) ;
      \path [fill=green] (0,0) to (4,2.3)  to  (4,0) to (0,0);
      \path [fill=green] (4,0) to (4,2.3)  to  (8,0) to (4,0);
  \draw [fill] (0,0) circle (3pt);
    \draw [fill]  (8,0) circle (3pt);
      \draw [fill] (4,6.9) circle (3pt);
        \draw [fill] (4,2.3) circle (3pt);
\draw (0,0) -- (8,0); \draw (0,0) -- (4,6.9); \draw (8,0) --
(4,6.9); \draw (0,0) -- (4,2.3); \draw (8,0) -- (4,2.3); \draw
(4,6.9) -- (4,2.3); \node at (-0.5,0) {$\mathbf v_1$}; \node at (8.6,0)
{$\mathbf v_2$}; \node at (4.2,7.4)
{$\mathbf v_3$}; \node at
(4.7,2.6) {$\mathbf w_3$};
\node at (3,3) {$\sigma_3$}; \node at (5,3.5) {$\sigma_2$};
\node at (4,1) {$\sigma_1$};
\end{tikzpicture}

 \caption{\label{Sigma3}  \small The fan $\Sigma_3$ of the partial resolution $Y_3$ and the associated planar graph}

\end{center}
\end{figure}

We consider the toric 3-fold $Y_3$ whose fan $\Sigma_3$ is generated by the 4 rays $\mathbf v_1$,  $\mathbf v_2$, $\mathbf v_3$,
$\mathbf w_3$, which is a partial resolution of $Y_0$.  This will appear
as the partial desingularization occuring at some of the walls of the $\zeta$ parameter space (space of stability conditions). The fan and the associated planar graph are shown in Figure \ref{Sigma3}. The cone $\sigma_1$ is singular, while $\sigma_2$ and $\sigma_3$ are smooth, i.e., $Y_3$ has one singular toric point.
By general theory we know that
$$ h_2(Y_3,\Q) = 2, \qquad h^2(Y_3,\Q) = h^2_c(Y_3,\Q) = h^4(Y_3,\Q) = h^2(Y_3,\Q) =1.$$

\subsubsection{Divisors}  
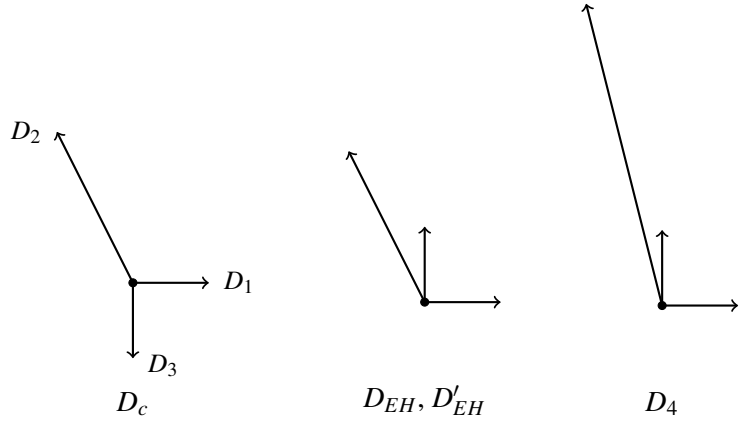
\begin{figure}
\begin{center}
\begin{tikzpicture}
\draw [thick,->] (0,0) -- (1,0) ;
\draw [thick,->] (0,0) -- (-1,2) ;
\draw [thick,->] (0,0) -- (0,-1) ;
  \draw [fill] (0,0) circle (1.5pt) ;
  \node at (0,-1.6) {$D_c$};
  \node at (0.4,-1.1) {\small $D_3$};
    \node at (1.4,0) {\small $D_1$};
    \node at (-1.4,2) {\small $D_2$};
\end{tikzpicture}
\hskip10mm
\begin{tikzpicture}
\draw [thick,->] (0,0) -- (1,0) ;
\draw [thick,->] (0,0) -- (-1,2) ;
\draw [thick,->] (0,0) -- (0,1);
  \draw [fill] (0,0) circle (1.5pt) ;
  \node at (0,-1.3) {$D_{EH}$, $D'_{EH}$};
\end{tikzpicture}
\hskip10mm
\begin{tikzpicture}
\draw [thick,->] (0,0) -- (1,0) ;
\draw [thick,->] (0,0) -- (-1,4) ;
\draw [thick,->] (0,0) -- (0,1);
  \draw [fill] (0,0) circle (1.5pt) ;
  \node at (0,-1.3) {$D_4$};
\end{tikzpicture}
\caption{\label{Div3}  \small The fans of the 4 toric divisors of $Y_3$}

\end{center}
\end{figure}

The divisors corresponding
to $\mathbf w_3$, $\mathbf v_1$, $\mathbf v_3$, $\mathbf v_2$ will be denoted $D_c$,  $D_{EH}$,
 $D'_{EH}$, $D_{4}$, respectively.  They are described in Table \ref{divisors3}. The corresponding fans are shown in Figure \ref{Div3}.

 For the variety $Y_3$, which is not smooth, the Picard group in the sequence \eqref{Pic} must be replaced by the class group $\operatorname{Cl}(Y_3)$,
 however after tensoring by the rationals the two groups coincide, so that we may ignore this fact.
 The group $\operatorname{Pic}(Y_3)\otimes \Q$ is generated by the class of $D_c$.
 The matrices $A$ and $B$ are now
  $$ A =\begin{pmatrix} 1 & 0 & 0 \\ -1 & 2 & 0 \\ 0 & -1 & 2  \\ 0 & 0 & 1 \end{pmatrix},\qquad
B =\begin{pmatrix}
 -\tfrac14 &  -\tfrac14 &  -\tfrac12  & 1
\end{pmatrix}
$$
(the order of the toric divisors is $D_{EH}$,  $D'_{EH}$, $D_4$, $D_c$). The relations we get among the divisor classes are
$$[D_{EH}]=[D'_{EH}] = -\tfrac14 [D_c],\qquad [D_4] = -\tfrac12 [D_c].$$
Again, this implies $[K_{Y_3}]=0$.

The integral generator of $\operatorname{Pic}(Y_3)$ is $D_4$. $D_{EH}$ and $D'_{EH}$
are linearly equivalent, and both generate the class group.

\begin{table}[ht]\renewcommand{\arraystretch}{1.50}
\caption{Toric Divisors in $Y_3$ \label{divisors3}} 
\vskip10pt
\centering 
\begin{tabular}{|c |c| c| c| c| c| } 
 \hline 
Ray & Divisor   & Fan  & Variety & Type & Component  \\ [1ex] 
\hline 
$\mathbf w_3$ & $D_c$  & \small $(1,0),\ (-1,2),\ (0,-1) $ & $\mathbb P[1,1,2]$ & Cartier & 1 \\   \hline 
$\mathbf v_1$ & $D_{EH}$  & \small$ (1,0),\ (-1,2),\ (0,1) $&  $\mbox{ALE}_{A_1}$  & Weil  &  $-\frac14$ \\   \hline
$\mathbf v_2$ & $D'_{EH}$ & \small$ (1,0),\ (-1,2),\ (0,1) $ &  $\mbox{ALE}_{A_1}$ & Weil  &  $-\frac14$ \\   \hline
$\mathbf v_3$ & $D_{4}$ &  \small$ (1,0),\ (-1,4),\ (0,1) $ & \small tot  $(\mathcal O(-4) \to \mathbb P^1$)& Cartier &  $-\frac12$ \\   [1ex] 
\hline 
\end{tabular}
\end{table}

\subsubsection{Toric curves and intersections}
Inspection of the planar graph in Figure \ref{Sigma3} shows that $Y_3$ has 3 toric compact curves, however we know that
there is only one independent class in $H_2(Y_3,\mathbb Q)$.
 The intersection numbers of the curves with the 4 toric divisors are shown in Table \ref{intersections3}.
The curves $C_1$, $C_2$, $C_4$ are the 3 compact curves. $C_3$
is a noncompact curve.

\begin{table}[ht]
\renewcommand{\arraystretch}{1.50}
\caption{Intersections among the toric curves and divisors in  $Y_3$. The toric curves are images of curves in $Y$ via the natural map $Y\to Y_3$, and we have used the same notation for a curve in $Y$ and its image in $Y_3$.  The curve $C_8$ is singular (it is actually the only singular curve among the  6 toric curves of $Y_3$).
Note that $C_8$ is indeed the strict transform of the $z$ axis, whose points have nontrivial isotropy,
and $Y_3$ does not resolve this singularity as $D_{nc}$ has been shrunk onto $C_8$. The curve $C_1$ of $Y$ has shrunk to a point. This will be explicitly checked in Section \ref{Y3Kallero} by noting that
the period of the K\"ahler form on $C_1$ goes to zero under the blow-down morphism
$Y\to Y_3$.
} 
\vskip10pt
\centering 
\begin{tabular}{|c|c |c| c| c| c| } 
 \hline 
 Face  & Curve & $D_c$    & $D_{EH}$   & $D'_{EH}$ & $D_4$  \\ \hline
$(\mathbf v_1\mathbf w_3)$ & $C_2$ & $-2 $ & $\frac12$  &  $\frac12$ & 1   \\ \hline
$(\mathbf v_2\mathbf w_3)$ & $C_4$ & $-2 $ & $\frac12$  &  $\frac12$ & 1   \\ \hline
$(\mathbf v_1\mathbf v_2)$ & $C_8$ & 1  &   &    & 0 \\ \hline
$(\mathbf v_3\mathbf w_3)$ & $C_5$ & $-4 $ & $1$  &  1 & 2   \\
 \hline 
\end{tabular}
\label{intersections3} 
\end{table}

\subsubsection{The class group}
The class group enters the exact sequence
$$ 0 \to M \xrightarrow{A} \operatorname{Div}{\mathbb T}(Y_3) \to \operatorname{Cl}(Y_3)\to 0.$$
So the problem is that of computing the quotient of two free abelian groups; it may have torsion.
The matrix $A$ can be put into a normal form called the Smith normal form \cite{smith}.
This is diagonal, and the diagonal entries determine the class group: a diagonal block equal to the identity  corresponds to a free summand of the appropriate rank, and if a value $m$ appears,   there is a summand $\Z_m$. For $Y_3$ the Smith normal form of the matrix $A$  is the identity matrix, so that the quotient is $\Z$, i.e.,
$$\operatorname{Cl}(Y_3) = \Z.$$
Comparing with Table \ref{divisors3} we see that the morphism $ \operatorname{Pic}(Y_3)\to \operatorname{Cl}(Y_3)$
is the multiplication by $2$ (indeed, $2[D'_{EH}]=[D_4]$).

\subsubsection{$Y_3$ as a line bundle over $\mathbb P[1,1,2]$}
Also $Y_3$ is the total space of a line bundle, in this case over $\mathbb P[1,1,2]$.
 The fan of $\mathbb P[1,1,2]$  is depicted on the left in Figure \ref{Div3}; it is generated by
the vectors $(1,0)$, $(-1,2)$, $(0,-1)$, corresponding respectively to the divisors $D_1$, $D_2$, $D_3$. The divisors $D_1$ and $D_2$ are Weil,
while $D_3$ is Cartier. In the class group, which is $\mathbb Z$, they are related by
$$ [D_3] = 2 [D_2] = 2 [D_1].$$
Each of $D_1$ and $D_2$ generates the class group, and $D_3$ generates the Picard group.

We study the line bundle $\cO_{\mathbb P[1,1,2]}(-2D_3) = \cO_{\mathbb P[1,1,2]}(-4)$. By applying the algorithm in \cite[\S 7.3]{CoxLS} we
see that its fan is that of $Y_3$, i.e., $Y_3=\mbox{tot}(\cO_{\mathbb P[1,1,2]}(-4))$. Again, since
$-2D_3$ is a canonical divisor of $Y_3$, $K_{Y_3}$ is trivial.
Again the toric divisors of $Y_3$ may be obtained from this description: $D_{EH}$ and $D'_{EH}$ are the inverse
images of $D_1$ and $D_2$, while $D_4$ is the inverse image of $D_3$; since $D_3\cdot D_3= 2$, then $D_4$ is the total space
of $\cO(-4)$ on $\mathbb P^1$. Moreover, $D_c$ is the image of the zero section.
Note that $Y_3$ is obtained by shrinking $D_{nc}$ to a $\mathbb P^1$; actually $D_{nc}$ is the total space of the trivial line bundle on
the divisor $E$ in $\mathbb F_2$, and $\mathbb P[1,1,2]$ is indeed obtained by shrinking that divisor to a point.

In Section \ref{Y3Kallero} using the generalized Kronheimer construction we shall calculate the K\"ahler potential, K\"ahler metric and
K\"ahler form of $Y_3$, verifying that the base of the bundle is indeed singular, as the
periods of the K\"ahler form on the cycle $C_1$ vanish. This means that $C_1$ shrinks to
a point, and since $C_1$ inside the compact exceptional divisor $\mathbb F_2$ is
the exceptional divisor of the blow-down $\mathbb F_2\to \mathbb P[1,1,2]$, the base variety becomes singular.

\subsection{The degeneration $Y_{EH}$}
Another degeneration occuring on some edges of the space of stability parameters
is the product $Y_{EH} =   \mbox{ALE}_{A_1} \times \mathbb C$.
\subsubsection{The fan}
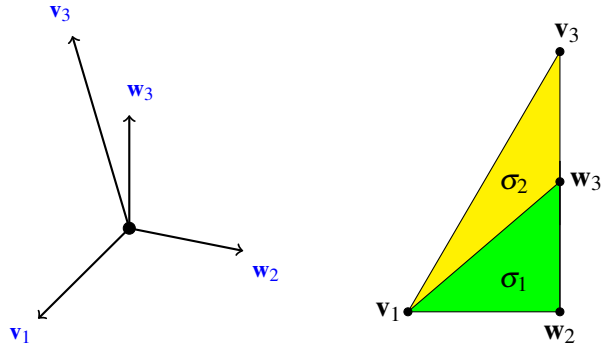
\begin{figure}
\begin{center}
\begin{tikzpicture}[scale=1.50]
\draw [thick,->] (0,0) -- (0,1) ;
\draw [thick,->] (0,0) -- (-0.5,1.7) ;
\draw [thick,->] (0,0) -- (-0.8,-0.8) ;
\draw [thick,->] (0,0) -- (1,-0.2);
\node at (1.2,-0.4){\blu{\footnotesize $\mathbf w_2$}};
\node at (0.1,1.2){\blu{\footnotesize $\mathbf w_3$}};
 \node at (-0.6,1.9){\blu{\footnotesize $\mathbf v_3$}};
 \node at (-0.95,-0.95){\blu{\footnotesize $\mathbf v_1$}};
  \draw [fill] (0,0) circle (1.5pt) ;

\end{tikzpicture}\hskip1cm
\begin{tikzpicture}[scale=0.50]
  \path [fill=yellow] (0,0) to (4,3.45) to  (4,6.9) to (0,0) ;
   \path [fill=green] (0,0) to (4,0)  to  (4,3.45) to (0,0);
  \draw [fill] (0,0) circle (3pt);
      \draw [fill] (4,6.9) circle (3pt);
\draw (0,0) -- (4,3.45);
\draw (0,0) -- (4,0); \draw (0,0) -- (4,6.9);    \draw
(4,6.9) -- (4,2.3); \node at (-0.5,0) {$\mathbf v_1$};   \node at (4.2,7.4)
{$\mathbf v_3$};  \node at (4.7,3.45)
{$\mathbf w_3$};

\node at
(4,-0.6) {$\mathbf w_2$}; \draw [fill]  (4,0) circle (3pt); \draw
(4,0) -- (4,4); \draw [fill]  (4,3.45) circle (3pt); \draw
(4,0) -- (4,4);

\node at (2.8,3.4) {$\sigma_2$}; \node at (2.8,0.8) {$\sigma_1$};
\end{tikzpicture}

 \caption{\label{EH}  \small The fan $\Sigma_{EH}$ of the product variety $Y_{EH}$ and the associated planar graph}

\end{center}
\end{figure}

We consider the toric 3-fold whose fan is generated by the rays
$\mathbf v_1$, $\mathbf w_2$, $\mathbf v_3$, $\mathbf w_3$.  The
three rays  $\mathbf w_2$, $\mathbf w_3$, $\mathbf v_3$ generate the
fan of the ALE space $\mbox{ALE}_{A_1}$  and are orthogonal to
$\mathbf v_1$, so that this is  indeed a product manifold $Y_{EH} =
\mbox{ALE}_{A_1} \times \mathbb C$. Its fan and planar graph are
shown in Figure \ref{EH}. All cones are contained in the cones of
$\Sigma_Y$ (this is easily visualized by noting that the planar
graph is a subgraph of that of $Y$), so that there is a morphism
$Y_{EH} \to Y$; on the contrary, there does not seem to be morphism
$Y \to Y_{EH}$, so that $Y_{EH}$ does not  appear to be a
degeneration of $ Y$, and  $Y_{EH} $ is not a desingularization of
$Y_0$.

\begin{remark} The fans generated by collections of rays  $\mathbf v_2$, $\mathbf w_2$, $\mathbf v_3$, $\mathbf w_3$,
and  $\mathbf v_1$, $\mathbf w_2$, $\mathbf v_2$, $\mathbf w_3$ describe the same variety.
\end{remark}

The cohomology of this variety is
$$h_2(Y_{EH},\Q) = 1, \qquad h^2_c(Y_{EH},\Q) = 0, \qquad h^2(Y_{EH},\Q) =1, \qquad h^4_c(Y_{EH},\Q) = 1, \qquad h^4(Y_{EH},\Q) = 0$$

\subsubsection{Divisors}
\begin{figure}
\begin{center}
\begin{tikzpicture}
\draw [thick,->] (0,0) -- (1,0) ;
\draw [thick,->] (0,0) -- (-1,0) ;
\draw [thick,->] (0,0) -- (0,1) ;
  \draw [fill] (0,0) circle (1.5pt) ;
  \node at (0,-1.3) {$D_3$};
\end{tikzpicture}
\hskip10mm
\begin{tikzpicture}
\draw [thick,->] (0,0) -- (1,0) ;
\draw [thick,->] (0,0) -- (-1,2) ;
\draw [thick,->] (0,0) -- (0,1);
  \draw [fill] (0,0) circle (1.5pt) ;
  \node at (0,-1.3) {$D_{EH}$};
\end{tikzpicture}
\hskip10mm
\begin{tikzpicture}
\draw [thick,->] (0,0) -- (1,0) ;
\draw [thick,->] (0,0) -- (0,1) ;
  \draw [fill] (0,0) circle (1.5pt) ;
  \node at (0,-1.3) {$D_0$};
\end{tikzpicture}
\hskip10mm
\begin{tikzpicture}
\draw [thick,->] (0,0) -- (1,0) ;
\draw [thick,->] (0,0) -- (-1,4) ;
\draw [thick,->] (0,0) -- (0,1);
  \draw [fill] (0,0) circle (1.5pt) ;
  \node at (0,-1.3) {$D_4$};
\end{tikzpicture}
\caption{\label{Div4}  \small The fans of the 4 toric divisors of $Y_{EH}$}
\end{center}
\end{figure}
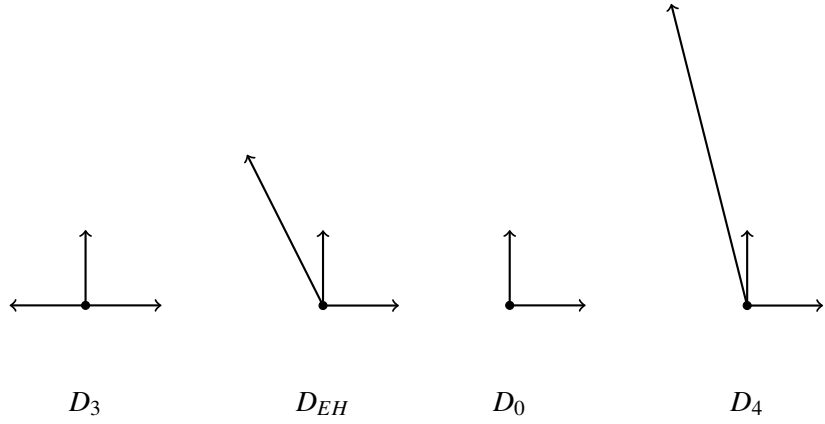

The 4 toric divisors corresponding to the rays generated by $\mathbf v_1$, $\mathbf w_3$, $\mathbf v_2$, $\mathbf v_3$ will be denoted
$D_{EH}$, $D_{3}$, $D_0$, $D_4$ respectively. They are described in Table \ref{divisors4}. They are all Cartier as $Y_{EH}$ is smooth. Their fans are depicted in Figure \ref{Div4}. The matrices $A$, $B$ in this case are
$$ A =\begin{pmatrix} 1 & 0 & 0 \\ 0 & 1 & 0 \\ 0 & -1 & 2  \\ 0 & 0  & 1\end{pmatrix},\qquad
B =\begin{pmatrix}
0 &  -\tfrac12 &  -\tfrac12  & 1
\end{pmatrix}
$$
with the divisors ordered as $D_{EH}$, $D_0$, $D_4$, $D_3$. The relations among the divisor classes are
$$[D_{EH}]=0,\qquad [D_0]=[D_4]=-\tfrac12 [D_3] .$$
Again, $[K_{Y_{EH}}]=0$. The fact that $D_{EH}$ is cohomologous to zero is consistent with
$\operatorname{Pic}(Y_{EH}) \simeq p_1^\ast \operatorname{Pic}(\mbox{ALE}_{A_1})$,
with $p_1\colon Y_{EH} \to \mbox{ALE}_{A_1}$ the projection onto the first factor.

\begin{table}[ht]\renewcommand{\arraystretch}{1.50}
\caption{Toric Divisors in $Y_{EH}$ } 
\vskip10pt
\centering 
\begin{tabular}{|c |c| c| c| c| } 
 \hline 
Ray & Divisor   & Fan  & Variety &   Component  \\ [1ex] 
\hline 
$\mathbf w_2$ & $D_{0} $  & \small $(1,0), \ (0,1) $ & $\mathbb C^2$  &  $-\frac12$  \\ \hline 
$\mathbf v_1$ & $D_{EH}$  & \small$ (1,0),\ (-1,2),\ (0,1) $&  $\mbox{ALE}_{A_1}$   &  0 \\   \hline
$\mathbf w_3$ & $D_{3}$ & \small $(1,0),\ (-1,0),\ (0,1) $  &  $\mathbb P^1\times\mathbb C$     &  1 \\   \hline
$\mathbf v_3$ & $D_{4}$ &  \small$ (1,0),\ (-1,4),\ (0,1) $ & \small tot  $(\mathcal O(-4) \to \mathbb P^1$)  & $ -\frac12 $\\   [1ex] 
\hline 
\end{tabular}
\label{divisors4} 
\end{table}

  \subsubsection{Toric curves and intersections}
  There is one compact toric curve, corresponding to the face
  $(\mathbf v_1 \mathbf w_3)$. It lies in $D_{EH}$ and $D_3$.
  The intersection numbers are shown in Table \ref{intersections4}.

  \begin{table}[ht]
\caption{Intersections among the toric curves and divisors in  $Y_{EH}$} 
\vskip10pt
\centering 
\begin{tabular}{|c |c| c| c| c| c|} 
 \hline 
  & $D_0$    & $D_{EH}$   & $D_3$ & $D_4$  \\ \hline
$C$ & 1 & $0  $  &  $ -2 $ & 1   \\
 \hline 
\end{tabular}
\label{intersections4} 
\end{table}

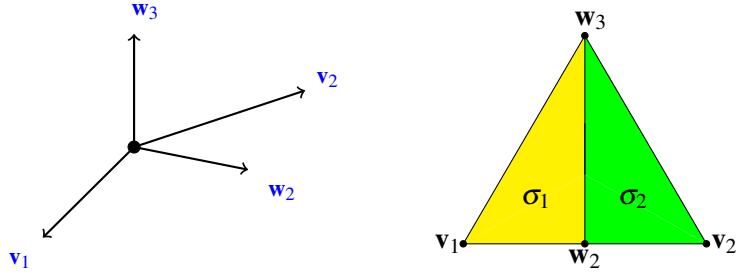
\begin{figure}[t]
\begin{center}
\begin{tikzpicture}[scale=1.50]
\draw [thick,->] (0,0) -- (0,1) ;
\draw [thick,->] (0,0) -- (1.5,0.5) ;
\draw [thick,->] (0,0) -- (-0.8,-0.8) ;
\draw [thick,->] (0,0) -- (1,-0.2);
\node at (1.3,-0.4){\blu{\footnotesize $\mathbf w_2$}};
\node at (-1,-1){\blu{\footnotesize $\mathbf v_1$}};
\node at (1.7,0.6){\blu{\footnotesize $\mathbf v_2$}};
\node at (0.1,1.2){\blu{\footnotesize $\mathbf w_3$}};

  \draw [fill] (0,0) circle (1.5pt) ;

\end{tikzpicture}\hskip1cm
\begin{tikzpicture}[scale=0.40]
  \path [fill=yellow] (0,0) to (4,2.3) to  (4,6.9) to (0,0) ;
    \path [fill=green] (8,0) to (4,2.3) to  (4,6.9) to (8,0) ;
      \path [fill=yellow] (0,0) to (4,2.3)  to  (4,0) to (0,0);
      \path [fill=green] (4,0) to (4,2.3)  to  (8,0) to (4,0);
  \draw [fill] (0,0) circle (3pt);
    \draw [fill]  (8,0) circle (3pt);
      \draw [fill] (4,6.9) circle (3pt);

\draw (0,0) -- (8,0); \draw (0,0) -- (4,6.9); \draw (8,0) --
(4,6.9);   \draw
(4,6.9) -- (4,2.3); \node at (-0.5,0) {$\mathbf v_1$}; \node at (8.6,0)
{$\mathbf v_2$}; \node at (4.2,7.4)
{$\mathbf w_3$};
\node at
(4,-0.5) {$\mathbf w_2$}; \draw [fill]  (4,0) circle (3pt); \draw
(4,0) -- (4,4);
\node at (5.6,1.5) {$\sigma_2$}; \node at (2.4,1.5) {$\sigma_1$};
\end{tikzpicture}

 \caption{\label{SigmaEH2}  \small The fan of the second realization of $Y_{EH}$  and the associated planar graph}

\end{center}
\end{figure}

 \subsubsection{A relation with the full resolution} \label{relation}

 Let $g=(\omega,\omega,\omega^2)$ be the generator of the action of $\Z_4$.
 The square $g^2$ leaves every point of the $z$ axis fixed, so that
 $$\mathbb C^3/ \langle g^2 \rangle \simeq \mathbb C^2/\Z_2\times\mathbb C.$$
 Blowing up we get $Y_{EH}$. Now $g$ still acts on $Y_{EH}$ producing a quotient which
 is singular along a $\mathbb P^1$.
 Blowing up we get $Y$; the corresponding exceptional divisor is $D_c$,
 the second Hirzebruch surface. So $Y_{EH}$ is the desingularization of a partial quotient of
 $\mathbb C^3$, and by a further quotient and subsequent desingularization it produces $Y$. The following diagram depicts this situation.
\begin{equation}\label{diagYEH} \xymatrix{
 & Y \ar[d] \\
 Y_{EH} \ar[r]\ar[d] & Y_{EH}/\Z_2 \ar[d] \\
 \mathbb C^3/\Z_2 \ar[r] & \mathbb C^3/\Z_4}
\end{equation}
 Note that $ Y_{EH}/\Z_2$ is therefore a partial desingularization of $\mathbb C^3/\Z_4$,
 and indeed it corresponds to the fan obtained by adding the ray $\mathbf w_2$ to the
 fan of $\mathbb C^3/\Z_4$. Note also that the composed map $Y_{EH}\to \C^3/\Z_4$ in diagram \eqref{diagYEH} is not
a resolution of singularities as it is 2:1.

 \subsubsection{$Y_{EH}$ as a fibration}
 As it happens also for $Y$ and $Y_3$, $Y_{EH}$ is the total space of a vector bundle, actually in two different way.
 In particular it is the pullback of the line bundle $\cO(-2)$ on $\mathbb P^1$ via the projection $\mathbb P^1\times \mathbb C \to \mathbb P^1$
 (namely, the total space of the canonical bundle of $\mathbb P^1\times \mathbb C $).

\subsection{Summary: the exceptional divisor, curves and homology $2$-cycles}
Some of the main upshots of the discussion and computations made in this Section
are the following.

$\bullet$ The general theory
encoded in the Ito-Reid theorem \cite{itoriddo}, confirmed by the
explicit toric constructions, tells us that the quotient ${\mathbb{C}^3}/{\mathbb{Z}_4}$ has a
crepant resolution of singularities $Y$. This may be computed using
toric geometry.  The exceptional divisor has a compact component,
$D_c$, isomorphic to the second Hirzebruch surface
$\mathbb F_2$, and a noncompact component $D_{nc}$,
isomorphic to $\mathbb C\times\mathbb P^1$. This agrees with the age
computations which show that the groups $H^{1,1}_{c}\left(Y\right)$
and $H^{2,2}\left(Y\right)$ are both one-dimensional.

\smallskip
$\bullet$ Let
$    \pi \colon Y  \to
\mathbb{C}^3/\mathbb{Z}_4 $
be the   blow-down  morphism. The compact exceptional
divisor $D_c$ is the inverse image $\pi^{-1}(0)$ of the fixed
point $0\in \mathbb{C}^3$,
while the noncompact component $D_{nc}$
is the preimage of the fixed line $
\{x=y=0\}$.

\smallskip
$\bullet$
Using this information it is immediate to identify the
equation of the compact exceptional divisor in the   open chart
associated with the cone $\sigma_1$ as
\begin{equation}\label{totointurchia} w=0. \end{equation}
This, together with  the substitutions \eqref{subberulle},
is all what we need to compute the periods of the first Chern
classes of the tautological bundles  represented by the following  closed $(1,1)$-forms:
\begin{equation}\label{giraldo}    \omega^{(1,1)}_{1,2,3} \, = \, \frac{i}{2\pi} \, {\partial} \,
    \bar\partial \, \log \left[ X_{1,2,3} \right]^{\alpha_{1,2,3}}
\end{equation}
As we explain further on, although we are not able to compute the forms
$\omega^{(1,1)}_{1,2,3}$ for the general case, yet we succeed in
calculating their periods on the basis of the homology cycles by
restricting the moment map equations to the latter, using to this
purpose the equations of such loci as derived from the toric
description.

 Again in the chart corresponding to the cone $\sigma_1$, the  noncompact
component $D_{nc}$ of the exceptional divisor has equation
\begin{equation}\label{totoinegitto} v=0. \end{equation}
 The two components
intersect along the curve
\begin{equation}\label{c2tilde}
 C_1 \, = \,\{u,v,w \mid w=0, \, v=0\}.
\end{equation}

 \smallskip
$\bullet$  Finally, we consider the curve
\begin{equation}\label{c1}
    C_2 \, = \,\{u,v,w \mid w=0,\, u=0\}.
\end{equation}
i.e., the intersection between $D_{EH}$ and
$D_c$. It corresponds to the face $xz$ of the fan of
$Y$. As a curve in $\mathbb F_2$, it corresponds to the ray
generated by $(1,0)$, squares to $-2$, and is the exceptional
divisor of the blowup $\mathbb F_2 \to \mathbb P^2/\mathbb Z_2$, and
a section of the fibration $\mathbb F_2\to\mathbb P^1$ (and of course it
is a copy of $\mathbb P^1$). It intersects $D_{nc}$ again
in the point $u=v=w=0$, which corresponds to the cone generated by
$(1,0)$ and $(0,1)$ in the fan of $D_{nc}$.  This is the
$\mathbb P^1$ which generates the Picard group of the Eguchi-Hanson
space $D_{EH}$.

 \smallskip
$\bullet$   Note that $\dim \, H_2(Y)=2$, and indeed the curves $ C_1$
and $ C_2$ provide a basis for $H_2(Y,\mathbb Z)$.

\section{Chamber Structure and the tautological bundles}
\label{camerataccademica}
\begin{figure}
\label{hilton1} \centering
\includegraphics[height=7cm]{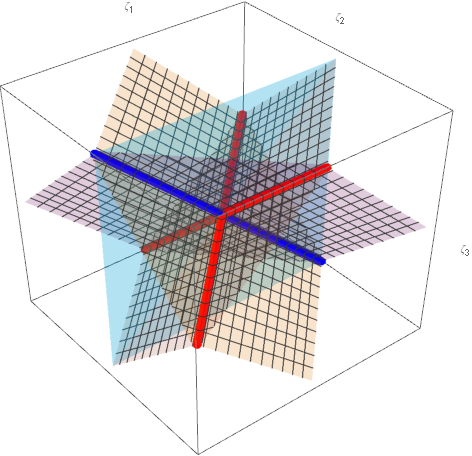}
\caption{\label{hilton1} The structure of the stability
chambers. The space $\mathbb{R}^3$ where the moment map equations
always admit real nonnegative solutions is divided in two halves by
the presence of a wall of type 0, named $\mathcal{W}_0$ which is
  defined by the equation $\zeta_2=0$ and is marked in
the figure as a cyan transparent surface without meshing. In
addition there are other three walls, respectively described by
$\mathcal{W}_1 \Leftrightarrow \zeta=\{x+y,x,y\}$,
$\mathcal{W}_2 \Leftrightarrow \zeta=\{x,x+y,y\}$ and
$\mathcal{W}_3 \Leftrightarrow \zeta=\{x,y,x+y\}$, where
$x,y\in \mathbb{R}$. The planes $\mathcal{W}_{1,3}$ are of type
0, while $\mathcal{W}_{2}$ is of type $1$. These three infinite
planes provide the partition of $\mathbb{R}^3$ into eight disjoint
chambers that are described in the main text. The three planes
$\mathcal{W}_{1,2,3}$ are marked in the figure as meshed surfaces
of three different colors. On the three intersections  of two of
these planes we find the already discussed  lines where the moment map
equations can be solved by radicals,
corresponding to the Eguchi-Hanson degeneration $ Y_{EH}\times
\mathbb{C}$ (blue line) and to the two Cardano degenerations (red
lines).}
\end{figure}
We can now compare the analytical results obtained from the K\"ahler
quotient \`a la Kronheimer  with the general predictions of the
resolution of the singularity provided by toric geometry. This
provides a concrete example of how the chamber structure of the
$\zeta$ parameter space controls the topology of the
resolutions of the singularity \cite{CrawIshii}.
In the following we evaluate the periods of the differential forms
arising from the Kronheimer construction on the cycles given by the
curves $C_1$, $C_2$ that both are contained in
$D_c$, namely in the compact component of the
exceptional divisor (actually we are pulling back the differential
forms from $\mathcal{M}_{a,b,c}$ to $Y$ via the relevant contraction
morphism $\gamma\colon Y \to \mathcal{M}_{a,b,c}$, and we use the
fact that the pullback is injective in cohomology; however we shall
understand that pullback in the notation). We succeed evaluating the
differential form periods on the considered curves by restricting
the moment map equations to the relevant exceptional divisor
$D_c$ and then to its relevant sub-loci.

According {to} the general theory of the Kronheimer-like
construction presented in \cite{Bruzzo:2017fwj}, there are three
tautological bundles; their first Chern classes are encoded in the
  {triple}  of (1,1)-forms given in eq.~\eqref{giraldo}.

\subsection{The stability chambers}
\label{camerataccademicaWeyl} The result of the  various calculations
presented in subsequent sections  singles out the structure of the
stability  chambers which is summarized in
figs.~\ref{hilton1} and \ref{hilton2}.
\begin{figure}
\label{hilton2} \centering
\includegraphics[height=7cm]{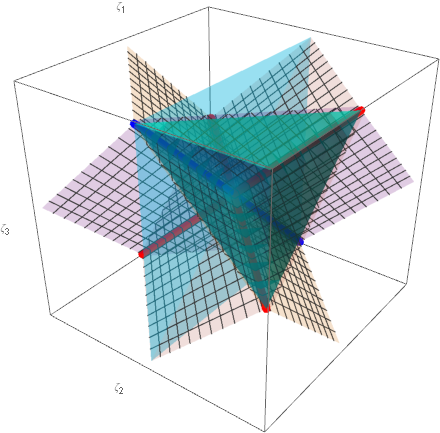}
\caption{\label{hilton2}  In fig.\ref{hilton1} we displayed  only
the walls defining the chamber structure. In the present picture
besides the walls we show also one of the eight chambers, namely
Chamber 1. It is marked as a transparent  greenish-blue colored portion of
three dimensional space. It is a convex cone delimited by the
aforementioned walls. }
\end{figure}
Let us illustrate this structure in detail. Understanding the
geometry of these pictures is a very useful guide through the
subsequent computations. The original data from which we start are
the following ones. To begin with we know that the entire
$\zeta$ space is just $\mathbb{R}^3$.

\begin{figure}
\centering
\includegraphics[height=9cm]{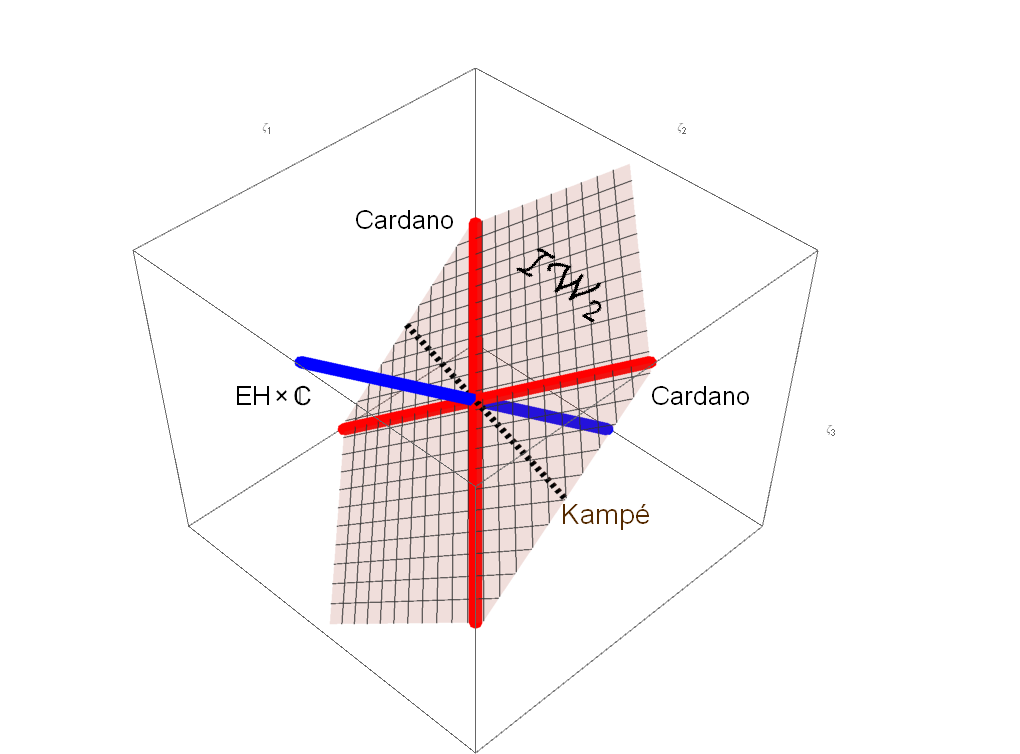}
\caption{ \label{pianoiwdue} This picture shows the type I plane
$\mathcal{W}_2$. The two Cardano manifolds (red lines) and the
Kamp\'{e} manifold (dashed black line) lay all in this plane, whose
generic point corresponds, as we are going to see, to the
degeneration Y3. The Eguchi-Hanson degeneration (blue line) is
instead out of this plane and intersects it only in the origin. }
\end{figure}

In figs.~\ref{hilton1} and \ref{hilton2} we have drawn  3 lines. These
lines correspond to the following 4 instances of degenerate spaces
where the algebraic system of the moment map equations becomes
solvable or partially solvable:
\begin{description}
\item[1)] Eguchi-Hanson case
\begin{equation}\label{ehcasus}
    \zeta_1 \, = \, s \quad ; \quad \zeta_2 \, = \, 0 \quad ; \quad
    \zeta_3 \, = \, s
\end{equation}
In pictures \ref{hilton1}, \ref{hilton2} this line is fat and drawn
in blue color:
\item[2)] Cardano 1
\begin{equation}\label{Carcasus1}
    \zeta_1 \, = \, s \quad ; \quad \zeta_2 \, = \, s \quad ; \quad
    \zeta_3 \, = \, 0
\end{equation}
In pictures \ref{hilton1}, \ref{hilton2} this line is solid fat and
drawn in red color. We remind the reader that the name Cardano is
due to the fact that the solution of the entire moment map equation
system reduces to the solution of a single algebraic equation of the
fourth order (see eq.~\eqref{baldop}).
\item[3)] Cardano 2
\begin{equation}\label{Carcasus2}
    \zeta_1 \, = \, 0 \quad ; \quad \zeta_2 \, = \, s \quad ; \quad
    \zeta_3 \, = \, s
\end{equation}
 In pictures \ref{hilton1}, \ref{hilton2} this line is solid fat and
drawn in red color.
\end{description}
In addition we have a 4th line that we have not drawn in fig.
\ref{hilton1} and \ref{hilton2}. This line entirely lies on one of
the walls to be described below.
\begin{description}
\item[4)] Kamp\'{e}
\begin{equation}\label{kampus1}
    \zeta_1 \, = \, s \quad ; \quad \zeta_2 \, = \, 2 \,s \quad ; \quad
    \zeta_3 \, = \, s
\end{equation}
In fig.~\ref{pianoiwdue} this line is dashed fat and drawn in black
color. We remind again the reader that the manifold was named
Kamp\'{e} because the solution of the moment map equations reduces
to finding the roots of a single algebraic equation of the sixth
order.
\end{description}
In all  these cases $s$ is a nonzero real number.

Since the exceptional solvable cases must lie on some walls we have
tried to conjecture which planar cones might partition the infinite
cube into chambers so that the exceptional lines could lie on such
planar cones and possibly be edges at some of their intersections.

\begin{figure}
\centering
\includegraphics[height=9cm]{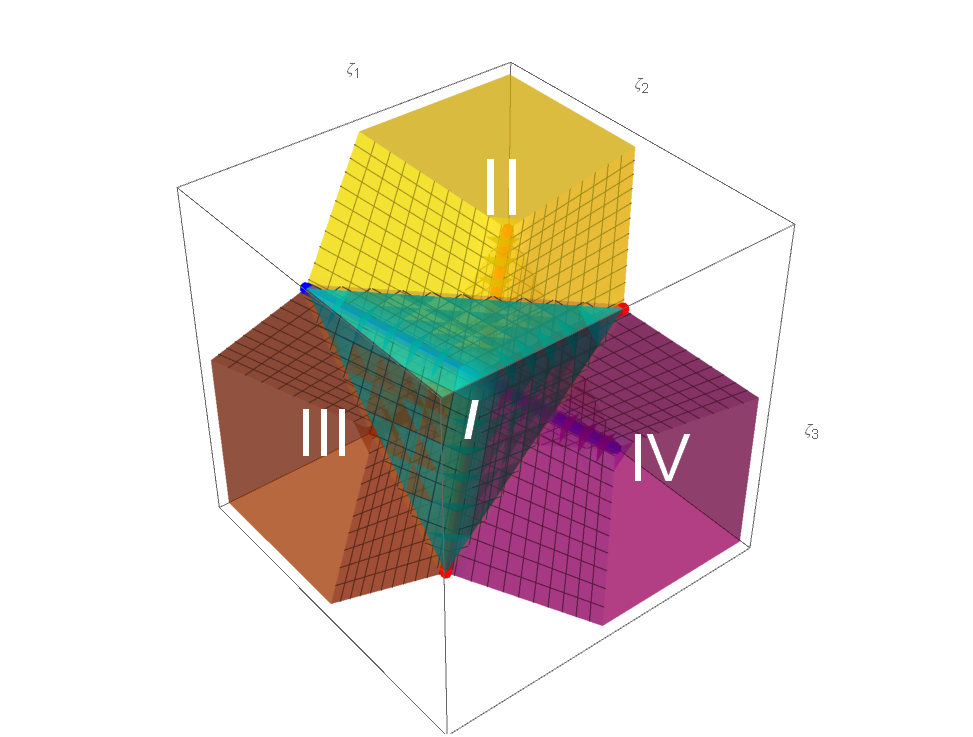}
\caption{ \label{pinetti} The partition of $\mathbb{R}^3$ into 8
convex cones. In the picture, out of the eight regions, we show only
four, marking them in different transparent colors.}
\end{figure}

With some ingenuity we introduced the following four planar walls
(here $x,y$ are real parameters)
\begin{align}
\mathcal{W}_1 :\qquad &  \{x+y,x,y\}  \label{IW1muro} \\
\mathcal{W}_2  : \qquad  &\{x,x+y,y\}  \label{IW2muro} \\
\mathcal{W}_3 : \qquad  &   \{x,y,x+y\} \label{IW3muro} \\
\mathcal{W}_0  : \qquad  & \{x,0,y\} \label{IW0muro}
\end{align}
that were depicted in fig.~\ref{hilton1}, \ref{hilton2}. and split
the space $\mathbb{R}^3$ into eight convex three--dimensional cones.
The list of the eight convex cones that provide as many  stability chambers is obtained through the following argument. The three
planes $\mathcal{W}_{1,2,3}$ are respectively orthogonal to the
following three vectors:
\begin{eqnarray}
  \pmb{n}_1 &=& \left\{-1,1,1 \right\} \\
  \pmb{n}_2 &=& \left\{1,-1,1 \right\} \\
  \pmb{n}_3 &=& \left\{1,1,-1 \right\} .\label{normalini}
\end{eqnarray}
The eight convex regions are defined by choosing the signs of the
projections $\pmb{n}_{1,2,3}\cdot \zeta$ in all possible ways.
In this way we obtain:
\begin{equation}
\begin{array}{rcl}
   \mbox{Chamber I}
  & \equiv &
  \, \left\{  \zeta_1\, -\,\zeta_2\,-\,\zeta_3\,> \,0
  \quad , \quad
    -\zeta_1\, +\,\zeta_2\,-\,\zeta_3\,> \,0 \quad , \quad
    -\zeta_1\, -\,\zeta_2\,+\,\zeta_3\,> \,0
   \right\}  \\
    \mbox{Chamber II}
  & \equiv &
  \, \left\{  \zeta_1\, -\,\zeta_2\,-\,\zeta_3\,> \,0
  \quad , \quad
    -\zeta_1\, +\,\zeta_2\,-\,\zeta_3\,> \,0 \quad , \quad
    -\zeta_1\, -\,\zeta_2\,+\,\zeta_3\,< \,0
\right\} \\
    \mbox{Chamber III}
  & \equiv &
  \, \left\{  \zeta_1\, -\,\zeta_2\,-\,\zeta_3\,> \,0
  \quad , \quad
    -\zeta_1\, +\,\zeta_2\,-\,\zeta_3\,< \,0 \quad , \quad
    -\zeta_1\, -\,\zeta_2\,+\,\zeta_3\,> \,0
   \right\}  \\
    \mbox{Chamber IV}
  & \equiv &
  \, \left\{  \zeta_1\, -\,\zeta_2\,-\,\zeta_3\,< \,0
  \quad , \quad
    -\zeta_1\, +\,\zeta_2\,-\,\zeta_3\,> \,0 \quad , \quad
    -\zeta_1\, -\,\zeta_2\,+\,\zeta_3\,> \,0
   \right\}  \\
    \mbox{Chamber V}
  & \equiv &
  \, \left\{  \zeta_1\, -\,\zeta_2\,-\,\zeta_3\,< \,0
  \quad , \quad
    -\zeta_1\, +\,\zeta_2\,-\,\zeta_3\,< \,0 \quad , \quad
    -\zeta_1\, -\,\zeta_2\,+\,\zeta_3\,> \,0
    \right\}  \\
    \mbox{Chamber VI}
  & \equiv &
  \, \left\{  \zeta_1\, -\,\zeta_2\,-\,\zeta_3\,< \,0
  \quad , \quad
    -\zeta_1\, +\,\zeta_2\,-\,\zeta_3\,> \,0 \quad , \quad
    -\zeta_1\, -\,\zeta_2\,+\,\zeta_3\,< \,0
    \right\}  \\
    \mbox{Chamber VII}
  & \equiv &
  \, \left\{  \zeta_1\, -\,\zeta_2\,-\,\zeta_3\,> \,0
  \quad , \quad
    -\zeta_1\, +\,\zeta_2\,-\,\zeta_3\,< \,0 \quad , \quad
    -\zeta_1\, -\,\zeta_2\,+\,\zeta_3\,< \,0
    \right\}  \\
    \mbox{Chamber VIII}
  & \equiv &
  \, \left\{  \zeta_1\, -\,\zeta_2\,-\,\zeta_3\,< \,0
  \quad , \quad
    -\zeta_1\, +\,\zeta_2\,-\,\zeta_3\,< \,0 \quad , \quad
    -\zeta_1\, -\,\zeta_2\,+\,\zeta_3\,< \,0
   \right\} \\
\end{array}
 \label{celle8}
 \end{equation}
Four of the above eight chambers are displayed in fig.~\ref{pinetti}.

\subsection{Edges} The special solvable cases that we have found
all sit at the intersection of two of the three walls
$\mathcal{W}_{1,2,3}$. In particular the Cardano manifolds are
edges at the intersection of the following walls:
\begin{equation}\label{cardanici}
    \text{Cardano 1} \, = \, \mathcal{W}_{1}\bigcap \mathcal{W}_{2} \quad ;
    \quad \text{Cardano 2} \, = \, \mathcal{W}_{2}\bigcap \mathcal{W}_{3}
\end{equation}
while the Eguchi-Hanson case is the intersection:
\begin{equation}\label{ansoniani}
    Y_{EH} \, = \, \mathcal{W}_{1}\bigcap
    \mathcal{W}_{3}
\end{equation}
Note also that  this edge lays entirely on the
wall $\mathcal{W}_0$. From this point of view the Eguchi-Hanson
case is similar to the Kamp\'{e} case, that lays entirely on the wall
$\mathcal{W}_2$. The difference however is that, as we advocate
below, the wall $\mathcal{W}_0$ is of type 0 while
$\mathcal{W}_{2}$ is of type 1. In the first case the Eguchi-Hanson
line is the only degeneracy pertaining to the wall $\mathcal{W}_0$,
while in the second case the Kamp\'{e} line yields an instance of
the degeneracy $Y_3$ as any other point of the same wall. Actually
also the Cardano cases that lay on the same wall correspond to a
different realization of $Y_3$.

\section{Periods of the Chern classes of the tautological bundles}
\label{Periodstaut}
The most appropriate instrument to verify the
degeneracy/non-degeneracy of the singularity resolutions provided by
the K\"ahler quotient with given $\zeta$ parameters is
provided by the calculation  of the period
matrix:
\begin{equation}\label{periodare}
  \pmb{\Pi} \, \equiv  \Pi_{i,J} \, = \, \int_{C_i} \omega_{J}
  \quad ; \quad
    i=(1,2), \quad J=(1,2,3)
\end{equation}
where $\omega_J$ are the first Chern classes of the tautological
bundles and the curves $C_i$ provide a basis of     homology
2-cycles. In particular the combination:
\begin{equation}\label{saccius}
    \pmb{Vol}_i \,  = \zeta_I \, \mathfrak{C}^{IJ} \,  \int_{C_i} \omega_J
    =
     \,  \tfrac{i}{2\pi}\, \zeta_I \, \mathfrak{C}^{IJ} \,
    \int_{C_i} \, \partial\bar\partial \log (X_J)^{\alpha_\zeta}\, = \, \int_{C_i}
    \mathbb{K}_\zeta
\end{equation}
is the volume of the cycle $C_i$  in the resolution
identified by the level parameters $\zeta$, having denoted by
$\mathbb{K}_\zeta$ the corresponding K\"ahler 2-form.\footnote{Let us remark that in agreement with
eq.~\eqref{caramboletta} the contribution
$\partial\bar{\partial}\mathcal{K}_0$ to the K\"ahler
2-form  is an exact form, whose integral on homology cycles therefore vanishes.
Hence the volume of the homology cycles  is  provided the linear
combination of periods specified in eq.~\eqref{saccius}.
We also remark that, as the volume of a nonzero cycle is always positive,
this is consistent with the positivity of the so-called Hodge line bundle
$\otimes_{J=1}^3 L_J^{\Theta(\mathcal{D}_J)}$.}  If the
volume of the two homology cycles yielding the homology basis is
nonzero there is no degeneration. Instead in case of degenerations at
least one of such volumes vanishes. This is precisely what happens
on the walls of type III, while in the interior of all chambers no
degeneration appears.

Through the calculations detailed in the following subsections we
have been able to compute the periods of the first Chern forms
$\omega_{1,2,3}^{(1,1)}$ on the basis of homology cycles
$C_{1,2}$ both for the interior points of all the
chambers and for all the walls. As far as the interior chamber case
is concerned our results are summarized in Table \ref{periodico}.
For the walls the results are instead summarized in Table
\ref{muraria}, while for the edges they are given in Table
\ref{spigolosa}.
\begin{table}
  \centering
  $$ \begin{array}{|c||c|}
  \hline
  \hline
          \text{Chamber 1} &\begin{array}{c|c|c|c}
\text{cycle} & \int \omega_1 & \int \omega_2 & \int \omega_3 \\
\hline \hline
C_1 & 1& 0 & 1\\
\hline
C_2 & 1& 1 & 1\\
  \end{array}\\
\hline \hline
          \text{Chamber 2} & \begin{array}{c|c|c|c}
          {\text{cycle}} & {\int \omega_1} & {\int \omega_2} & {\int \omega_3} \\
          \hline
C_1 & 1& 0 & 1\\
\hline
C_2 & 1& 1 & 1\\
  \end{array} \\
          \hline
          \hline
         \text{Chamber 3} &  \begin{array}{c|c|c|c}
           {\text{cycle}} &  {\int \omega_1} &  {\int \omega_2} &  {\int \omega_3} \\
           \hline
C_1 & 1& 0 & 1\\
\hline
C_2 & 0 & -1 & 0 \\
  \end{array} \\
         \hline
          \hline
         \text{Chamber 4} & \begin{array}{c|c|c|c}
           {\text{cycle}} &  {\int \omega_1} &  {\int \omega_2} &  {\int \omega_3} \\
           \hline
C_1 & 1& 0 & 1\\
\hline
C_2 & 1& 1 & 1\\
  \end{array} \\
         \hline
          \hline
         \text{Chamber 5} & \begin{array}{c|c|c|c}
           {\text{cycle}} &  {\int \omega_1} &  {\int \omega_2} &  {\int \omega_3} \\
           \hline
C_1 & 1& 0 & 1\\
\hline
C_2 & 0 & -1 & 0 \\
  \end{array} \\
         \hline
          \hline
         \text{Chamber 6} & \begin{array}{c|c|c|c}
           {\text{cycle}} &  {\int \omega_1} &  {\int \omega_2} &  {\int \omega_3} \\
           \hline
C_1 & 1& 0 & 1\\
\hline
C_2 & -1 & -1 & -1 \\
  \end{array} \\
         \hline
          \hline
         \text{Chamber 7} & \begin{array}{c|c|c|c}
           {\text{cycle}} &  {\int \omega_1} &  {\int \omega_2} &  {\int \omega_3} \\
           \hline
C_1 & 1& 0 & 1\\
\hline
C_2 & 0 & -1 & 0 \\
  \end{array} \\
         \hline
          \hline
        \text{Chamber 8} & \begin{array}{c|c|c|c}
           {\text{cycle}} &  {\int \omega_1} &  {\int \omega_2} &  {\int \omega_3} \\
           \hline
C_1 & 1& 0 & 1\\
\hline
C_2 & 0 & -1 & 0 \\
  \end{array} \\
        \hline
          \hline
          \end{array}
   $$
  \caption{The periods of the tautological bundle first Chern classes on the basis of homological cycles
  calculated in the interior points of all the chambers.}\label{periodico}
\end{table}
\begin{table}\renewcommand{\arraystretch}{1.50}
  \centering
  $$ \begin{array}{|c||c|}
  \hline
  \hline
          \text{Wall $\mathcal{W}_0$} &\begin{array}{c|c|c|c}
\text{cycle} & \int \omega_1 & \int \omega_2 & \int \omega_3 \\
\hline \hline
C_1 & 3 & 0 & 3\\
\hline
C_2 & 2 & 0 & -2 \\
  \end{array}\\
\hline \hline
          \text{Wall $\mathcal{W}_1$} & \begin{array}{c|c|c|c}
          {\text{cycle}} & {\int \omega_1} & {\int \omega_2} & {\int \omega_3} \\
          \hline
C_1 & 1 & 0 & 1 \\
\hline
C_2 &0 & -2 & 0 \\
  \end{array} \\
          \hline
          \hline
         \text{Wall $\mathcal{W}_2$} &  \begin{array}{c|c|c|c}
           {\text{cycle}} &  {\int \omega_1} &  {\int \omega_2} &  {\int \omega_3} \\
           \hline
C_1 & 0 & 0 & 0 \\
\hline
C_2 & 1 & 4 & 1 \\
  \end{array} \\
         \hline
          \hline
         \text{Wall $\mathcal{W}_3$} & \begin{array}{c|c|c|c}
           {\text{cycle}} &  {\int \omega_1} &  {\int \omega_2} &  {\int \omega_3} \\
           \hline
C_1 & 1 & 0 & 1 \\
\hline
C_2 &0 & -2 & 0 \\
  \end{array} \\
         \hline
          \hline
          \end{array}
   $$
  \caption{The periods of the tautological bundle first Chern Classes on the basis of homological cycles
  calculated on the 4 walls. For the walls $\mathcal{W}_0$ and $\mathcal{W}_2$
  we have chosen $\alpha=4$ instead of $\alpha=2$ to get integer values for the periods.
  \label{muraria}}
\end{table}

\begin{table}\renewcommand{\arraystretch}{1.50}
  \centering
  $$ \begin{array}{|c||c|}
  \hline
  \hline
          \text{Cardano $\zeta = \{0,s,s\}$} &\begin{array}{c|c|c|c}
\text{cycle} & \int \omega_1 & \int \omega_2 & \int \omega_3 \\
\hline \hline
C_1 & 0 & 0 & 0 \\
\hline
C_2 & -1  & -1 & 0 \\
  \end{array}\\
\hline \hline
          \text{Cardano $\zeta = \{s,s,0\}$} & \begin{array}{c|c|c|c}
          {\text{cycle}} & {\int \omega_1} & {\int \omega_2} & {\int \omega_3} \\
          \hline
C_1 &0 & 0 & 0 \\
\hline
C_2 &0 &  -1  & -1  \\
  \end{array} \\
          \hline
          \hline
         \text{Eguchi-Hanson $\zeta = \{s,0,s\}$} &  \begin{array}{c|c|c|c}
           {\text{cycle}} &  {\int \omega_1} &  {\int \omega_2} &  {\int \omega_3} \\
           \hline
C_1 &  -1  & 0 & \-1  \\
\hline
C_2 & 0 & 0 & 0 \\
  \end{array} \\
         \hline
          \hline
          \end{array}
   $$
  \caption{The periods of the tautological bundle first Chern Classes on the basis of homological cycles
  calculated on the edges. Again, on the Eguchi-Hanson edge we have taken $\alpha=4$.
   }\label{spigolosa}
\end{table}

Let us discuss the results in Table \ref{periodico}. The three leftmost columns display the degrees of the tautological line bundles
$\mathcal R_I$ restricted to the curves $C_1$, $C_2$; as the classes of the latter provide a basis of $H_2(Y,\Z)$, these numbers give the
first Chern classes of the line bundles over the integral basis of the Picard group $\operatorname{Pic}(Y)$ given by the divisors
$D_{EH}$, $D_4$, dual to the previously mentioned basis of $H_2(Y,\Z)$. Note that the three columns correspond to the compact junior conjugacy class, the noncompact junior class, and the senior class respectively, and the Poincar\'e duality between $H^2_c(Y) $ and
$H^4(Y)$ explains why   in all chambers $\mathcal R_1$ and $\mathcal R_3$ have the same Chern classes, and are therefore isomorphic.

We know from the McKay correspondence that the cohomology of $Y$ is generated by algebraic classes which are in a one-to-one correspondence with the elements of $\Z_4$. One issue is how these classes are expressable in terms of the  Chern characters of the taulological bundles.
In general, this correspondence is highly nontrivial and is governed by a complicated combinatorics \cite{Reid-Kino,Craw-Hilb}. This is indeed exemplified by our calculations. For instance, one can check that in the chambers 1, 2 and 4 one has
\begin{equation}\label{chern} \langle \operatorname{ch}_2(\mathcal R_1), w\rangle = \langle\operatorname{ch}_2(\mathcal R_3) ,w\rangle= 1, \qquad
\langle\operatorname{ch}_2(\mathcal R_2), w\rangle = 0 ,\end{equation}
(the triangular brackets are the pairing between cohomology and homology in dimension 4),\footnote{In general, homology and cohomology are not dual to each other, but are rather related by the exact sequence
$$ 0 \to \operatorname{Ext}^1_\Z(H_{n-1}(Y,\Z),\Z) \to H^n(Y,\Z) \to H_n(X,\Z)^\vee \to 0 .$$
However in our case the Ext groups are zero as the homology groups are free over $\Z$ (see e.g.~\cite[Thm.~3.2]{Hatch}).}
 where $w$ is the generator of $H_4(Y,\Z)$ given by the divisor $D_4$. So   the second Chern character of $ \mathcal R_1\simeq \mathcal R_3$ does provide a $\Z$-basis for $H^4(Y,\Z)$,
and similary in chamber 6 with $-1$ instead of 1. On the other hand
in    the  chambers 3, 5, 7 and 8 all line bundles have zero second Chern character, and one needs to take a suitable combination of them to get a generator.

In order to illustrate the procedure which leads to the results
presented in the mentioned tables there are several steps one has to
take that are explained in the following subsections.

\subsection{The fundamental algebraic system and a dense
toric chart covering the variety}\label{ballavantana} The main
difficulty with any explicit calculation within the framework of the
K\"ahler quotient \`{a} la Kronheimer is that, for this case as for
the majority of all cases, the moment map equation system is
algebraic of higher order and explicit analytic solutions are mostly
out of reach.

However, in the case under consideration and similarly in most of
the other cases, the homology cycles on which we would like to
integrate our differential forms are entirely contained in the
compact component of the exceptional divisor $D_c$.
This is not too surprising since all the homology is produced by the
resolution and disappears in the original orbifold case. Hence the
first step in a strategy that leads to the calculation of the period
matrix \eqref{periodare} necessarily foresees a reduction of the
moment map algebraic system \eqref{sistemico} to the exceptional
divisor in the hope that there it becomes  of lower effective degree
and therefore manageable.

In view of the above observation we consider the transcription of
the algebraic system \eqref{sistemico} in terms of toric coordinates
that expose the presence of the exceptional divisor. There are 4
toric charts (see eq. \eqref{subberulle}), yet from the point of
view of the algebraic system of moment map equations there are only
two distinguishable charts, namely the chart $X_{\sigma _1}$ and
 $X_{\sigma _4}$. Indeed, at this level, chart
$X_{\sigma _2}$ is identical  with chart $X_{\sigma _1}$ and
chart $X_{\sigma _3}$ is identical with chart $X_{\sigma _4}$.
We choose chart $X_{\sigma _1}$  since, as it is evident from the planar diagram in
Fig.~\ref{SigmaY}, the charts 1 and 2 are the only ones that
contain both   curves $C_1$ and $C_2$.
\subsubsection{Chart $X_{\sigma _1}$}
In order to perform our calculations we find it convenient to
introduce the following notation. With reference to the coordinates
$u_1$,$v_1$,$w_1$ we set:
\begin{equation}
\lambda = \sqrt{\left|w_1\right|} \quad ; \quad \sigma  =
\left|v_1\right| \quad ; \quad \delta
=\left(1+\left|u_1|^2\right.\right)\label{cavedanorosso}
\end{equation}
which yields
\begin{equation}
\Sigma =\delta  \lambda  \sigma \quad ; \quad U=\lambda ^2
\label{nocdue}
\end{equation}
The next point in the algorithm leading to the calculation of the
period matrix on the compact exceptional divisor consists of the
replacement of the variables $X_{1,2,3}$  with new ones
$T_{1,2,3}$ related to the previous ones in the following way:
\begin{equation}
X_1=\frac{\sqrt{\lambda } T_1}{\sqrt{\sigma }} \quad ;\quad
X_2=\lambda ^2 T_{2 }\quad ;\quad X_3=\frac{\sqrt{\lambda }
T_3}{\sqrt{\sigma }}\label{3subia}
\end{equation}
The rationale of the transformation \eqref{3subia} is as follows.
The reduction to the compact exceptional divisor consists of setting
$w_1=0$ and hence $\lambda \to 0$. From the point of view of the
(1,1)-forms $\omega_I$ defined in eq.~\eqref{giraldo},
multiplication of $X_I$ by any power of the modulus square of any
complex coordinate is uneffective because of the logarithm. In other
words, instead of eq.~\eqref{giraldo} we can write equally
well:
\begin{equation}\label{carnacina}
    \omega_I^{(1,1)} \, = \, \frac{i}{2\pi} \, \partial \,
    \bar{\partial} \, \log\left[T_I\right]^{\alpha_I}
\end{equation}
The specific choice of powers of $\lambda$ and $\sigma$ utilized in
\eqref{3subia} is the result of a search of the appropriate values
that lead to a finite algebraic system with nontrivial solutions in
the two limits $\lambda\to 0$ and $\sigma\to 0$, corresponding
respectively to the compact and noncompact exceptional divisor.

 After these replacements in eqs.~\eqref{sakerdivoli}, the
final form of the system of the moment map equations is turned into
the following one: {\small
\begin{eqnarray}
&&\left(
\begin{array}{c}
 -T_1^2 \left(1+\lambda ^4 T_2^2\right)-\delta  T_1^3 T_3+\left(1+\lambda ^4 T_2^2\right) T_3^2+T_1 T_3
 \left(\delta  T_3^2+T_2 \left(-\zeta _1+\zeta
_3\right)\right) \\
 \delta  \lambda ^4 \sigma ^2 T_2^3-\delta  T_1 T_3 \left(T_1^2+T_3^2\right)
 +T_2 \left(\delta  \sigma ^2-T_1 T_3 \left(\zeta _1-\zeta _2+\zeta _3\right)\right)
\\
 -\left(-1+\lambda ^4 T_2^2\right) \left(T_1^2+\delta  \sigma ^2 T_2+T_3^2\right)-T_1 T_2 T_3 \zeta _2 \\
\end{array}
\right)\, = \,\left(
\begin{array}{c}
 0 \\
 0 \\
 0 \\
\end{array}
\right)\label{brillo}
\end{eqnarray}
}
\subsubsection{Equations of the two homology cycles}
In order to calculate the periods of the differential forms
\eqref{carnacina} on the basis of the homology curves
$C_1$ and $C_2$ we need the equations of
such loci in the coordinates $\lambda $, $\sigma$, $\delta $ defined
in equation \eqref{cavedanorosso}. We have
\begin{equation}
\begin{array}{lclclcl}
\text{Curve} \ C_1 & : & \sigma  = 0 &; &\lambda
\rightarrow 0
& ; & \delta = (1+\Delta )\\
\text{Curve} \ C_2 &: & \sigma  = \sqrt{\Delta }& ;
&\lambda \rightarrow  0 & ; & \delta = 1\\
\end{array}
\end{equation}
Conventionally, we denote $\Delta = \vert t\vert^2$  the modulus
squared of the complex coordinate $t$ spanning the curve, whatever it
might be.  In the case $C_1$ we have $\Delta = \vert u\vert^2$, while in the case $C_2$ we have
$\Delta=  \vert v\vert^2$.

In the sequel we use also polar coordinates, setting:
\begin{eqnarray}
\begin{array}{rclcrcl}
 \rho & \equiv & \sqrt{\left|u\right|^2}& ;& r &\equiv&
\sqrt{\left|v\right|^2} \nonumber\\
u & = & \rho \, \exp [i \theta] & ; & v & = & r \, \exp [i \psi]\\
\end{array}
\label{brchefreddo}
\end{eqnarray}
\subsubsection{Reduction to the compact exceptional divisor}
Furthermore it is convenient to introduce the following
variables, already utilized in eq.~\eqref{mhovarpi}:
\begin{equation}\label{ciuciosardo}
    \varpi \, \equiv \, \delta^2 \, \sigma^2 \quad ; \quad \mho \,
    \equiv \, \lambda^4
\end{equation}
and the following relation between the unknowns $T_{1,2,3}$ which
follows from linear combinations of the equations in the system
\eqref{brillo}
\begin{equation}\label{banalata}
    T_3\, = \, T_1 \sqrt{\frac{\zeta _2+\zeta _3 \left(T_2^2 \mho -1\right)}{\zeta _2+\zeta _1 \left(T_2^2 \mho
   -1\right)}}
\end{equation}
We see that in the limit $\mho \to 0$, which is the equation of the
compact exceptional divisor $D_c$, the expression
for $T_3$ is proportional to that for $T_1$ in any point of
$\zeta$ space. This has the nontrivial consequence that the
periods of $\omega_1$ and $\omega_3$ are always equal. In other
words the first Chern classes of the first and third tautological
bundles are always cohomologous.

Inserting eqs.~\eqref{banalata}, \eqref{ciuciosardo} into
\eqref{brillo} and performing the limit $\mho \to 0$ we obtain the
new system:
\begin{equation}\label{sacrabusta}
    \left(
\begin{array}{c}
 -\frac{\left(\zeta _1-\zeta _3\right) T_1^2 \left(\delta  \sqrt{\frac{\zeta _3-\zeta _2}{\zeta _1-\zeta _2}}
   T_1^2+\left(\zeta _1-\zeta _2\right) \sqrt{\frac{\zeta _3-\zeta _2}{\zeta _1-\zeta _2}} T_2+1\right)}{\zeta
   _1-\zeta _2} \\
 T_2 \left(\varpi -\sqrt{\frac{\zeta _3-\zeta _2}{\zeta _1-\zeta _2}} \left(\zeta _1-\zeta _2+\zeta _3\right)
   T_1^2\right)-\frac{\delta  \left(\zeta _1-2 \zeta _2+\zeta _3\right) \sqrt{\frac{\zeta _3-\zeta _2}{\zeta
   _1-\zeta _2}} T_1^4}{\zeta _1-\zeta _2} \\
 \frac{\left(\zeta _2-\zeta _3\right) T_1^2}{\zeta _2-\zeta _1}-\zeta _2 \sqrt{\frac{\zeta _3-\zeta _2}{\zeta
   _1-\zeta _2}} T_2 T_1^2+T_2 \varpi +T_1^2 \\
\end{array}
\right)\, = \,\left(
                \begin{array}{c}
                  0 \\
                  0 \\
                  0 \\
                \end{array}
              \right)
\end{equation}
which is the appropriate reduction to the exceptional divisor
$D_c$ of the moment map equations. By construction
the three equations in \eqref{sacrabusta} are linear dependent and
can be solved for the two unknowns $T_{1,2}$ in terms of the
variables $\delta,\varpi$. Indeed they are quadratic, cubic, or
quartic and can be solved by radicals.

\subsection{Periods inside the
chambers.} Let us begin with the periods in the interior points of
the chambers, the result of whose calculation was summarized in
Table \ref{periodico}. We found it convenient to choose eight
rational points as representatives of the eight chambers. Explicitly we
utilize the following ones:
\begin{equation}\label{pirillini}
    \begin{array}{|l|ccc||}
    \hline
 \text{Chamber 1}&\zeta _1\to -\frac{1}{2} & \zeta _2\to -\frac{3}{4} & \zeta _3\to -\frac{1}{2} \\[3pt]
 \hline
 \text{Chamber 2}&\zeta _1\to -\frac{1}{16} & \zeta _2\to -\frac{3}{16} & \zeta _3\to -\frac{3}{4} \\[3pt]
\hline
  \text{Chamber 3}&\zeta _1\to -\frac{1}{16} & \zeta _2\to -\frac{3}{4} & \zeta _3\to -\frac{3}{16} \\[3pt]
 \hline
\text{Chamber 4} &\zeta _1\to -\frac{3}{4} & \zeta _2\to -\frac{1}{16} & \zeta _3\to -\frac{3}{16} \\[3pt]
\hline
\text{Chamber 5}& \zeta _1\to -\frac{1}{4} & \zeta _2\to -\frac{1}{2} & \zeta _3\to \frac{1}{2} \\[3pt]
\hline
\text{Chamber 6}& \zeta _1\to -\frac{1}{4} & \zeta _2\to \frac{1}{2} & \zeta _3\to -\frac{1}{2} \\[3pt]
\hline
\text{Chamber 7}& \zeta _1\to \frac{3}{4} & \zeta _2\to -\frac{1}{4} & \zeta _3\to -\frac{1}{2} \\[3pt]
\hline
\text{Chamber 8}& \zeta _1\to \frac{3}{4} & \zeta _2\to \frac{1}{2} & \zeta _3\to \frac{3}{4} \\[3pt]
\hline
\end{array}
\end{equation}
Inserting the values \eqref{pirillini} into the system
\eqref{sacrabusta} we obtain eight algebraic systems that we
specialize to either the $C_1$ or the $C_2$ cycle by setting:
\begin{equation}\label{gramellino}
    \begin{array}{|l|rclcrcl|}
    \hline
       C_1 & \varpi & = & 0 & ; & \delta & = & \left(1+\rho^2\right) \\
       \hline
       C_2 & \varpi & = & r^2 & ; & \delta & = & 1\\
       \hline
     \end{array}
\end{equation}
The 16 linear systems obtained in this way can be solved for
$T_1,T_2$, and thanks to the relation \eqref{banalata}, each solution
for $T_{1,2}$ yields also a solution for $T_3$. Excluding the
spurious solutions $ T_1 \to 0,\, T_2\to 0$ and $T_1\to 0$ in
the case of the $C_1$ systems we find 2 branches, while
in the case of the $C_2$ systems we find 4 branches. We
know that in each case only one branch is the limit on the
considered curve of the unique positive real solution of the full
system \eqref{brillo} yet in absence of the analytic form of the
solution of \eqref{brillo} we do not know which is the right branch.
We circumvent this difficulty in the following way. First we remark
that in the case of a form:
\begin{equation}\label{grimaldello}
    \Omega \, = \, \tfrac{i}{2\pi} \partial \, \bar{\partial}\, \mathrm{J}(t,\bar{t})
\end{equation}
where $t=\mathfrak{r}\, e^{i \xi}$ is the complex coordinate
written in polar form, of a $\mathbb{P}^1$ , when
$\mathrm{J}(t,\bar{t})$ is a function only of the modulus
$\mathfrak{r}$
\begin{equation}\label{batuffolo}
    \mathrm{J}(t,\bar{t}) \, = \, {J}(\mathfrak{r})
\end{equation}
we have:
\begin{equation}\label{fioredicampo}
    \Omega \, = \, \tfrac{1}{4\pi} \,
    \left(\frac{d{J}(\mathfrak{r})}{d\mathfrak{r}}+\mathfrak{r}\,\frac{d^2{J}(\mathfrak{r})}{d\mathfrak{r}^2}\right)
    \, d\mathfrak{r} \wedge d \xi
\end{equation}
Correspondingly for the integral of $\Omega$ on the supporting space
we find:
\begin{equation}\label{giugurta}
    \int_{\mathbb{P}^1} \, \Omega \, = \, \tfrac{1}{2}\,
    \int_0^\infty \left(\frac{d{J}(\mathfrak{r})}{d\mathfrak{r}}+\mathfrak{r}\,
    \frac{d^2{J}(\mathfrak{r})}{d\mathfrak{r}^2}\right)\,
    d\mathfrak{r} \, = \, \left. \mathfrak{r} \,
    \frac{d{J}(\mathfrak{r})}{d\mathfrak{r}} \, \right|^{\infty}_{0}
\end{equation}
Utilizing this idea we calculate $\rho \,
\frac{d{J}_{1,2,3}(\rho)}{d\rho}$ in the case when
\begin{equation}\label{Jrho}
    \mathrm{J}_{1,2,3}(u,\bar{u}) \, = \,
    \log[T_{1,2,3}(\rho)]\,\mid_{C_1}
\end{equation}
is defined in term of a  solution $T_{1,2,3}(\rho)$ of  the
moment map equations reduced to $C_1$, and $r \,
\frac{d{J}_{1,2,3}(r)}{dr}$ in the case when
\begin{equation}\label{Jerre}
    \mathrm{J}_{1,2,3}(v,\bar{v}) \, = \,
    \log[T_{1,2,3}(r)]\,\mid_{C_2}
\end{equation}
is defined in term of a  solution $T_{1,2,3}(r)$ of  the moment map
equations reduced to $C_2$. In both cases we used all the
available nontrivial branches.
\paragraph{Cycle $C_1$.} In this case the result is very
simple and uniform. For all chambers and for all branches we always
have:
\begin{equation}\label{garatusco}
   \rho \frac{d{J}_{1,3}(\rho)}{d\rho} \, =\, -\frac{1}{ \left(\rho
   ^2+1\right)} \quad ; \quad \rho \frac{d{J}_{2}(\rho)}{d\rho} \,
   =\, 0
\end{equation}
This implies that in all chambers we  have the periods:
\begin{equation}\label{risultoC1}
    \int_{C_1} \, \omega_1 \, = \, 1 \quad ; \quad
    \int_{C_1} \, \omega_2 \, = \, 0 \quad ; \quad \int_{C_1} \, \omega_3 \, = \,1
\end{equation}
which is the result shown in Table \ref{periodico}. There is however
an implication of this universal result on the factor
$\alpha_\zeta$. Utilizing \eqref{risultoC1} in
eq.\eqref{saccius} we obtain that the volume of the cycle
$C_1$ is given by
\begin{equation}\label{caramellamu}
\left.\pmb{Vol}_1\right|_{\text{Chamber k}} \, = \,
\alpha _k \, \left(\zeta _1-\zeta _2+\zeta _3\right) \, = \,
\alpha_k \, \,\pmb{n}_2\,\cdot\,\zeta
\end{equation}
Hence in order for the volume of the cycle $C_1$ to be
positive in every chamber the factor $\alpha _k$ has to change sign
so as to compensate the negative sign of $
\pmb{n}_2\,\cdot\,\zeta$ when this occurs. Specifically we have:
\begin{equation}\label{alfetta}
    \begin{array}{|c|c|c|c|c|c|c|c|}
    \hline
       \text{Ch. I} & \text{Ch. II} & \text{Ch. III} & \text{Ch. IV} & \text{Ch. V} & \text{Ch. VI} &
       \text{Ch. VII} & \text{Ch. VIII} \\
       \hline
       \alpha_1 < 0 & \alpha_2 < 0 & \alpha_3 > 0 & \alpha_4 < 0 & \alpha_5 > 0 & \alpha_6 < 0 & \alpha_7 > 0 & \alpha_8
       > 0\\
       \hline
     \end{array}
\end{equation}
As we have previously mentioned, in the interiors of the chambers we always take
$\vert\alpha\vert=2$, while on some walls we have taken $\vert\alpha\vert=4$.

\paragraph{Cycle $C_2$.} In the case of the second cycle,
the result is more complex since, as we already mentioned, we have
four branches. The explicit analytic forms of $r \,
J^\prime_{1,2,3}(r)$ for all the branches and in all the chambers is
displayed in Table \ref{integraluschi}.
\begin{table}
  \centering
  $$ \begin{array}{|c||l|}
  \hline
  \hline
          \text{Chamber 1} & \begin{array}{l|ccc}
\null & -r \, J^\prime_1(r) & -r \, J^\prime_2(r)&
-r\,J^\prime_3(r)\\
\hline
 \text{br. 1} &\frac{4 r^2-1}{4 \sqrt{16 r^4-40 r^2+1}} &
\frac{4 r^2}{\sqrt{16 r^4-40 r^2+1}} & \frac{4 r^2-1}{4 \sqrt{16
r^4-40
   r^2+1}} \\
\text{br. 2} & \frac{4 r^2-1}{4 \sqrt{16 r^4-40 r^2+1}} & \frac{4
r^2}{\sqrt{16 r^4-40 r^2+1}} & \frac{4 r^2-1}{4 \sqrt{16 r^4-40
   r^2+1}} \\
\text{br. 3} & \frac{1-4 r^2}{4 \sqrt{16 r^4-40 r^2+1}} & -\frac{4
r^2}{\sqrt{16 r^4-40 r^2+1}} & \frac{1-4 r^2}{4 \sqrt{16 r^4-40
   r^2+1}} \\
\text{br. 4} & \frac{1-4 r^2}{4 \sqrt{16 r^4-40 r^2+1}} & -\frac{4
r^2}{\sqrt{16 r^4-40 r^2+1}} & \frac{1-4 r^2}{4 \sqrt{16 r^4-40
   r^2+1}} \\
\end{array} \\
\hline \hline
          \text{Chamber 2} & \begin{array}{l|ccc}
\text{br. 1} & \frac{5-8 r^2}{4 \sqrt{64 r^4+32 r^2+25}} & -\frac{8
r^2}{\sqrt{64 r^4+32 r^2+25}} & \frac{5-8 r^2}{4 \sqrt{64 r^4+32
   r^2+25}} \\
 \text{br. 2} &\frac{5-8 r^2}{4 \sqrt{64 r^4+32 r^2+25}} & -\frac{8 r^2}{\sqrt{64 r^4+32 r^2+25}}
 & \frac{5-8 r^2}{4 \sqrt{64 r^4+32
   r^2+25}} \\
\text{br. 3} & \frac{8 r^2-5}{4 \sqrt{64 r^4+32 r^2+25}} & \frac{8
r^2}{\sqrt{64 r^4+32 r^2+25}} & \frac{8 r^2-5}{4 \sqrt{64 r^4+32
   r^2+25}} \\
\text{br. 4} & \frac{8 r^2-5}{4 \sqrt{64 r^4+32 r^2+25}} & \frac{8
r^2}{\sqrt{64 r^4+32 r^2+25}} & \frac{8 r^2-5}{4 \sqrt{64 r^4+32
   r^2+25}} \\
\end{array} \\
          \hline
          \hline
         \text{Chamber 3} & \begin{array}{l|ccc}
 \text{br. 1}&\frac{2 r^2+1}{4 \sqrt{4 r^4-16 r^2+1}} & \frac{2 r^2}{\sqrt{4 r^4-16 r^2+1}} &
 \frac{2 r^2+1}{4 \sqrt{4 r^4-16
   r^2+1}} \\
\text{br. 2}& \frac{2 r^2+1}{4 \sqrt{4 r^4-16 r^2+1}} & \frac{2
r^2}{\sqrt{4 r^4-16 r^2+1}} & \frac{2 r^2+1}{4 \sqrt{4 r^4-16
   r^2+1}} \\
\text{br. 3}& \frac{-2 r^2-1}{4 \sqrt{4 r^4-16 r^2+1}} & -\frac{2
r^2}{\sqrt{4 r^4-16 r^2+1}} & \frac{-2 r^2-1}{4 \sqrt{4 r^4-16
   r^2+1}} \\
\text{br. 4}& \frac{-2 r^2-1}{4 \sqrt{4 r^4-16 r^2+1}} & -\frac{2
r^2}{\sqrt{4 r^4-16 r^2+1}} & \frac{-2 r^2-1}{4 \sqrt{4 r^4-16
   r^2+1}} \\
\end{array}\\
         \hline
          \hline
         \text{Chamber 4} & \begin{array}{l|ccc}
\text{br. 1}& \frac{7-8 r^2}{4 \sqrt{64 r^4+96 r^2+49}} & -\frac{8
r^2}{\sqrt{64 r^4+96 r^2+49}} & \frac{7-8 r^2}{4 \sqrt{64 r^4+96
   r^2+49}} \\
\text{br. 2}& \frac{7-8 r^2}{4 \sqrt{64 r^4+96 r^2+49}} & -\frac{8
r^2}{\sqrt{64 r^4+96 r^2+49}} & \frac{7-8 r^2}{4 \sqrt{64 r^4+96
   r^2+49}} \\
\text{br. 3}& \frac{8 r^2-7}{4 \sqrt{64 r^4+96 r^2+49}} & \frac{8
r^2}{\sqrt{64 r^4+96 r^2+49}} & \frac{8 r^2-7}{4 \sqrt{64 r^4+96
   r^2+49}} \\
\text{br. 4}& \frac{8 r^2-7}{4 \sqrt{64 r^4+96 r^2+49}} & \frac{8
r^2}{\sqrt{64 r^4+96 r^2+49}} & \frac{8 r^2-7}{4 \sqrt{64 r^4+96
   r^2+49}} \\
\end{array}\\
         \hline
          \hline
         \text{Chamber 5} & \begin{array}{l|ccc}
\text{br. 1}& \frac{4 r^2+3}{4 \sqrt{16 r^4-56 r^2+9}} & \frac{4
r^2}{\sqrt{16 r^4-56 r^2+9}} & \frac{4 r^2+3}{4 \sqrt{16 r^4-56
   r^2+9}} \\
\text{br. 2}& \frac{4 r^2+3}{4 \sqrt{16 r^4-56 r^2+9}} & \frac{4
r^2}{\sqrt{16 r^4-56 r^2+9}} & \frac{4 r^2+3}{4 \sqrt{16 r^4-56
   r^2+9}} \\
\text{br. 3}& \frac{-4 r^2-3}{4 \sqrt{16 r^4-56 r^2+9}} & -\frac{4
r^2}{\sqrt{16 r^4-56 r^2+9}} & \frac{-4 r^2-3}{4 \sqrt{16 r^4-56
   r^2+9}} \\
\text{br. 4}& \frac{-4 r^2-3}{4 \sqrt{16 r^4-56 r^2+9}} & -\frac{4
r^2}{\sqrt{16 r^4-56 r^2+9}} & \frac{-4 r^2-3}{4 \sqrt{16 r^4-56
   r^2+9}} \\
\end{array} \\
         \hline
          \hline
         \text{Chamber 6} & \begin{array}{l|ccc}
\text{br. 1}& \frac{5-4 r^2}{4 \sqrt{16 r^4+72 r^2+25}} & -\frac{4
r^2}{\sqrt{16 r^4+72 r^2+25}} & \frac{5-4 r^2}{4 \sqrt{16 r^4+72
   r^2+25}} \\
\text{br. 2}& \frac{5-4 r^2}{4 \sqrt{16 r^4+72 r^2+25}} & -\frac{4
r^2}{\sqrt{16 r^4+72 r^2+25}} & \frac{5-4 r^2}{4 \sqrt{16 r^4+72
   r^2+25}} \\
\text{br. 3}& \frac{4 r^2-5}{4 \sqrt{16 r^4+72 r^2+25}} & \frac{4
r^2}{\sqrt{16 r^4+72 r^2+25}} & \frac{4 r^2-5}{4 \sqrt{16 r^4+72
   r^2+25}} \\
\text{br. 4}& \frac{4 r^2-5}{4 \sqrt{16 r^4+72 r^2+25}} & \frac{4
r^2}{\sqrt{16 r^4+72 r^2+25}} & \frac{4 r^2-5}{4 \sqrt{16 r^4+72
   r^2+25}} \\
\end{array} \\
         \hline
          \hline
         \text{Chamber 7} & \begin{array}{l|ccc}
\text{br. 1}& \frac{-2 r^2-1}{4 \sqrt{4 r^4-8 r^2+1}} & \quad
-\frac{2 r^2}{\sqrt{4 r^4-8 r^2+1}} \quad & \frac{-2 r^2-1}{4
\sqrt{4 r^4-8
   r^2+1}} \\
\text{br. 2}& \frac{-2 r^2-1}{4 \sqrt{4 r^4-8 r^2+1}} & -\frac{2
r^2}{\sqrt{4 r^4-8 r^2+1}} & \frac{-2 r^2-1}{4 \sqrt{4 r^4-8
   r^2+1}} \\
\text{br. 3}& \frac{2 r^2+1}{4 \sqrt{4 r^4-8 r^2+1}} & \frac{2
r^2}{\sqrt{4 r^4-8 r^2+1}} & \frac{2 r^2+1}{4 \sqrt{4 r^4-8 r^2+1}}
   \\
\text{br. 4}& \frac{2 r^2+1}{4 \sqrt{4 r^4-8 r^2+1}} & \frac{2
r^2}{\sqrt{4 r^4-8 r^2+1}} & \frac{2 r^2+1}{4 \sqrt{4 r^4-8 r^2+1}}
   \\
\end{array} \\
         \hline
          \hline
        \text{Chamber 8} & \begin{array}{l|ccc}
\text{br. 1}& \quad\frac{-r^2-1}{4 \sqrt{r^4+1}}\quad &\quad\quad
-\frac{r^2}{\sqrt{r^4+1}}\quad\quad &
\quad \frac{-r^2-1}{4 \sqrt{r^4+1}}\quad \\
\text{br. 2}& \frac{-r^2-1}{4 \sqrt{r^4+1}} & -\frac{r^2}{\sqrt{r^4+1}} & \frac{-r^2-1}{4 \sqrt{r^4+1}} \\
\text{br. 3}& \frac{r^2+1}{4 \sqrt{r^4+1}} & \frac{r^2}{\sqrt{r^4+1}} & \frac{r^2+1}{4 \sqrt{r^4+1}} \\
\text{br. 4}& \frac{r^2+1}{4 \sqrt{r^4+1}} & \frac{r^2}{\sqrt{r^4+1}} & \frac{r^2+1}{4 \sqrt{r^4+1}} \\
\end{array} \\
        \hline
          \hline
          \end{array}
   $$
  \caption{The indefinite integrals for the calculation of periods of the tautological Chern Classes along
  the cycle $C_2$.}\label{integraluschi}
\end{table}
Utilizing eq. \eqref{giugurta} and the results of Table
\ref{integraluschi} we obtain the candidate values of the periods
that are displayed in Table \ref{mamertino}.
\begin{table}\small
\renewcommand{\arraystretch}{1.20}
  \centering
  $$ \begin{array}{|c||l|}
  \hline
  \hline
          \text{Chamber 1} & \begin{array}{l|ccc}
          \null & \int_{C_2}\, \omega_1 & \int_{C_2}\,
          \omega_2& \int_{C_2}\, \omega_3\\
          \hline
 \text{branch}_1 & -1 & -2 & -1 \\
 \text{branch}_2 & -1 & -2 & -1 \\
 \text{branch}_3 & 1 & 2 & 1 \\
 \text{branch}_4 & 1 & 2 & 1 \\
\end{array}\\
\hline \hline
          \text{Chamber 2} & \begin{array}{l|ccc}
          \null & \int_{C_2}\, \omega_1 & \int_{C_2}\,
          \omega_2& \int_{C_2}\, \omega_3\\
          \hline
 \text{branch}_1 & 1 & 2 & 1 \\
 \text{branch}_2 & 1 & 2 &  1 \\
 \text{branch}_3 & -1 & -2 & -1 \\
 \text{branch}_4 & -1 & -2 & -1 \\
\end{array} \\
          \hline
          \hline
         \text{Chamber 3} & \begin{array}{l|ccc}
         \null & \int_{C_2}\, \omega_1 & \int_{C_2}\,
          \omega_2& \int_{C_2}\, \omega_3\\
          \hline
 \text{branch}_1 & 0 & -2 & 0 \\
 \text{branch}_2 & 0 & -2& 0 \\
 \text{branch}_3 & 0 & 2 & 0 \\
 \text{branch}_4 & 0 & 2 & 0 \\
\end{array}\\
         \hline
          \hline
         \text{Chamber 4} & \begin{array}{l|ccc}
         \null & \int_{C_2}\, \omega_1 & \int_{C_2}\,
          \omega_2& \int_{C_2}\, \omega_3\\
          \hline
 \text{branch}_1 & 1 & 2 & 1 \\
 \text{branch}_2 & 1 & 2 & 1 \\
 \text{branch}_3 & -1 & -2 & -1 \\
 \text{branch}_4 & -1 & -2 & -1\\
\end{array} \\
         \hline
          \hline
         \text{Chamber 5} & \begin{array}{l|ccc}
         \null & \int_{C_2}\, \omega_1 & \int_{C_2}\,
          \omega_2& \int_{C_2}\, \omega_3\\
          \hline
 \text{branch}_1 & 0 & -2 & 0 \\
 \text{branch}_2 & 0 & -2 & 0 \\
 \text{branch}_3 & 0 & 2 & 0 \\
 \text{branch}_4 & 0 & 2 & 0 \\
\end{array} \\
         \hline
          \hline
         \text{Chamber 6} & \begin{array}{l|ccc}
         \null & \int_{C_2}\, \omega_1 & \int_{C_2}\,
          \omega_2& \int_{C_2}\, \omega_3\\
          \hline
 \text{branch}_1 & 1 & 2 & 1 \\
 \text{branch}_2 & 1 & 2 & 1 \\
 \text{branch}_3 & -1 & -2 & -1 \\
 \text{branch}_4 & -1& -2 & -1\\
\end{array}\\
         \hline
          \hline
         \text{Chamber 7} & \begin{array}{l|ccc}
         \null & \int_{C_2}\, \omega_1 & \int_{C_2}\,
          \omega_2& \int_{C_2}\, \omega_3\\
          \hline
 \text{branch}_1 & 0 & 2 & 0 \\
 \text{branch}_2 & 0 & 2 & 0 \\
 \text{branch}_3 & 0 & -2& 0 \\
 \text{branch}_4 & 0 & -2 & 0 \\
\end{array} \\
         \hline
          \hline
        \text{Chamber 8} & \begin{array}{l|ccc}
        \null & \int_{C_2}\, \omega_1 & \int_{C_2}\,
          \omega_2& \int_{C_2}\, \omega_3\\
          \hline
 \text{branch}_1 & 0 & 2 & 0 \\
 \text{branch}_2 & 0 & 2& 0 \\
 \text{branch}_3 & 0 & -2 & 0 \\
 \text{branch}_4 & 0 & -2 & 0 \\
\end{array} \\
        \hline
          \hline
          \end{array}
   $$
  \caption{The candidate period integrals  of the tautological Chern classes along
  the cycle $C_2$.}\label{mamertino}
\end{table}
As one sees from that table, in every chamber there is only the
ambiguity of an overall sign. In view of the result \eqref{alfetta}
for the factor $\alpha_k$ relative to the various chambers, in each
of them we choose the overall sign for the $C_2$ periods
 that leads to a positive value for the volume of that
cycle, according to equation \eqref{saccius}. Performing such a
choice, one finally arrives at the result displayed in Table
\ref{periodico}.

\subsection{Periods on the walls and on the edges}
Utilizing the same algorithm as in the case of the interior of the
chambers, we have calculated the periods also on the walls and on the
edges obtaining the results displayed in Tables \ref{muraria} and
\ref{spigolosa}. We skip the details for the type 0 walls
$\mathcal{W}_{1,3}$, since these calculations are identical with
those presented in the previous subsection, simply with different
values of the $\zeta$ parameters. For the case of the wall
$\mathcal{W}_2$ the detailed calculation of the Kamp\'{e} case
presented in section \ref{Y3Kallero}  produces the periods displayed in Table \ref{muraria}. Additional care is
instead required while treating the case of the type 0 wall
$\mathcal{W}_0$ and the Cardano edges.

\subsubsection{The type 0 wall $\mathcal{W}_0$}
If we choose:
\begin{equation}\label{cramenio}
    \zeta \, = \, \left\{p,0,q\right\} \quad ; \quad p,q\in \mathbb{R}
\end{equation}
we are, by definition,  on the wall $\mathcal{W}_0$. The main
property of this wall is that the moment map algebraic system
\eqref{sistemico} implies
\begin{equation}\label{xdueuno}
    X_2 \, = \, 1\,.
\end{equation}
Hence the rescaling ansatz \eqref{3subia} is not appropriate on this
wall for the reduction of the moment map equations to the
exceptional compact divisor $D_c$. Another rescaling
scheme is required.

Choosing \eqref{cramenio} and implementing eq.~\eqref{xdueuno} the
third of eqs.~\eqref{sistemico} is automatically satisfied and we
are left with
\begin{equation}\label{carmencitaabitaqui}
    \left(
\begin{array}{c}
 X_3 X_1 \left(-p+q+X_3^2 \sqrt[4]{\mho } \sqrt{\delta  \varpi }\right)+X_3 X_1^3 \sqrt[4]{\mho } \left(-\sqrt{\delta
    \varpi }\right)-2 X_1^2 \sqrt{\mho }+2 X_3^2 \sqrt{\mho } \\
 2 \sqrt[4]{\mho } \sqrt{\delta  \varpi }-X_3 X_1 \left(p+q+X_3^2 \sqrt[4]{\mho } \sqrt{\delta  \varpi }\right)+X_3
   X_1^3 \sqrt[4]{\mho } \left(-\sqrt{\delta  \varpi }\right) \\
 0 \\
\end{array}
\right) \, = \, \left(
                  \begin{array}{c}
                    0 \\
                    0 \\
                    0 \\
                  \end{array}
                \right)
\end{equation}
where we used the same real variables $\delta,\varpi,\mho$
utilized in previous sections. Given eq.~\eqref{carmencitaabitaqui},
an appropriate rescaling that assures a finite limit $\mho\to 0$ is
provided by the one below:
\begin{equation}\label{iconica}
    X_1 \, = \,  T_1 \mho ^{3/8} \quad , \quad X_3\, = \,  \frac{T_3}{\sqrt[8]{\mho
    }}
\end{equation}
Implementing the above substitution in the system
\eqref{carmencitaabitaqui} and factoring out a common factor
$\mho^{1/4}$ we can perform the limit $\mho \to 0$ and we get:
\begin{equation}\label{risculando}
  \left(
\begin{array}{c}
 T_3 \left(T_1 \left(-p+q+T_3^2 \sqrt{\delta  \varpi }\right)+2 T_3\right) \\
 2 \sqrt{\delta  \varpi }-T_1 T_3 \left(p+q+T_3^2 \sqrt{\delta  \varpi }\right) \\
 0 \\
\end{array}
\right) \, = \, \left(
                  \begin{array}{c}
                    0 \\
                    0 \\
                    0 \\
                  \end{array}
                \right)
\end{equation}
The limiting system \eqref{risculando} is solvable by radicals and
it has four branches that can be treated exactly as in the case of
the previous section, calculating the derivatives with respect to the
$\rho$ and $r$ variables that lead to the evaluation of the K\"ahler
form and of its integrals. An additional care which is required in
this case is that one has to calculate first the derivatives and
then set either $r$ or $\rho$ to zero. Doing the operations in the
opposite order one meets undefined limits for either $T_1$ or $T_2$.
Just as above we have to choose the right branch in order to get a
positive volume for the two homology cycles. Through these steps one
arrives at the result displayed for $\mathcal{W}_0$ in Table
\ref{muraria}.
\subsubsection{Periods of the Cardano manifold}
The case of the Cardano manifold was treated in section
\ref{kallusquidam}. In particular it was shown that the K\"ahler
potential for the two instances of this manifold is given by
eq.~\eqref{croccus}, where $X=\mathfrak{X}(\varpi,\mho)$ is the 4th
root of the quartic equation \eqref{baldoppo}, explicitly written
down in eq.~\eqref{caponatasiciliana}. At the same time the triplet
of (1,1)-forms $\omega_I$ is defined as
\begin{eqnarray}
\begin{array}{cccccccc}
  \zeta=\left\{1,1,0\right\} & \omega_1=0 & \omega_2 = \Omega & \omega_3= \Omega & ; &\Omega & = &
  \frac{i}{2\pi} \, \partial \bar{\partial} \,\mathfrak{X}(\varpi,\mho)^2  \\
  \zeta=\left\{0,1,1\right\} & \omega_1=\Omega & \omega_2 = \Omega & \omega_3= 0 & ; & \Omega & = &
  \frac{i}{2\pi} \, \partial \bar{\partial} \,\mathfrak{X}(\varpi,\mho)^2
\end{array}
\end{eqnarray}
Developing the function $\mathfrak{X}(\varpi,\mho)$ in power series
of the variable $\mho$ we find
\begin{equation}\label{sviluserio}
 \mathfrak{X}(\varpi,\mho) \, = \,  \frac{1}{2} \left(\sqrt{\varpi }+\sqrt{\varpi +4}\right)
 +\frac{1}{12} \left(-\frac{6 \left(\varpi ^2+3 \varpi
   \right)}{\sqrt{\varpi +4}}-6 \sqrt{\varpi } (\varpi +1)\right) \sqrt{\mho } +\mathcal{O}\left(\mho\right)\,.
\end{equation}
In this case the reduction of the forms to the exceptional compact
divisor can be performed in complete analytic safety
restricting $\mathfrak{X}(\varpi,\mho)$ to its zeroth order term in
$\mho$. In this way we obtain the precise expression of the
(1,1)-form $\Omega$ whose integrals are then easily calculated. From
eq.~\eqref{sviluserio} and from the definition of the variable
$\varpi$ we arrive at the conclusion that
\begin{eqnarray}\label{Omegone}
  \Omega & = &   \frac{i}{2\pi} \, \partial \bar{\partial}
  \,\mathfrak{Q}(\rho,r) \nonumber\\
\mathfrak{Q}(\rho,r)& = & \log \left(\frac{\left(\rho ^2+1\right)^2
r^2+\sqrt{\left(\rho ^2+1\right)^2 r^2} \sqrt{\left(\rho
^2+1\right)^2
   r^2+4}+4}{2 \sqrt{\left(\rho ^2+1\right)^2 r^2+4}}\right)
\end{eqnarray}
From the above result it immediately follows that:
\begin{equation}\label{integomegone}
   \left. \Omega \right|_{C_1}  \, = \, 0 \quad ; \quad
   \int_{C_2} \Omega \, = \, -1\,.
\end{equation}
This  yields the values of the periods as displayed in
Table \ref{spigolosa}.

It appears that the Cardano manifold is another realization of the
$Y_3$ degeneration of the full resolution $Y$ just as the Kamp\'{e}
manifold to be discussed in section \ref{Y3Kallero}, see in particular subsection
\ref{interpretazia}. The same arguments presented there apply here
as well and lead to the same conclusion. Hence also the Cardano
manifold is a line bundle over the  singular weighted projective
space $\mathbb{P}[1,1,2]$.


\subsection{Summary of the chamber and wall structure}
We have 8 chambers and 4 walls.

\begin{itemize} \item All interiors of the 8 chambers correspond
to the full resolution $Y$. This is consistent with Theorem 1.1 in \cite{CrawIshii}, which states that there exists
a chamber where the variety corresponding to the generic point is a resolution of singularities, and with
the fact that according to toric geometry there appears to be only one full resolution of singularities.
\item Among the four walls $\mathcal W_i$, only $\mathcal W_2$ is if type III, while the others are type 0.
The generic point on $\mathcal W_2$ corresponds to the partial resolution $Y_3$ (note that $Y_3$ is obtained
from $Y$ by  collapsing the noncompact exceptional divisor to a line).
\item The triple intersection $\mathcal W_0\cap\mathcal W_1 \cap \mathcal W_3$
is an edge   whose generic point corresponds to
the smooth variety $Y_{EH}$. This is quite interesting as $Y_{EH}$ is not a resolution of singularities
of $\C^3/\Z_4$; a morphism $Y_{EH}\to \C^3/\Z_4$ exists, but it is 2:1, as we discussed in Section
\ref{relation}.
\end{itemize}

\section{The resolved variety $Y$}
\label{Ysezia} As stressed in the previous Section about the chamber
structure, for all points of the $\zeta$ space that do not lay
on walls, the topological and algebraic character of the resolution
obtained from the K\"ahler quotient \`{a} la Kronheimer is always the
same, namely the variety we named $Y$. Hence, in order to describe
the  K\"ahler Geometry of the resolved variety $Y
 \rightarrow  \mathbb{C}^3/\mathbb{Z}_{4 }$, we can utilize any preferred convenient point in $\zeta$ space
that avoids the walls. Furthermore we utilize the crucial
information that $Y$ is the total space of the canonical line bundle over the second Hirzebruch surface $\mathbb{F}_2$. \\
The strategy that we adopt to find the explicit form of the K\"ahler  geometry of the variety $Y$ is based on the following steps:
\begin{enumerate}
\item  We choose the realization of the $Y$ space provided by the Kronheimer construction
in the plane $\zeta _1 = \zeta _3 =a$ and $\zeta _2 =b \neq
2a$. In this case, as we know, two of the tautological fiber bundles
are identified having   $X_1=X_3$ and the moment
map equations are simpler.
\item We reduce the moment map equations to the compact exceptional divisor (the second Hirzebruch surface)
and we obtain a system solvable by radicals whose explicit solution
is particularly simple.
\item We  obtain the complete solution for the full variety starting from the solution
on the Hirzebruch surface and expressing the required fiber metrics
$T_{1,2}$ in terms of a unique function of two variables that is
defined as a particular root of a sextic equation.
\end{enumerate}
\subsection{The two addends of the K\"ahler potential}
First we write the restriction to the hypersurface
$\mathcal{N}_{\zeta }$  of the  K\"ahler potential of the flat
variety  $\text{Hom}(Q\otimes R,R)^{\Z_4}$.  It is the following
object.
\begin{equation}
\mathcal{K}_0=\frac{U \left(X_2^2+1\right)
\left(X_1^2+X_3^2\right)+\Sigma \left(X_2^3+X_2+X_1 X_3
\left(X_1^2+X_3^2\right)\right)}{X_1 X_2 X_3} \label{Kappazero}
\end{equation}
As we know from eq.~\eqref{caramboletta} the final K\"ahler potential
of the resolved variety is of the form
\begin{equation}
\mathcal{K}=\mathcal{K}_0\,+\,\mathcal{K}_{\log }\,\\
\end{equation}
where
\begin{equation}
\mathcal{K}_{\log }\, \equiv \, \sum _{I=1}^3 \, \sum _{J=1}^3\zeta
_I \, \mathfrak{C}^{IJ} \, \log\left[X_J\right]^{\alpha_\zeta}\,
 \label{kappalogatto}
\end{equation}
is the logarithmic part of the K\"ahler potential that contains the
information on the tautological bundle Chern classes. The matrix
$\mathfrak{C}^{IJ}$ is defined by
\begin{equation}
\mathfrak{C}^{IJ}= \text{Tr}\left(\tau ^I\tau ^J\right) \, =
\, \left(
     \begin{array}{ccc}
       2 & -1 & 0 \\
       -1 & 2 & -1 \\
       0 & -1 & 2 \\
     \end{array}
   \right) \, = \, \text{Cartan matrix of $\mathfrak{a}_3$}
\end{equation}
where $\tau ^I$ are the generators of the U(1)$\times $U(1)$\times
$U(1) gauge group $\mathcal{F}_{\mathbb{Z}_4}$ in the 4$\times $4
dimensional representation corresponding to the regular
representation of $\mathbb{Z}_4$ advocated by the Kronheimer
construction.
Next, according to the general strategy outlined at the beginning
of this Section we extract the K\"ahler geometry induced on the
compact exceptional divisor by the Kronheimer--like crepant
resolution of the orbifold singularity.

Explicitly, we derive the K\"ahler potential of the Hirzebruch surface
from the K\"ahler quotient construction when
\begin{equation}
\zeta _1=\zeta _3=a \quad ; \quad  \zeta _2 =b \neq 2a.
\end{equation}
\subsubsection{Reduction to the compact exceptional
divisor of the moment map equations} The final form of the moment
map equations in the  coordinates \eqref{cavedanorosso} was given
above in eq.~\eqref{brillo}. After some experiments we found that a
convenient point in   chamber number 2
 is
 \begin{equation}
\zeta _1 =\zeta _3=\frac{1}{2} \quad ; \quad \zeta _2=2\,.
\end{equation}
If we choose such values for the moment map levels and furthermore
if we set $T_3 = T_1,$ as indeed we must do in this case, the
 system \eqref{brillo} reduces to
\begin{equation}
\left(
\begin{array}{c}
 0 \\
 -2 \delta  T_1^4+T_1^2 T_2+\delta  \sigma ^2 T_2 \left(1+\lambda ^4 T_2^2\right) \\
 -2 T_1^2 T_2-\left(2 T_1^2+\delta  \sigma ^2 T_2\right) \left(-1+\lambda ^4 T_2^2\right) \\
\end{array}
\right)=\left(
\begin{array}{c}
 0 \\
 0 \\
 0 \\
\end{array}
\right)\,.
\end{equation}
\subsection{Exact solution of the moment map equations} Let us
consider the following six order algebraic equation for an unknown
$F$:

\begin{eqnarray}
&&\mathfrak{P}(F) \equiv 2+F (-5-\varpi )+F^2 (3-2 \mho )+2 F^6 \mho
^3+F^3 (2 \mho +2 \varpi  \mho )\nonumber\\
&&+F^4 \left(\mho -2 \mho ^2\right)+F^5 \left(3 \mho ^2-\varpi  \mho
^2\right) = 0 \label{prillini}
\end{eqnarray}
 The coefficients of the above equations are written in terms of
the two quantities:
\begin{equation}\label{prillini}
\varpi  = \delta ^2 \sigma ^2 = \left(1+\left|u|^2\right. \right)^2
|v|^2 \quad ; \quad \mho  = \lambda ^4 =\left|w|^2 \right.
\end{equation}
 which depend on the toric coordinates $u,v,w$ of the variety $Y$.
Since $w=0$ is the equation of the compact component of the
exceptional divisor, which is isomorphic to the second Hirzebruch
surface $\mathbb{F}_2$, it follows that $u,v$ are coordinates of
$\mathbb{F}_2$. Furthermore taking into account the fibered
structure of $\mathbb{F}_2$$\longrightarrow $ $\mathbb{P}_1$ which
is a $\mathbb{P}_1$-bundle over $\mathbb{P}_1$, $u$ is a coordinate
for the $\mathbb{P}_1$ basis while $v$ is a coordinate for the
$\mathbb{P}_1$ fiber.
\par
Equation \eqref{prillini} has six roots that implicitly { }define as many
functions of { }$\varpi $, $\mho $. The sextic { }polynomial
$\mathfrak{P}$(F) has the important property that for $\varpi >0 $
and $\mho >0 $ there are always two real roots and two pairs of
complex conjugate roots. Hence the largest real root is a unique
identification of one of the six roots. We unambiguously define a
function $\mathfrak{F}$($\varpi $,$\mho $) by saying that it is the
largest real root of equation (4)
\begin{equation}
\mathfrak{F}(\varpi ,\mho ) \, \equiv \, \text{largest real  root
 of } \, \mathfrak{P}(F)\,.
\end{equation}
\par
The exact solution of the moment map equations can be written as
follows: {
\begin{eqnarray}
T_1&=& \sqrt{\frac{\mathfrak{F}}{2 \delta }} \left(-\varpi
+\mathfrak{F}^2 (3+2 \varpi ) \mho +2 \mathfrak{F}^5 \mho ^3-2
(1+\mho )+\mathfrak{F}^4 \mho ^2 (3-\varpi +2 \mho
)\right.\nonumber\\
&&\left.+\mathfrak{F}^3 \mho (1+\mho -\varpi  \mho )+\mathfrak{F}
(3+\mho +\varpi  \mho
)\right)^{\frac{1}{2}}\nonumber\\
T_2 &=&\mathfrak{F} \label{esatamence}
\end{eqnarray}
} where by $\mathfrak{F}$ we obviously mean $\mathfrak{F}( \varpi ,
\mho )$.
\subsubsection{Properties of the function $\mathfrak{F}(\varpi,\mho)$}
 The function $\mathfrak{F}$($\varpi $,$\mho $) is
well defined and it can be developed in power series of the
parameter $\mho $:
\begin{equation}
\mathfrak{F}(\varpi ,\mho )\, = \, \sum _{n=0}^{\infty }
\mathfrak{F}_n(\varpi )\, \mho ^n
\end{equation}
We display the first two terms of this development:
\begin{eqnarray}
\mathfrak{F}_0(\varpi )&=&\frac{1}{6} \left(5+\varpi +\sqrt{1+10
\varpi +\varpi ^2}\right)\nonumber\\
\mathfrak{F}_1(\varpi )&=&-\left(\left(\left(5+\varpi
+\sqrt{1+\varpi (10+\varpi )}\right)^2 \left(7+11 \sqrt{1+\varpi
(10+\varpi )}\right.\right.\right.\nonumber\\
&&\left.\left.\left. +\varpi  \left(46+7 \varpi +7 \sqrt{1+\varpi
(10+\varpi )}\right)\right)\right)/\left(648 \sqrt{1+\varpi
(10+\varpi )}\right)\right) \label{Fgotsvilup}
\end{eqnarray}
The function $\mathfrak{F}(\varpi , \mho)$ can also be plotted and
its behavior is displayed in fig. \ref{effegotica}.
\begin{figure}
\label{effegotica} \vskip -1.5 cm
\begin{center}
\includegraphics[height=8cm]{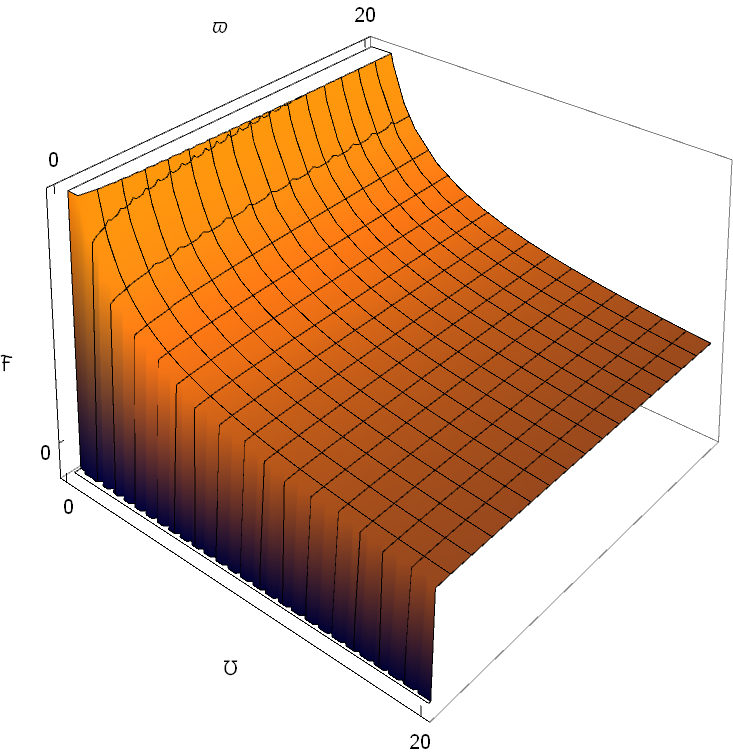}
\caption{\label{effegotica}{ Plot of the function
$\mathfrak{F}(\varpi , \mho)$ }}
\end{center}
\end{figure}
\subsection{Induced K\"ahler geometry of the exceptional divisor $D_c \sim
\mathbb{F}_2$} \label{inducedF2}
 Next we study the K\"ahler geometry of the compact
exceptional divisor induced by the Kronheimer construction of the
K\"ahler geometry of $Y$.
\subsubsection{Solution of the moment map equations reduced to the
compact exceptional divisor}
If we perform the limit $\lambda \rightarrow $ 0 in the moment map
equations \eqref{prillini} we reduce them to compact exceptional divisor, namely
to second Hirzebruch surface
\begin{equation}
\left(
\begin{array}{c}
 0 \\
 -2 \delta  T_1^4+\left(\delta  \sigma ^2+T_1^2\right) T_2 \\
 \delta  \sigma ^2 T_2-T_1^2 \left(-2+2 T_2\right) \\
\end{array}
\right)=\left(
\begin{array}{c}
 0 \\
 0 \\
 0 \\
\end{array}
\right) \label{sistema12}
\end{equation}
The system \eqref{sistema12} has five different solutions but the
only one that is positive real both for $T_1$ and $T_2$ is the
following one:
\begin{eqnarray}
T_1 &=&\frac{1}{2} \sqrt{\frac{1+\delta ^2 \sigma ^2+\sqrt{1+10
\delta ^2 \sigma ^2+\delta ^4 \sigma ^4} }{\delta }}\nonumber\\
T_2 &=&\frac{1}{6} \left(5+\delta ^2 \sigma ^2+\sqrt{1+10 \delta ^2
\sigma ^2+\delta ^4 \sigma ^4}\right)
\end{eqnarray}
This  in terms of the invariants reads as follows:
\begin{eqnarray}
T_1|_{\mathbb{F}_2} &=&\frac{1}{2} \sqrt{\frac{1+\varpi +\sqrt{1+10
\varpi +\varpi ^2} }{\delta }}\text{    };\text{
}\nonumber\\
T_2|_{\mathbb{F}_2}&=&\frac{1}{6} \left(5+\varpi +\sqrt{1+10 \varpi
+\varpi ^2}\right) = \mathfrak{F}_0(\varpi ) \label{soluziacinque}
\end{eqnarray}
Comparing with eqs. \eqref{Fgotsvilup} and \eqref{esatamence} we see
that indeed the reduction to the compact exceptional divisor of the
solution corresponds to the zero-th order terms in the series
expansion in $\mho $.
\subsubsection{Derivation of the K\"ahler potential of
$\mathbb{F}_2$.} In order to study the K\"ahler geometry we have
first to calculate the K\"ahler potential. Performing  the
substitutions \eqref{3subia} and \eqref{nocdue} in $\mathcal{K}_0$
as defined by equation \eqref{Kappazero}, substituting the solution
\eqref{esatamence} for $T_{1,2}$ and performing the limit $\lambda
\rightarrow $0, namely $\mho \, \rightarrow\, 0$, we obtain:
\begin{equation}
\mathcal{K}_0|_{\mathbb{F}_2}=\frac{3+7 \varpi +3 \sqrt{1+10 \varpi
+\varpi ^2}}{1+\varpi +\sqrt{1+10 \varpi +\varpi ^2}}\,.
\end{equation}
For the logarithmic part of the K\"ahler potential defined in
eq. \eqref{kappalogatto} we have instead
\begin{equation}
\mathcal{K}_{\log }|_{\mathbb{F}_2}\equiv \frac{\kappa _1}{2}
\log\left[\frac{1}{2} \sqrt{\frac{1+\varpi +\sqrt{1+10 \varpi
+\varpi ^2} }{\delta }}\right]+\kappa _2\log\left[\frac{1}{6}
\left(5+\varpi +\sqrt{1+10 \varpi +\varpi ^2}\right) \right]\,.
\end{equation}
The parameters $\kappa _{1,2}$ have been introduced to keep track of
the consequences of  the pairing matrix $\mathfrak{C}_{IJ}$ choice.
\paragraph{The K\"ahler potential in polar coordinates.}
The final outcome of the above construction is that the K\"ahler
potential of the metric induced on the second Hirzebruch surface is
the following one
\begin{equation}
\mathcal{K}_{\mathbb{F}_2}=J\left(\rho ,r,\kappa _1,\kappa _2\right)
\label{samotracia}
\end{equation}
where $\rho$ and $r$, defined in eq.~\eqref{brchefreddo} are the
norms of the two complex coordinates and the function $J\left(\rho
,r,\kappa _1,\kappa _2\right)$ is the following explicit one:
\begin{eqnarray}
J(\rho ,r,\kappa _1,\kappa _2)&=&\frac{3+7r^2\left(1+\rho
^2\right)^2+3\sqrt{1+10r^2\left(1+\rho ^2\right)^2+r^4\left(1+\rho
^2\right)^4}}{1+r^2\left(1+\rho
^2\right)^2+\sqrt{1+10r^2\left(1+\rho ^2\right)^2+r^4\left(1+\rho
^2\right)^4}}\nonumber\\
&&+\frac{\kappa
_1}{2}\log\left[\frac{1}{2}\sqrt{\frac{1+r^2\left(1+\rho
^2\right)^2+\sqrt{1+10r^2\left(1+\rho ^2\right)^2+r^4\left(1+\rho
^2\right)^4}}{1+\rho ^2}}\right]\nonumber\\
&&+\kappa _2 \,\log\left[\frac{1}{6}\left(5+r^2\left(1+\rho
^2\right)^2+\sqrt{1+10r^2\left(1+\rho ^2\right)^2+r^4\left(1+\rho
^2\right)^4}\right)\right]\nonumber\\
\label{Jpolare}
\end{eqnarray}

\subsubsection{Calculation of the  K\"ahler metric and of the Ricci tensor of $\mathbb{F}_2$}
From the above data we can calculate the K\"ahler metric, the K\"ahler
2-form and also the determinant of the metric which finally yields the
Ricci tensor and the Ricci 2-form. We performed this calculation
utilizing a {\sc mathematica} code and in the following lines we present
these results. The best way to display them is in terms of polar
coordinates, namely performing the transformation
\begin{equation}
u=\exp[i \theta ]\,\rho \quad ;\quad v= \exp[i \psi ]\, r
\end{equation}
\paragraph{The K\"ahler 2-form.}
We do not display the explicit form of the K\"ahler metric since it
is too long and messy. As usual it is just given by the derivatives
with respect to the original complex coordinates of the K\"ahler
potential:
\begin{equation}
g_{IJ^*} = \partial _i\partial _{j^*}J(\rho
,r,\kappa_1,\kappa_2) \quad ; \quad z_i \equiv  \{u,v\} \quad ;
\quad \bar{z}_{j^*} \equiv
 \left\{\bar{u},\bar{v}\right\}\,;\label{Kmetricozza}
\end{equation}
once transformed to the polar coordinates $\rho,r,\theta,\psi$ the K\"ahler form has
the following structure:
\begin{eqnarray}
\mathbb{K}& = &   \mathbb{K}_{\text{r$\theta $}} \text{dr}\wedge
\text{d$\theta $}+\mathbb{K}_{\text{r$\rho $}} \text{dr}\wedge
\text{d$\rho $}-\mathbb{K}_{\text{$\psi $r}} \text{dr}\wedge
\text{d$\psi $} \nonumber\\ & + & \mathbb{K}_{\theta \rho } \text{d$\theta $}\wedge
\text{d$\rho $}-\mathbb{K}_{\psi \theta } \text{d$\theta $}\wedge
\text{d$\psi $}+\mathbb{K}_{\rho \psi } \text{d$\rho $}\wedge
\text{d$\psi $}
\label{Kformafredda}
\end{eqnarray}
where the explicit form of the components  calculated by the
{\sc mathematica} code are also too long and   messy to be displayed.
\paragraph{The Ricci tensor and the Ricci 2- form}
Using the same MATHEMATICA code we have calculated the Ricci tensor
of the above metric defined by:
\begin{equation}
R_{ij^*} = \partial _i\partial_{j^*}\log[\text{Det}[g]]
\end{equation}
and the Ricci 2-form defined by
\begin{equation}
\text{$\mathbb{R}$ic}= - \, \tfrac{i}{2\pi }R_{ij^*}
\text{dz}^i\wedge d\bar{z}^{j^*} \, = \, \tfrac{i}{2\pi }
\bar{\partial} \, \partial \log[\text{Det}[g]]\label{F2riccioide}
\end{equation}
Once transformed to polar coordinates the Ricci 2-form has the same
structure as the K\"ahler 2-form, namely
\begin{eqnarray}
\text{$\mathbb{R}$ic} &=&\text{Ric}_{\text{r$\theta $}}
\text{dr}\wedge \text{d$\theta $}+\text{Ric}_{\text{r$\rho $}}
\text{dr}\wedge \text{d$\rho $}-\text{Ric}_{\text{$\psi $r}}
\text{dr}\wedge \text{d$\psi $}+\text{Ric}_{\theta \rho }
\text{d$\theta $}\wedge \text{d$\rho $}\nonumber\\
& - & \text{Ric}_{\psi \theta } \text{d$\theta $}\wedge \text{d$\psi
$}+\text{Ric}_{\rho
\psi } \text{d$\rho $}\wedge \text{d$\psi $}
\label{Ricformozza}
\end{eqnarray}
The explicit form of the components of $\text{$\mathbb{R}$ic}$ is
even more massive than those of the K\"ahler 2-form and cannot be
exhibited. An important issue is whether the constructed K\"ahler
metric might be a K\"ahler-Einstein metric, namely whether the Ricci
tensor might be proportional to the metric coefficients.  However it
is well known that Hirzebruch surfaces cannot carry K\"ahler-Einstein metrics \cite{Besse}.
An easy way of checking this fact is to note that, since the Ricci form represents the
first Chern class of the tangent bundle to the Hirzebruch surface, we have
\begin{equation} \int_{C_1} \mathbb{R}\text{ic} =  \int_{C_1} c_1(T_{\mathbb F_2})
= 2 H\cdot C_1 = 0\,, \label{periodRicci1}\end{equation} where $H$
is the divisor  described in Sections \ref{coorcurves} and
\ref{Ylinebundle}, while on the other hand the integral of K\"ahler
form on the curve $C_1$ is of course positive; note that one also
must have:
\begin{equation} \int_{C_2} \mathbb{R}\text{ic} = 2 H \cdot C_2 = 2.\label{periodRicci2} \end{equation}
To check the robustness of our computations we verified explicitly these equations, as we show below.

\paragraph{Reduction to the
homology cycle $C_1$.} The reduction to the homology cycle
$C_1$ is obtained by setting r = $\psi $ = 0 together
with the vanishing of their differentials. Applying such a procedure
to the K\"ahler 2-form and to the Ricci 2-form we obtain:
\begin{equation}
\mathbb{K}|_{C_1}= -\kappa _1\frac{ \rho  \,
d\rho\wedge d\theta}{4 \pi  \left(1+\rho ^2\right)^2} \quad
;\quad \text{$\mathbb{R}$ic}|_{C_1}= 0
\end{equation}
This  confirms eq.~\eqref{periodRicci1}.
\paragraph{Period of the K\"ahler two form on $C_1$.}
Next we can calculate the period of the K\"ahler form on
$C_1$ and we obtain:
\begin{equation}
\int _{C_1}\mathbb{K} = -  \int _0^{2\pi }d\theta
\int_0^{\infty } \frac{ \kappa _1 \rho }{4 \pi  \left(1+\rho
^2\right)^2} \, d\rho = -\frac{\text{  }\kappa _1 }{4}
\end{equation}
\paragraph{Period of the K\"ahler two form on $C_2$.}
Here we calculate the restriction of the K\"ahler 2-form to the
homology cycle $C_2$ and we obtain
\begin{equation}
\mathbb{K}|_{C_2}\, = - \,\left(f_0(r)+\kappa_1
\,f_1(r)+\kappa_2 \,f_2(r)\right)\, dr\wedge d\psi
\end{equation}
where {\small
\begin{eqnarray}
 f_0(r) &=& \frac{r \left(r^6-\left(\sqrt{r^4+10 r^2+1}-15\right) r^4+\left(27-10
   \sqrt{r^4+10 r^2+1}\right) r^2-\sqrt{r^4+10 r^2+1}+5\right)}{2 \pi
   \left(r^4+10 r^2+1\right)^{3/2}}\nonumber \\
 f_1(r) &=& \frac{3 r \left(r^2+1\right)}{4 \pi  \left(r^4+10 r^2+1\right)^{3/2}} \nonumber\\
 f_2(r) &=& \frac{r \left(5 r^2+1\right)}{\pi  \left(r^4+10 r^2+1\right)^{3/2}}
\end{eqnarray}
} We find
\begin{equation}
\int_0^{\infty } f_1(r) \, dr \,=\, -\frac{1}{4}\quad; \quad
\int_0^{\infty } f_2(r) \, dr\, = \, -1 \quad ; \quad \int_0^{\infty
} f_0(r) \, dr \,=\,0
\end{equation}
So that the period of the K\"ahler 2-form on $C_2$ is
the following one:
\begin{equation}
\int _{C_2}\mathbb{K} = \frac{\text{  }\kappa _1
}{4} + \kappa _2\,.
\end{equation}
Recalling eq.~\eqref{kappalogatto}, and with the present choice of $\zeta$, the
values of $\kappa_1$ and $\kappa_2$ are
$$\kappa_1 = -4\alpha, \quad \kappa_2 = 3\alpha, $$
so that the volumes of $C_1$ and $C_2$ are
$$\int_{C_1}\mathbb K = \alpha,\qquad
 \int_{C_2}\mathbb K = 2\alpha, $$
As we have previosly discussed, we choose $\alpha=2$ so that the
periods of the tautological bundles are all integral.
\par
In a similar way we calculated the period of the Ricci 2-form on
$C_2$ and as it was expected we found:

\subsection{The isometry group of the second Hirzebruch surface and of the full resolution $Y$}
The K\"ahler potential is function only of the following combination
\begin{equation}
\varpi  = |v|^2\left(1+\left|u|^2 \right)^2\right)\,.
\end{equation}
Given an element of $\mathrm{SU (2)}$, namely a 2$\times $2
matrix
\begin{equation}
\gamma =\left(
\begin{array}{cc}
 a & b \\
 c & d \\
\end{array}
\right) ;\quad ad-bc  \, ; \quad \gamma
^{\dagger }=\gamma ^{-1}\,,
\end{equation}
the object $\varpi $ is invariant under the holomorphic
transformation
\begin{equation}
(u,v)\longrightarrow  \left(\frac{a u + b}{ c u + d\text{  }},v (c
u+d)^2\right)\,. \label{pretarabaccus}
\end{equation}
According to the description we give in the Appendix, the coordinate $(u,
t)$ transform as follows:
\begin{equation}
(u,t)\longrightarrow  \left(\frac{a u + b}{ c u + d\text{  }},t (c
u+d)^{-2}\right) \,.\label{tarabaccus}
\end{equation}
Hence we conclude that the coordinate $t$ spanning the fiber in the
Hirzebruch surface is related with the toric coordinate $v$  by \begin{equation}
t= \frac{1}{v}
\end{equation}
It is also evident that the complete isometry group of the K\"ahler metric $\mathbb K$ is
$\mathrm{SU(2)\times U(1)}$, where  $\text{SU}(2)$ acts as in eq.~\eqref{pretarabaccus}, while
$\text{U}(1)$ is simply the phase transformation of $v$.
This  is   inherited by the
K\"ahler metric of the full variety $Y$. Actually in the case of $Y$
the isometry group extends by means of an extra
$\mathrm{U(1)}$ factor corresponding to the phase transformation of
the coordinate $w$ spanning the fiber of the line bundle
$Y\longrightarrow \mathbb{F}_2$:
\begin{equation}\label{balordus}
    \mathrm{Iso}_{Y} \, = \, \mathrm{SU(2)\times U(1)\times U(1)}\,.
\end{equation}

\subsection{Ricci-flat metrics on Y} All smooth resolutions of singularities of $\C^3/\Gamma$, where
$\Gamma$ acts as a subgroup of $\operatorname{SU}(3)$, carry
Ricci-flat K\"ahler metrics, as proved in \cite{Joyce-QALE}.
Therefore, the variety $Y$ carries a Ricci-flat metric --- actually,
as $\dim H^2(Y,\Q)=2$,   it carries a 2-parameter family of such
metrics. However {\em one should not expect} the metric coming from
the generalized Kronheimer construction to be one of these metrics;
we shall check this point in calculations that will appear in a
future publication \cite{conmasbia}. Moreover, as the action of
$\Z_4$ on $\C^3-\{0\}$ is not free  (all points of the $z$ axis have
a $\Z_2$ isotropy, compare discussion in Section \ref{relation}),
the Ricci-flat metrics are not ALE; they do have a suitable
asymptotically Euclidean behaviour away from the singular locus, but
as the latter is not compact, their asymptotics is more complicated
than that of an ALE metric (these metrics have been called ``QALE''
by Joyce).

\section{The partial resolution $Y_3$ and its K\"ahler geometry}
\label{Y3Kallero} In this Section we   construct the K\"ahler
Geometry of the partial desingularization $P_3$ of the quotient
$\mathbb{C}^3/\mathbb{Z}_{4 }$ that occurs at some walls of the
$\zeta$ space; the computations suggest that the partial
desingularization $P_3$  is  again the total space of a line bundle,
this time over a singular variety $Q_2$. Actually $P_3$
 is  $ Y_3$, one of the degenerations arising in our toric analysis,
 and   the base variety $Q_2$ is   the weighted projective space  $\mathbb{P}$[1,1,2].

The strategy that we adopt is analogous to that utilized for the
nondegenerate variety $Y$ and it goes
along the following steps:
\begin{enumerate}
  \item We choose the partial desingularization  space $P_3$ provided by the Kronheimer
construction in the plane $\zeta _1$ = $\zeta _3 = a$, $ \zeta
_2 = 2a$ .
  \item Reducing the moment map equations to the compact exceptional
  divisor (the singular surface $Q_2$) we obtain a system
  solvable by radicals and the explicit solution is particularly simple.
\item We  obtain the complete solution for the
full variety starting from the solution on the $Q_2$
surface and expressing the sought-for fiber metrics $T_{1,2}$ as
power series in the coordinate $w$ that represents the section of
the line bundle over $Q_2$.
\end{enumerate}
\subsection{Construction of the K\"ahler geometry of $Q_2$}
With the same logic utilized in the case of the full resolution we
begin with the analysis of the K\"ahler Geometry of the singular base
manifold $Q_2$ of the line bundle $P_3\longrightarrow
Q_2$. Our main weapon in this analysis is the reduction of
the algebraic system of moment map equation to the exceptional
divisor by means of the limit $\lambda \longrightarrow  0$. We
construct the K\"ahler potential of $Q_2$ performing the
limit $\lambda \longrightarrow  0$ both on the nonlogarithmic part
of the K\"ahler potential of the resolution and on the logarithmic
one.
\subsubsection{Construction of the K\"ahler potential}
We choose the following special point on the chamber wall
$\mathcal{W}_1$
\begin{equation}\label{briscola}
    \zeta _1= 1, \quad \zeta _3= 1, \quad \zeta _2= 2
\end{equation}
so that the system in eq.~\eqref{brillo} becomes
\begin{equation}\label{eqqa1}
    \left(
\begin{array}{c}
 \left(-1-\lambda ^4 T_2^2-\delta  T_1 T_3\right) \left(T_1^2-T_3^2\right) \\
 \delta  \left(\sigma ^2 T_2+\lambda ^4 \sigma ^2 T_2^3-T_1 T_3 \left(T_1^2+T_3^2\right)\right) \\
 -2 T_1 T_2 T_3-\left(-1+\lambda ^4 T_2^2\right) \left(T_1^2+\delta  \sigma ^2 T_2+T_3^2\right) \\
\end{array}
\right)=\left(
\begin{array}{l}
 0 \\
 0 \\
 0 \\
\end{array}
\right)
\end{equation}
According with what we discussed in Section \ref{ballavantana} and
eq.~\eqref{raschiotto} we perform the replacement
\begin{equation}\label{eqqa2}
T_1=\sqrt{\frac{\sigma }{\lambda }} \frac{
\sqrt[4]{Z^3+Z}}{\sqrt[4]{2}} \quad;\quad  T_2=\frac{1}{\lambda ^2}
Z \quad ;\quad T_3=\sqrt{\frac{\sigma }{\lambda }}\frac{
\sqrt[4]{Z^3+Z}}{\sqrt[4]{2}}
\end{equation}
and   introduce the appropriate rescaling
\begin{equation}\label{eqqa3}
    Z= \lambda ^2 z\,.
\end{equation}
In this way the system \eqref{eqqa1} becomes
\begin{equation}\label{eqqa4}
    \left(
\begin{array}{c}
 0 \\
 0 \\
 -\sqrt{2} \left(-1+z+z^2 \lambda ^4\right) \sqrt{z+z^3 \lambda ^4} \sigma +\left(z \delta -z^3 \delta  \lambda ^4\right) \sigma ^2 \\
\end{array}
\right)=\left(
\begin{array}{c}
 0 \\
 0 \\
 0 \\
\end{array}
\right)\,.
\end{equation}
Then we reduce the system to the exceptional divisor performing the
limit $\lambda \to  0$ and we obtain the algebraic equation
\begin{equation}\label{eqqa5}
    \sqrt{z} \sigma  \left(\sqrt{2}-\sqrt{2} z+\sqrt{z} \delta
\sigma \right) =0
\end{equation}
whose unique  everywhere positive real solution is  \begin{equation}\label{eqqa6}
    z =  \frac{1}{4} \left(4+\delta ^2 \sigma ^2+\delta \sigma
\sqrt{8+\delta ^2 \sigma ^2}\right)\,.
\end{equation}
\subsection{The K\"ahler potential addends}
As in the case of the fully resolved variety $Y$, we begin by writing
the restriction to the hypersurface $\mathcal{N}_{\zeta }$ of the
K\"ahler potential of the flat variety $\text{Hom}(Q\otimes R,R)^{\Z_4}$. Inserting the above choices in
eq.~\eqref{balordus} we obtain
\begin{equation}\label{eqqa7}
    \mathcal{K}_0 = \frac{2+2 z^2 \lambda ^4+2 \sqrt{2}\text{
}\sqrt{z+z^3 \lambda ^4} \delta  \sigma }{z}
\end{equation}
For the logarithmic part of the K\"ahler potential we have
\begin{equation}\label{eqqa8}
    \mathcal{K}_{\log }\equiv  \zeta_I\,  \mathfrak{C}^{IJ } \,
\log \left[X_J\right]^{\alpha_{\zeta_J}}
\end{equation}
with the above choices and disregarding addends $\log[\lambda $]
since $\lambda $ is a modulus square of a holomorphic function,we
find:
\begin{equation}\label{eqqa9}
\mathcal{K}_{\log } = 2  \alpha_{\zeta_J}\log\left[T_2\right] = 2 \alpha_{\zeta_2} \log[z]
\end{equation}
so that
\begin{equation}\label{eqqa10}
    \mathcal{K} = \frac{2+2 z^2 \lambda ^4+2 \sqrt{2}
\delta  \sigma
 \sqrt{z+z^3 \lambda ^4} }{z}+ 2 \alpha_{\zeta_2}\log[z]
\end{equation}
This determines the K\"ahler geometry of the   singular  variety
$P_3$, provided we are able to write the appropriate positive real
solution $z=\hat{\mathfrak{F}}(\lambda ,\delta \sigma )$ of the
moment map equation \eqref{eqqa4}.
\subsubsection{The K\"ahler potential of $Q_2$}
Performing the limit $\lambda \to  0$  and substituting for
$z$ the solution of the reduced moment map equations presented in
eq.~\eqref{eqqa6} we obtain the K\"ahler potential of the divisor
$Q_2$:
\begin{eqnarray}\label{eqqa11}
    \mathcal{K} |_{Q_2}&=& \frac{2 \left(4+2 \sqrt{2} \delta  \sigma  \sqrt{4+\delta  \sigma  \left(\delta  \sigma +\sqrt{8+\delta
^2 \sigma ^2}\right)}\right) }{4+\delta  \sigma  \left(\delta \sigma
+\sqrt{8+\delta ^2 \sigma
^2}\right)}\nonumber\\
&&+2 \alpha_{\zeta_J} \log\left[\frac{1}{4} \left(4+\delta
 \sigma  \left(\delta  \sigma +\sqrt{8+\delta ^2 \sigma ^2}\right)\right)\right]
\end{eqnarray}
Naming
\begin{equation}\label{eqqa12}
   T= \delta  \sigma \,=\, r \left(1+\rho ^2\right)\quad ;\quad
   W = \frac{1}{4} \left(4+T
\left(T+\sqrt{8+T^2}\right)\right)
\end{equation}
where, just as before,
\begin{equation}\label{eqqa13}
    r= |v| \quad ; \quad \rho = |u|
\end{equation}
the result \eqref{eqqa11} can be rewritten in the following much
simpler form:
\begin{equation}\label{eqqa14}
    \mathcal{K} |_{\mathbb{Q}_{2 }}= 4-\frac{2}{W}+2 \alpha_{\zeta_2}\log[W]
\approx  -\frac{2}{W}+2 \alpha_{\zeta_2}\log[W]
\end{equation}
From these data one can compute the K\"ahler metric; when  $ \alpha_{\zeta_2} =1$
this takes the
particularly simple  form
\begin{equation}\label{eqqa15}
    g_{ij^*}\, = \,\left(
\begin{array}{cc}
\displaystyle \frac{2 r \left(r+\frac{2 \sqrt{2}}{\sqrt{W}}\right)}{1+W}
 & \displaystyle  \frac{2 \sqrt{2} e^{-i (\theta -\psi )} \rho }{\sqrt{W} (1+W)} \\[10pt]
\displaystyle   \frac{2 \sqrt{2} e^{i (\theta -\psi )} \rho }{\sqrt{W} (1+W)}
 &  \displaystyle  \frac{2 (-1+W)}{r^2 \left(W+W^2\right)} \\
\end{array}
\right)
\end{equation}
where $\theta $ and $\psi $ are the phases of $u$ and $v$,
respectively. This is enough for our purposes since the periods we are interested in
scale multiplicatively with respect to $ \alpha_{\zeta_2}$.
\subsubsection{The determinant of the K\"ahler metric and the Ricci tensor}
Calculating the determinant of $g_{ij^*}$ we find
\begin{equation}\label{eqqa16}
    \det [g] = \frac{4}{W+W^2}
\end{equation}
then calculating the Ricci tensor we obtain:
\begin{eqnarray}
\text{Ric}_{\text{11}^*}&=& -\frac{(-1+W) W (1+5 W)}{r^2 (1+W)^4}\nonumber\\
\text{Ric}_{\text{12}^*}&=& -\frac{\sqrt{2} e^{i (\theta -\psi )}
W^{3/2} (1+5 W)
\rho }{(1+W)^4}\nonumber\\
\text{Ric}_{\text{21}^*}&=& -\frac{\sqrt{2} e^{-i (\theta -\psi )}
W^{3/2}
(1+5 W) \rho }{(1+W)^4}\nonumber\\
\text{Ric}_{\text{22}^*}&=& \frac{1}{\sqrt{2} \sqrt{W} (1+W)^5}r
(140-4 W (70
+W (-34+W (6+5 W)))\nonumber\\
&&+\left.\sqrt{2} r \sqrt{W} \left(140+W \left(-139+W \left(4+W-2
W^2\right)\right)\right)\right.\nonumber\\
&&\left.-70 r^2 W \left(-1+\rho ^4\right)\right)\label{eqqa17}
\end{eqnarray}
Again, from here  we can see that  the constructed metric is not K\"ahler-Einstein.

Next considering the transformation to polar coordinates
\begin{equation}\label{eqqa18}
    u=e^{i \theta } \rho \quad ; \quad v=e^{i \psi } r
\end{equation}
we write out the  form of the K\"ahler 2-form explicitly:
\begin{eqnarray}\label{eqqa19}
\mathbb{K}&=&\frac{2 \sqrt{2} \rho ^2 \text{dr}\wedge \text{d$\theta
$}}{\pi  \sqrt{W} (1+W)}+\frac{2 (-1+W) \text{dr}\wedge \text{d$\psi
$}}{\pi
 r W (1+W)}\nonumber\\
 &&-\frac{2 r \left(2 \sqrt{2}+r \sqrt{W}\right) \rho  \text{d$\theta $}\wedge
 \text{d$\rho $}}{\pi  \sqrt{W} (1+W)}+\frac{2 \sqrt{2} r \rho
 \text{d$\rho $}\wedge \text{d$\psi $}}{\pi  \sqrt{W} (1+W)}
\end{eqnarray}
\subsubsection{Calculations of periods of the K\"ahler 2-form
on  homology cycles} Starting from eq.~\eqref{eqqa18} we
can calculate the periods of the K\"ahler 2-form on  the
homology cycles $C_1$ and $C_2$. This
allows one to get a clear picture of the degeneration of the $Y$ variety,
showing which cycles in $Y_3$ shrink to a vanishing volume.
\paragraph{Cycle $C_1$.}
Setting $r=0$ and $\psi =0$  we obtain the reduction of the K\"ahler
2-form to the cycle $C_1$. Expanding in power series of
$r $ around $ r=0$  we  obtain\begin{equation}\label{eqqa20}
    \mathbb{K} = -\frac{2 \left(\sqrt{2} \rho  \text{d$\theta $}\wedge
\text{dr}\right) r}{\pi }+\mathcal{O}\left(r^2\right)
\end{equation}
It follows that for $r = 0$ (equation of the cycle
$C_1$) the K\"ahler 2-form goes to zero, namely the
$C_1$ cycle shrinks to zero.
\paragraph{Cycle $C_2$.}
The reduction to the cycle $C_2$ is obtained in the
limit $\rho \rightarrow $0, $\theta $$\rightarrow $0. We obtain
\begin{equation}\label{eqqa21}
\mathbb{K}|_{C_2}=\frac{\left(4+r^2-r\sqrt{8+r^2}\right)\text{dr}
\wedge\text{d$\psi $}}{2\pi \sqrt{8+r^2}}
\end{equation}
so that for $\alpha_{\zeta_J}=1$ we get
\begin{equation}\label{eqqa22}
    \int _{C_2}\mathbb{K}\text{           }=\text{ }2\pi
\int_0^{\infty } \frac{\left(4+r^2-r \sqrt{8+r^2}\right) }{2 \pi
\sqrt{8+r^2}} \, dr = 2
\end{equation}
and the result of a generic $\alpha_{\zeta_J}$ is therefore
$$ \mathbb{K}|_{C_2} =2 \alpha_{\zeta_J}.$$
It follows that for $\rho $ = 0 (equation of the cycle
$C_2$) the K\"ahler form has period $2\alpha_{\zeta_J}$.
\subsection{Interpretation}
\label{interpretazia}
The interpretation of the above results is sufficiently clear. The
cycle $C_1$ is the intersection of the two components of the
exceptional divisor $D_c$ and $D_{nc}$, where $D_c$ is the
Hirzebruch surface $\mathbb{F}_2$ and $D_{nc}$ = $\mathbb{P}^1
\times \mathbb{C}$. The vanishing of the cycle $C_1$ means that the
exceptional divisor $D_{nc}$ has disappeared,  while the compact one
$D_c$ remains in the form of the singular variety
$\mathbb{P}[1,1,2]$. This means that $P_3$ is precisely the partial
resolution of the orbifold singularity called  $Y_3$  in Section
\ref{Y3sezia}.

\subsubsection{Periods of the   Chern forms $\omega _{1,2,3}$.}
The fiber metrics of the three tautological bundles are for the
chosen point of the $\zeta $ space given by
\begin{equation}\label{eqqa24}
\mathcal{H}_{1,2,3}=\left\{
\alpha_{\zeta_J}\, \log\left[\frac{\left(Z+Z^3\right)^{1/4}}{2^{1/4}}\right],\
\alpha_{\zeta_J}\, \log[Z],\
\alpha_{\zeta_J}\, \log\left[\frac{\left(Z+Z^3\right)^{1/4}}{2^{1/4}}\right]\right\}\;
\end{equation}
substituting $Z\, \to \, \lambda ^2 \, z$ we obtain:
\begin{equation}\label{eqqa25}
\mathcal{H}_{1,2,3}  = \left\{\frac{1}{4}
\alpha_{\zeta_J}\, \log\left[\frac{1}{2} \left(z \lambda ^2+z^3 \lambda
^6\right)\right],\ \alpha_{\zeta_J}\, \log\left[z \lambda ^2\right],\ \frac{1}{4} \alpha_{\zeta_J}\,
\log\left[\frac{1}{2} \left(z \lambda ^2+z^3 \lambda
^6\right)\right]\right\}
\end{equation}
Performing the reduction to compact exceptional divisor we have;
\begin{eqnarray}\label{eqqa26}
    \mathcal{H}_{1,2,3}
&=&\left\{\alpha_{\zeta_J}\, \frac{\log[z]}{4}, \
\alpha_{\zeta_J}\, \log[z],\ \alpha_{\zeta_J}\, \frac{\log[z]}{4}\right\}\nonumber\\
z &=&\frac{1}{4} \left(4+r^2 \left(1+\rho ^2\right)^2+r \left(1+\rho
^2\right) \sqrt{8+r^2 \left(1+\rho ^2\right)^2}\right)\,.
\end{eqnarray}
So that for the period integrals, whatever is the supporting cycle,
we obtain:
\begin{equation}\label{eqqa27}
\left\{\int \, \omega_1,\, \int \, \omega_2,\, \int \,
\omega_3\right\} \, = \,\left\{ \frac 14 \, \alpha_{\zeta_J}\, \int \Omega , \ \alpha_{\zeta_J}\,  \int
\Omega , \ \frac 14 \, \alpha_{\zeta_J}\, \int \Omega \right\}\quad ; \quad \Omega =
\frac{i}{2\pi }\partial \bar{\partial } \log[z]
\end{equation}
The periods of $\Omega $ are very easily calculated since in
cohomology we have:
\begin{equation}\label{eqqa28}
   \alpha_{\zeta_J}\,  \left[\Omega \right] = \frac{1}{2}\, \left[\mathbb{K}\right]
\end{equation}
where $\mathbb{K}$ is the K\"ahler 2-form. The above equation follows
from eq.~\eqref{eqqa10} and the observation that the non-logarithmic
part of the K\"ahler potential gives rise to a cohomologically
trivial addend in $\mathbb{K}$.

Hence we have:
\begin{equation}\label{eqqa29}
\int _{C_1}\Omega  = 0 \quad;\quad
\int_{C_2}\Omega = 1
\end{equation}
In conclusion we have that there is only one independent
tautological bundle corresponding to $\omega _2$ and we have:
\begin{equation}\label{eqqa30}
\int _{C_1}\omega _{1,2,3}= 0 \quad ; \quad
\int_{C_2}\omega _2 =\alpha_{\zeta_J} \quad ; \quad
\int_{C_2}\omega _{1,3 }= \frac{\alpha_{\zeta_J}}{4}\quad ; \quad
\left[\mathbb{K}\right]= 2\left[\omega _2\right]\,.
\end{equation}
\subsection{The K\"ahler geometry of the singular   variety $Y_3$}
\begin{figure}
\begin{center}
\includegraphics[height=8cm]{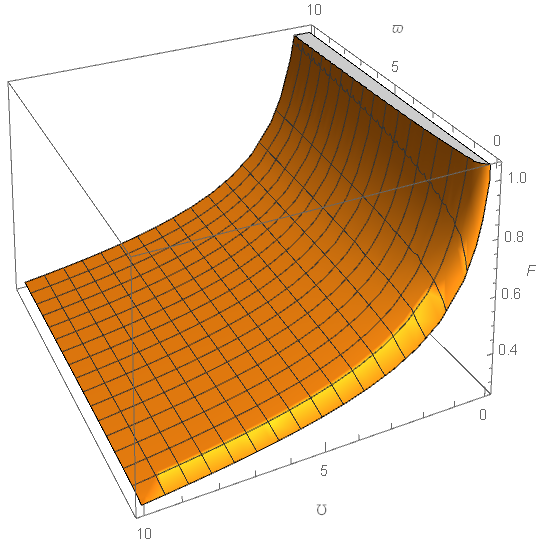}
\caption{\label{gigotica}  Plot of the function
$\mathfrak{G}(\varpi , \mho)$. }
\end{center}
\end{figure}

Let us finally derive the K\"ahler geometry of the singular threefold
$Y_3$ which is the total space of a line bundle over the singular
exceptional divisor $\mathbb{P}$[1,1,2]. To this
effect let us consider the following sextic equation for an unknown
F, where the coefficients are functions of the same invariants
$\varpi $ and $\mho $ previously defined in eqs.~\eqref{prillini}:
\begin{eqnarray}
\mathcal{P}(F) &\equiv& 2+F (-4-\varpi )+F^2 (2-2 \mho )+2 F^3
\varpi \mho +2 F^6 \mho ^3+F^4 \left(2 \mho -2 \mho
^2\right)\nonumber\\
&&+F^5 \left(4 \mho ^2-\varpi
 \mho ^2\right) \, = \, 0 \label{eqqa31}
\end{eqnarray}
Just as in the case of the full variety $Y$, the sextic polynomial
$\mathcal{P}$(F) has the property that, for all positive values of
the parameters $\varpi $ and $\mho $, it has two positive real roots
and four complex roots arranged in two pairs of complex conjugate
roots. Hence the largest real root is uniquely identified and
singles out a precise function $\mathfrak{G}$($\varpi $,$\mho $) of
the parameters:
\begin{equation}\label{eqqa32}
\mathfrak{G}(\varpi ,\mho ) \equiv \text{largest real root of }
\mathcal{P}(F)\,.
\end{equation}
The function $\mathfrak{G}$($\varpi $,$\mho $) can be developed in
power series of $\mho $ and we have:  \begin{eqnarray}
\mathfrak{G}(\varpi ,\mho )&=&\frac{1}{4}
\left(4+\varpi +\sqrt{\varpi  (8+\varpi )}\right)\nonumber\\
&&-\frac{\left(\left(4+\varpi +\sqrt{\varpi  (8+\varpi )}\right)^2
\left(3 \varpi ^2+4 \sqrt{\varpi  (8+\varpi )}+\varpi \left(16+3
\sqrt{\varpi (8+\varpi )}\right)\right)\right) \mho }{64
\sqrt{\varpi  (8+\varpi )}}\nonumber\\&&+O[\mho ]^2 \label{eqqa33}
\end{eqnarray}
In terms of this function and of the above variables the K\"ahler
potential of the complete $Y_3$ variety takes the   form
\begin{equation}\label{eqqa34}
\mathcal{K}_{\text{Y3}} = \frac{2 \left(1+\mathfrak{G}^2 \mho
+\sqrt{2} \sqrt{\mathfrak{G} \varpi  \left(1+\mathfrak{G}^2 \mho
\right)}+\mathfrak{G} \log[\mathfrak{G}]\right)}{\mathfrak{G}}
\end{equation}
The function $\mathfrak{G}(\varpi , \mho)$  is displayed in Figure \ref{gigotica}.
\section{Summary of the Chamber Structure Discussion} \label{summatheologica}
We can now try to summarize our long and detailed discussion of the
chamber structure pertaining to the K\"ahler quotient resolution
\`{a} la Kronheimer of the singularity $\mathbb{C}^3/\mathbb{Z}_4$.

First of all let us stress that the chamber structure is one of the
most relevant aspects of the entire construction from the point of
view of any physical application in the context of the gauge/duality
correspondence. Indeed the $\zeta$ parameters are the
Fayet-Iliopoulos parameters in the gauge theory side of the
correspondence, while they should be retrievable as fluxes of
suitable $p$-forms on the supergravity side of the correspondence
and hence as \textit{parameters of the Ricci flat K\"ahler metric}
existing on the same resolved smooth manifold. Therefore, loosely
speaking, the chamber structure is a mathematical synonymous of
\textit{Phase Diagram} of the Gauge Theory.
\par
Having clarified the physical relevance of the topic, let us state
what   results we found.
The toric analysis of the problem has revealed several possible
degenerations of the full resolution
$Y\,\longrightarrow\mathbb{C}^3/\mathbb{Z}_4$. Only three of such
manifolds are realized by the K\"ahler quotient:
\begin{description}
  \item[a)] The complete resolution $Y$, which happens to be the
  total space of the canonical  line bundle of the compact exceptional divisor
  $D_c\,\simeq \, \mathbb{F}_2$ where $\mathbb{F}_2$
  is the second Hirzebruch surface. In this case the exceptional
  divisor has an additional non-compact component $D_{nc}\,\simeq \, \mathbb{P}^1\times \mathbb{C}$
  \item[b)] The partial resolution $Y_3$ which happens to be   the
  total space of  the canonical line bundle over the  singular compact exceptional
  divisor ${D}_c\,\simeq \,
  \mathbb{P}[1,1,2]$. In this case the non-compact exceptional
  divisor $D_{nc}$ disappears since its compact factor
  $\mathbb{P}^1$ shrinks to zero.
  \item[c)] The partial resolution $Y_{EH} = \text{ALE}_{A_1}\times\C$,
  which again can be seen as the total
  space of  the canonical line bundle of  the noncompact exceptional divisor
  $D_{nc}$.
  In this case it is the compact exceptional divisor that
  disappears.
\end{description}
The three aforementioned manifolds are realized in
$\zeta$ space $\mathbb{R}^3$ in the way we summarize
below.

There are four intersecting planar walls $\mathcal{W}_{0,1,2,3}$
that partition the entire $\zeta$ space into eight disjoint
convex cones (stability chambers).
\begin{enumerate}
\item
The smooth space $Y$ is realized in all interior points of all the
chambers and in all generic points of the three walls of type 0,
namely $\mathcal{W}_0$, $\mathcal{W}_1$, $\mathcal{W}_3$.
\item The singular space $Y_3\longrightarrow \mathbb{P}[1,1,2]$ is
realized in all generic points of the type $1$ wall $\mathcal{W}_2$. The
latter contains the Kamp\'{e} line and the two Cardano edges that
are also realization of $Y_3$.
\item The smooth space $Y_{EH}$ is realized on
the homonymous edge, which is the intersection of the wall
$\mathcal{W}_0$ with the wall $\mathcal{W}_2$.
\end{enumerate}
\paragraph{Wall Crossing.} Crossing a wall is the mathematical
analog of a physical phase transition. When the wall we cross is
of type 0, we simply go from one realization of the $Y$ manifold
to another one. On the two sides of the wall the  supporting
variety is the same, yet the tautological bundles may be
different. On the walls of type 0  we have a third realization of
$Y$, but the configuration of the tautological line bundles may vary.

Most dramatic is the crossing of a wall of type III. In this case we
go from one realization of the $Y$ manifold to another one passing
trough a degenerate singular manifold that is located on the wall.
There is a simple numerical procedure to visualize such a
phenomenon. Let us explain how it works.

 Considering the three-dimensional Euclidean space with coordinates $X_1,X_2,X_3$, a
solution of the moment map system \eqref{sistemico} defines a
two-dimensional surface in this space. This is so because the
coefficients of the system depend on the two parameters $\Sigma$, $U$,
or alternatively $\varpi,\mho$. Creating a grid of
$\varpi,\mho$-values we can obtain a picture of the aforementioned
two-dimensional surface by interpolating the numerical solutions of
the system \eqref{sistemico} in all points of the grid at fixed
$\zeta$ parameters. Such a picture is  like a photogram of a movie.
The other photograms of the this movie are provided by repeating the
same procedure with other $\zeta$ parameters. The effect of
crossing is better visualized when all the photograms are arranged
along a line in $\zeta$ space that crosses the wall and is
orthogonal to it.

\begin{figure}
\begin{center}
\includegraphics[height=8cm]{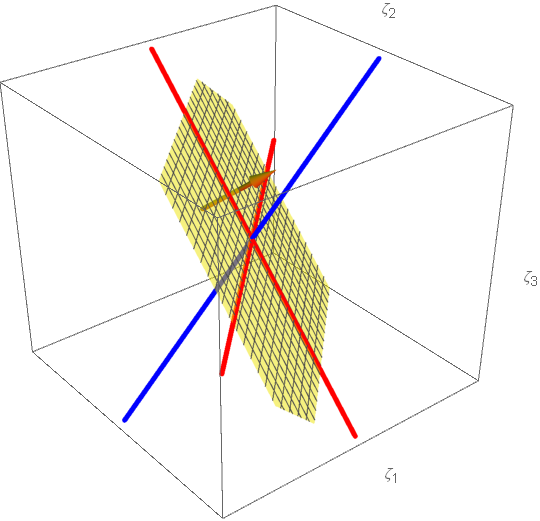}
\caption{\label{direttoria}{Path crossing the $\mathcal{W}_2$ wall.
The line with an arrowhead shown in this figure is the one we have
chosen to create a movie of wall crossing.  }}
\end{center}
\end{figure}

\begin{figure}
\begin{center}
\includegraphics[height=5cm]{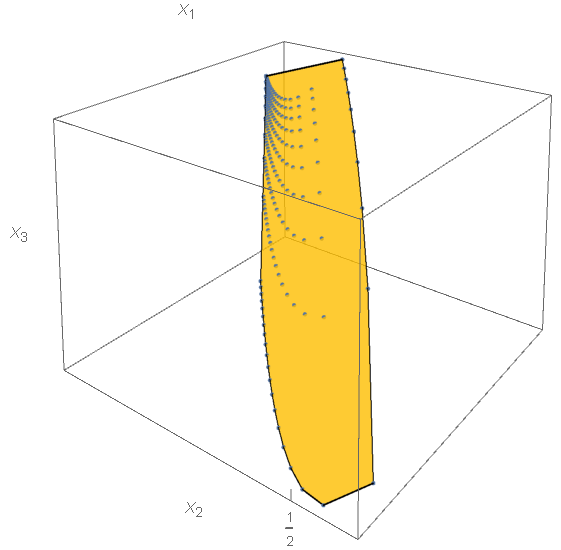}
\includegraphics[height=5cm]{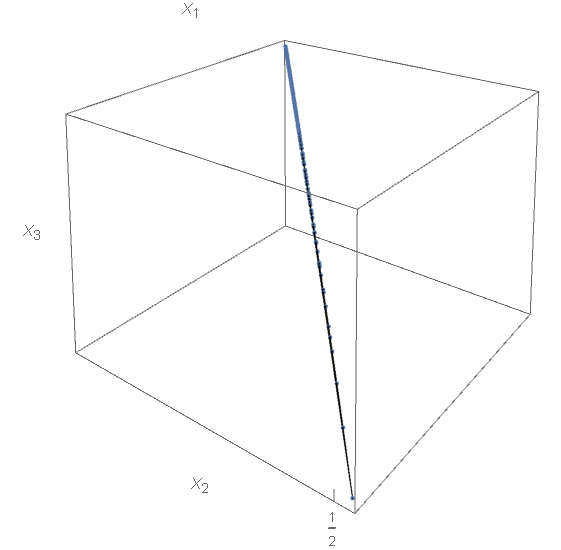}
\includegraphics[height=5cm]{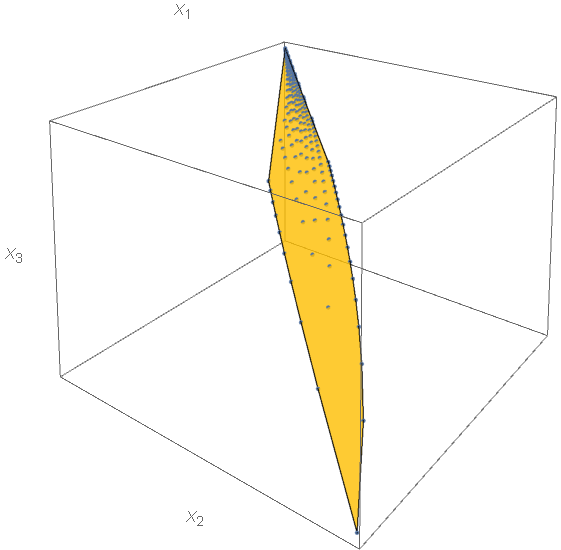}
\caption{\label{fotogrammi}{Crossing the $\mathcal{W}_2$ wall.
Before crossing the solution of the algebraic system
\eqref{sistemico} traces a surface in $\pmb{X}$ space. After
crossing the solution traces another surface. Just on the wall the
solution traces a line rather than a surface. This is just the
symptom of degeneration of the corresponding manifold. }}
\end{center}
\end{figure}

Let us consider the unique type III wall $\mathcal{W}_2$ and the
line that crosses it orthogonally, displayed in
fig.\ref{direttoria}.
Along this line we have numerically constructed a few photograms as
previously described. In fig.~\ref{fotogrammi} we display three of
them. One before crossing, one on the wall and one after crossing.
As it is evident from the figure, just on the wall the surface
representing the solution is substituted by a line. This happens
because the solution for $X_1,X_2,X_3$ is expressed in terms of
simple functions of single function $Z(\varpi,\mho)$ of the two
variables. This obviously yields the parametric equation
 of a line.  We already know this for the Kamp\'{e}
line, where we have eq.~\eqref{raschiotto} and the function $Z$ is
defined as a positive real root of the sextic equation
\eqref{sesticina}. Actually a similar result applies to all points of
the $\mathcal{W}_2$ plane, namely for:
\begin{equation}\label{circopedestre}
    \zeta \, = \, \left\{x,x+y,y\right\}
\end{equation}
Indeed we can easily prove by direct substitution that for such a
choice of the level parameters the solution of the moment map
equations \eqref{sistemico} can be written as follows:
\begin{eqnarray}
  X_1 &=& \sqrt[4]{\frac{Z \left(x Z^2+y\right)^{3/2}}{(x+y) \sqrt{x+y
   Z^2}}} \nonumber \\[5pt]
  X_2 &=& Z \label{portapannolini} \\
  X_3 &=& \frac{\sqrt[4]{\frac{Z}{x+y}} \left(x+y
   Z^2\right)^{3/8}}{\sqrt[8]{x Z^2+y}} \nonumber
\end{eqnarray}
where $Z$ is, by definition, the real positive solution of the
following equation:
\begin{equation}
    \frac{U \left(Z^4-1\right) \sqrt{Z (x+y)} \left(x
   Z^2+y\right)^{3/4}}{\sqrt[4]{x+y Z^2}}+Z \left(x Z^2+y\right)
   \left(\sqrt{Z (x+y)} \sqrt[4]{\left(x Z^2+y\right) \left(x+y
   Z^2\right)}+\Sigma  \left(Z^2-1\right)\right) \, = \,
   0\label{ekaterina}
\end{equation}
Note that the solution \eqref{portapannolini}, \eqref{ekaterina}
becomes that of the Kamp\'{e} manifold
\eqref{raschiotto}, \eqref{sesticina} when $x=y=s$. Similarly for
either $x=0$ or $y=0$ the solution
\eqref{portapannolini}, \eqref{ekaterina} degenerates in either version
of the solution defining the Cardano manifold. In that case the
degree of the equation reduces from six to four allowing for the use
of Cardano's formula. This is the conclusive prove that at any point
of the type III  wall we realize the same manifold $Y_3$. Why is it
degenerate? The answer is that the very fact that we express all the
$X_I$ in terms of a single function $Z(\varpi,\mho)$ implies that
the Chern classes of all the tautological bundles are cohomologous
and conversely that the homology group is of dimension $1$ rather
than two.
\section{Conclusions}
What we have done in this paper was already described at the end of
the introductory Section \ref{introito} and we do not repeat it
here. We rather point out the perspectives opened up by our results
and what are the necessary steps that have still to be taken in the
development of our program.

As we have advocated, for  the sake of the conceivable physical
applications in the context of brane theory, of the quotient
singularity resolutions, it is quite relevant to understand the
explicit analytic structure of the exceptional divisors and of the
homology 2-cycles, having, at the same time, command on the explicit
form of the  forms $\omega^{(1,1)}$   representing the first Chern
classes of the tautological bundles. In this paper we succeeded in
these two tasks because we had two very strong allies:
\begin{enumerate}
  \item Toric Geometry, applicable to the case of abelian cyclic
  groups $\Gamma$,  that allows one to identify the possible full
  or partial resolutions of the quotient singularity and study in great detail their
  geometry,
     determining also the appropriate coordinate
  transformations leading to the equations of the divisors.
  \item The technique of localizing the moment map equations on
  divisors and curves, which allows for their solutions on these loci.
\end{enumerate}
Furthermore, we were able to clarify the Chamber Structure which, as
already stressed, encodes the phase diagram of the corresponding
gauge theory.

However it is on the two fronts mentioned above that mathematical
work is still needed. Ito-Reid theorem applies in general and in
particular to nonabelian groups. An urgent task is the extension of
the sort of explicit results obtained in this paper to nonabelian
cases. A second urgent task which is equally important for abelian
and nonabelian cases is the explicit solution of the moment map
equations in cases where the algebraic equation is of degre $d\, >
\, 4$. To this effect a quite valuable help might come from the
recent developments in algebraic geometry that represent roots of
higher order algebraic equations in terms of theta constants
associated with suitable hyperelliptic Riemann surfaces
\cite{gicovo6istica}. We plan to study that case.

From the physical side, as we already stated, we plan to utilize the
present explicit singularity resolution to study the corresponding
supergravity brane solutions and dual gauge theories. In this
context the geometrical results obtained here might be the origin of
new interesting D3-brane and M2-brane physics. Such applications to
the holographic scheme demand the construction of \textit{Ricci flat
metrics} on the same manifolds $Y$ that are derived from the
Kronheimer construction and the precise identification of the their
deformation parameters with the Fayet-Iliopoulos parameters $\zeta$.
These questions and issues will be addressed in forthcoming
publications \cite{conmasbia}.
\section*{Acknowledgements}
We acknowledge with great pleasure important and illuminating
discussions with our colleagues and close friends, Mario Trigiante,
Massimo Bianchi, Francisco Morales and Francesco Fucito.
U.B.~likes to thank CEMPI (Centre Europ\'een pour les Math\'ematiques,
la Physique et leur Interactions) for supporting his visit to Universit\'e
de Lille in February-March 2019 (Labex CEMPI ANR-11-LABX-0007-01).
U.B.'s research is also supported by GNSAGA-INdAM and by the PRIN
project ``Geometria delle Variet\`a Algebriche."

\addcontentsline{toc}{section}{Appendix: structure of the Hirzebruch surfaces}
\section*{Appendix: structure of the Hirzebruch surfaces}
\label{hirzegeo}
Let us give some details about the geometry of the second Hirzebruch surface
$\mathbb F_2$, which appears as the compact component of the exceptional
divisor of the resolution $Y\to \mathbb C^3/Z_4$.

Let $(U,V)$ be   homogeneous   coordinates on
$\mathbb{P}^1$   and $(X,Y,Z)$   homogeneous
  coordinates of  $\mathbb{P}^2$.
\begin{definizione}
The $n$-th Hirzebruch surface $\mathbb F_n$ is defined as
the locus cut out in $\mathbb{P}^1\times \mathbb{P}^2$ by the
following  equation of degree $n+1$:
\begin{equation}\label{irgobrutto}
    0 \, = \, \mathcal{P}_n(U,V,X,Y,Z) \, = \, X \, U^n \, - \, Y\,
    V ^n
\end{equation}
\end{definizione}
It is convenient to describe the Hirzebruch surface in terms of
inhomogeneous coordinates choosing open charts both for
$\mathbb{P}^1$ and for $\mathbb{P}^2$. We can
cover $\mathbb{P}^1$ with two charts: that where $V\neq 0$ and that
where $U\neq 0$. For $\mathbb{P}^2$, we need instead three charts
respectively corresponding to $Z \neq 0$, $Y\neq 0$ and $X\neq 0$.
Hence we have a total of six charts.
\paragraph{Description of $\mathbb F_n$  in the chart $V\neq 0, Z\neq 0$.} According
with the chosen conditions we set:
\begin{equation}\label{nonmolli1}
    s=\frac{U}{V} \quad ; \quad v= \frac{X}{Z} \quad ; \quad w = \frac{Y}{Z}
\end{equation}
and from eq.~\eqref{irgobrutto} we obtain:
\begin{equation}\label{irgobruttissimo}
    0 \, = \,  v \, s^n \, - \, w.
\end{equation}
Introducing a new complex variable $t$, a simple parametric
description of the locus satisfying the constraint
\eqref{irgobruttissimo} is provided by setting:
\begin{equation}\label{compiacio}
    v(u,t)=  tu^{-1} \quad ; \quad w(u,t) = t \, u^{n-1} \quad ;
    \quad s(u,t) \, = \, u
\end{equation}
Hence $(u,t)$ can be identified with a system of local coordinates
on $\mathbb F_n$. Let us now recall that the group $\mathrm{SU(2)}$
is the isometry group of $\mathbb{P}^1$ equipped  with the
Fubini-Study K\"ahler metric. Given a group element
\begin{equation}\label{gelemento}
    \mathbf{g} = \left(
          \begin{array}{cc}
            a & b \\
            c & d \\
          \end{array}
        \right) \, \in \, \mathrm{SU(2)}
\end{equation}
its action on the coordinate $u$ (regarded as a coordinate of
$\mathbb{P}^1$) is given by the corresponding fractional linear
transformation
\begin{equation}\label{marescalzo}
    \mathbf{g}(u) \, = \, \frac{a \, u + b}{c\, u + d}
\end{equation}
The action of $\mathbf{g}$ on the coordinate $u$ can be extended
also to the variable $t$ in such a way that the image point still
belongs to the Hirzebruch surface $\mathbb F_n$. We just
set:
\begin{equation}\label{ciabattarotta}
    \forall \mathbf{g} \in \mathrm{SU(2)} \quad : \quad
    \mathbf{g}\left(u,t\right) \, = \,\left(\frac{a \, u + b}{c\, u +
    d}, \quad t \, \left(c \,u+d\right)^{-n}\right)
\end{equation}
and we easily verify:
\begin{description}
 \item[a)] The transformation \eqref{ciabattarotta} respects the
  group product and provides a nonlinear representation since:
  \begin{equation}\label{belan}
    \mathbf{g}_1\left(\mathbf{g}_2\left(u,t\right)\right) \, =
    \,\mathbf{g_1}\cdot\mathbf{g}_2\left(u,t\right)
  \end{equation}
 \item[b)] The transformation \eqref{ciabattarotta} maps points of
  the Hirzebruch surface into points of the same surface since:
\begin{equation}\label{gerogammo}
    v\left(\mathbf{g}(u,t)\right) \, \left(s\left(\mathbf{g}(u,t)\right)\right)^n -
    w\left(\mathbf{g}(u,t) \right) \, = \,0
\end{equation}
if:
\begin{equation}\label{gerogammus}
    v\left(u,t\right) \, \left(s(u,t)\right)^n -
    w\left(u,v \right) \, = \,0
\end{equation}
\end{description}
On the Hirzebruch surface, as described in the current open chart
(which is dense in the manifold), we can introduce  nice K\"ahler
metrics that are invariant under the action of $\mathrm{SU(2)}$ as
given in eq.~\eqref{ciabattarotta}. In Section \ref{inducedF2} we
derive one of them (see eqs.~\eqref{samotracia} and \eqref{Jpolare})
induced on the compact exceptional divisor by  the K\"ahler quotient
construction.

\bigskip \frenchspacing
%

\addcontentsline{toc}{section}{References}
 \begin{thebibliography}{10}

\bibitem{Aharony:2008ug}
{\sc O.~Aharony, O.~Bergman, D.~L. Jafferis, and J.~Maldacena}, {\em {N=6}
  superconformal {C}hern-{S}imons-matter theories, {M2}-branes and their
  gravity duals}, Journal of High Energy Physics, 2008 (2008), p.~091.
\newblock doi:10.1088/1126-6708/2008/10/091 [arXiv:0806.1218 [hep-th]].

\bibitem{mango}
{\sc D.~Anselmi, M.~Bill\`o, P.~Fr\'e, L.~Girardello, and A.~Zaffaroni}, {\em
  {ALE manifolds and conformal field theories}}, Int. J. Mod. Phys., A9 (1994),
  pp.~3007--3058.

\bibitem{Bertolini:2002pr}
{\sc M.~Bertolini, V.~L. Campos, G.~Ferretti, P.~Salomonson, P.~Fr\'e, and
  M.~Trigiante}, {\em {BPS three-brane solution on smooth ALE manifolds with
  flux}}, Fortsch. Phys., 50 (2002), pp.~802--807.

\bibitem{Besse}
{\sc A.~L. Besse}, {\em Einstein manifolds}, Classics in Mathematics,
  Springer-Verlag, Berlin, 2008.
\newblock Reprint of the 1987 edition.

\bibitem{conmasbia}
{\sc M.~Bianchi, U.~Bruzzo, P.~Fr\'e,  P.~A. Grassi and D.
Martelli}, {\em work in
  progress}, to appear,  (2019).

\bibitem{Bianchi:2007wy}
{\sc M.~Bianchi, F.~Fucito, and J.~F. Morales}, {\em {D-brane instantons on the
  T**6 / Z(3) orientifold}}, JHEP, 07 (2007), p.~038.

\bibitem{Bianchi:2009bg}
\leavevmode\vrule height 2pt depth -1.6pt width 23pt, {\em {Dynamical
  supersymmetry breaking from unoriented D-brane instantons}}, JHEP, 08 (2009),
  p.~040.

\bibitem{Bianchi:1995ad}
{\sc M.~Bianchi, F.~Fucito, and G.~Rossi}, {\em {Instanton effects in
  supersymmetric Yang-Mills theories on ALE gravitational backgrounds}}, Phys.
  Lett., B359 (1995), pp.~56--61.

\bibitem{Bianchi:1994gi}
{\sc M.~Bianchi, F.~Fucito, G.~Rossi, and M.~Martellini}, {\em {ALE instantons
  in string effective theory}}, Nucl. Phys., B440 (1995), pp.~129--170.

\bibitem{Bianchi:1996zj}
\leavevmode\vrule height 2pt depth -1.6pt width 23pt, {\em {Explicit
  construction of Yang-Mills instantons on ALE spaces}}, Nucl. Phys., B473
  (1996), pp.~367--404.

\bibitem{Bianchi:2000de}
{\sc M.~Bianchi and J.~F. Morales}, {\em {Anomalies \& tadpoles}}, JHEP, 03
  (2000), p.~030.

\bibitem{ringoni}
{\sc M.~Bill\`o, D.~Fabbri, P.~Fr{\'e}, P.~Merlatti, and A.~Zaffaroni}, {\em
  Rings of short {$\mathcal{N}=3$} superfields in three dimensions and
  {M}-theory on {$\mathrm{AdS}_4\times \mathrm{N}^{010}$}}, Classical and
  Quantum Gravity, 18 (2001), p.~1269.
\newblock doi:10.1088/0264-9381/18/7/310 [hep-th/0005219].

\bibitem{Billo:1993rd}
{\sc M.~Bill\`o and P.~Fr\'e}, {\em {N=4 versus N=2 phases, hyperKahler
  quotients and the 2-d topological twist}}, Class. Quant. Grav., 11 (1994),
  pp.~785--848.

\bibitem{Billo:1992uw}
{\sc M.~Bill\`o, P.~Fr\'e, L.~Girardello, and A.~Zaffaroni}, {\em {Stringy
  gravitational instantons, the H map and N=4 moduli deformations}}, in
  {International Workshop on String Theory, Quantum Gravity and the Unification
  of Fundamental Interactions Rome, Italy, September 21-26, 1992}, 1992,
  pp.~28--40.

\bibitem{Billo:1992ei}
\leavevmode\vrule height 2pt depth -1.6pt width 23pt, {\em {Gravitational
  instantons in heterotic string theory: The H map and the moduli deformations
  of (4,4) superconformal theories}}, Int. J. Mod. Phys., A8 (1993),
  pp.~2351--2418.

\bibitem{Billo:1992zv}
{\sc M.~Bill\`o, P.~Fr\'e, A.~Zaffaroni, and L.~Girardello}, {\em {Heterotic
  vacua including gravitational instantons}}, in {10th Italian Conference on
  General Relativity and Gravitational Physics (It will include 4 workshops to
  take place in parallel sessions) Bardonecchia, Italy, September 1-5, 1992},
  1992, pp.~601--606.

\bibitem{bouchard}
{\sc V.~Bouchard}, {\em Orbifold {G}romov-{W}itten invariants and topological
  strings}, in Modular forms and string duality, vol.~54 of Fields Inst.
  Commun., Amer. Math. Soc., Providence, RI, 2008, pp.~225--246.

\bibitem{Bruzzo:2017fwj}
{\sc U.~Bruzzo, A.~Fino, and P.~Fr\'e}, {\em {The K\"ahler Quotient Resolution
  of $\mathbb{C}^3/\Gamma$ singularities, the McKay correspondence and
  $D\!=\!3$ $\mathcal{N}\!=2\!$ Chern-Simons gauge theories}}, Commun. Math.
  Phys., 365 (2019), pp.~93--214.

\bibitem{sergiotorino}
{\sc A.~Ceresole, G.~Dall'Agata, R.~D'Auria, and S.~Ferrara}, {\em Spectrum of
  type {IIB} supergravity on {$AdS_5\times T^{11}$}: predictions on
  {$\mathcal{N}=1$} {SCFT}'s}, Physical Review D, 61 (2000), p.~066001.
\newblock [hep-th/9905226].

\bibitem{CoxLS}
{\sc D.~A. Cox, J.~B. Little, and H.~K. Schenck}, {\em Toric varieties},
  vol.~124 of Graduate Studies in Mathematics, American Mathematical Society,
  Providence, RI, 2011.

\bibitem{crawthesis}
{\sc A.~Craw}, {\em The {M}c{K}ay correspondence and representations of the
  {M}c{K}ay quiver}, PhD thesis, Warwick University, United Kingdom, 2001.

\bibitem{Craw-Hilb}
\leavevmode\vrule height 2pt depth -1.6pt width 23pt, {\em An explicit
  construction of the {M}c{K}ay correspondence for {$A$}-{H}ilb {$\Bbb C^3$}},
  J. Algebra, 285 (2005), pp.~682--705.

\bibitem{CrawIshii}
{\sc A.~Craw and A.~Ishii}, {\em Flops of {$G$}-{H}ilb and equivalences of
  derived categories by variations of {GIT} quotient}, Duke Math. J., 124
  (2004), pp.~259--307.

\bibitem{degeratu}
{\sc A.~Degeratu and T.~Walpuski}, {\em Rigid {HYM} connections on tautological
  bundles over {ALE} crepant resolutions in dimension three}, SIGMA Symmetry
  Integrability Geom. Methods Appl., 12 (2016), pp.~Paper No. 017, 23.

\bibitem{eguccio}
{\sc T.~Eguchi and A.~J. Hanson}, {\em {Selfdual Solutions to Euclidean
  Gravity}}, Annals Phys., 120 (1979), p.~82.

\bibitem{Fabbri:1999hw}
{\sc D.~Fabbri, P.~Fr{\'e}, L.~Gualtieri, C.~Reina, A.~Tomasiello,
  A.~Zaffaroni, and A.~Zampa}, {\em 3{D} superconformal theories from
  {S}asakian seven-manifolds: new non-trivial evidences for {$\mathrm{AdS}_4 /
  \mathrm{CFT}_3$}}, Nuclear Physics B, 577 (2000), pp.~547--608.
\newblock [hep- th/9907219].

\bibitem{Fabbri:1999ay}
{\sc D.~Fabbri, P.~Fr\'e, L.~Gualtieri, and P.~Termonia}, {\em
  {$\mathrm{Osp(N|4)}$} supermultiplets as conformal superfields on
  {$\partial\mathrm{AdS}_4$} and the generic form of {$\mathcal{N}=2$}, {D=3}
  gauge theories}, Classical and Quantum Gravity, 17 (2000), p.~55.
\newblock [hep-th/9905134].

\bibitem{Ferrara:1998jm}
{\sc S.~Ferrara and C.~Fronsdal}, {\em Gauge fields as composite boundary
  excitations}, Physics Letters B, 433 (1998), pp.~19--28.
\newblock doi:10.1016/S0370-2693(98)00664-9 [hep-th/9802126].

\bibitem{Ferrara:1998ej}
{\sc S.~Ferrara, C.~Fronsdal, and A.~Zaffaroni}, {\em On {$\mathcal{N}=8$}
  supergravity in {AdS}$_5$ and {$\mathcal{N}=4$} superconformal {Y}ang-{M}ills
  theory}, Nuclear Physics B, 532 (1998), pp.~153--162.
\newblock doi:10.1016/S0550-3213(98)00444-1 [hep-th/9802203].

\bibitem{pappo1}
{\sc P.~Fr\'e and P.~A. Grassi}, {\em {The Integral Form of D=3 Chern-Simons
  Theories Probing ${\mathbb C}^n/\Gamma$ Singularities}}, Fortsch. Phys., 65
  (2017), p.~1700040.

\bibitem{Fre1999xp}
{\sc P.~Fr{\'e}, L.~Gualtieri, and P.~Termonia}, {\em The structure of
  {$\mathcal{N}=3$} multiplets in {$\mathrm{AdS}_4$} and the complete
  {$\mathrm{Osp(3|4)}\times \mathrm{SU(3)}$} spectrum of {M}-theory on
  {$\mathrm{AdS}_4\times \mathrm{N}^{0,1,0}$}}, Physics Letters B, 471 (1999),
  pp.~27--38.
\newblock [hep-th/9909188].

\bibitem{advancio}
{\sc P.~G. Fr{\'e}}, {\em {Advances in Geometry and Lie Algebras from
  Supergravity}}, Theoretical and Mathematical Physics book series, Springer,
  2018.

\bibitem{Gaiotto:2009tk}
{\sc D.~Gaiotto and D.~L. Jafferis}, {\em {Notes on adding D6 branes wrapping
  $\mathbb{RP}^3$ in $AdS(4) \times \mathbb {CP}^3$}}, JHEP, 11 (2012), p.~015.

\bibitem{Gaiotto:2007qi}
{\sc D.~Gaiotto and X.~Yin}, {\em Notes on superconformal
  {C}hern-{S}imons-{M}atter theories}, Journal of High Energy Physics, 2007
  (2007), p.~056.
\newblock doi:10.1088/1126-6708/2007/08/056 [arXiv:0704.3740 [hep-th]].

\bibitem{Gibbons:1979zt}
{\sc G.~W. Gibbons and S.~W. Hawking}, {\em {Gravitational Multi -
  Instantons}}, Phys. Lett., 78B (1978), p.~430.

\bibitem{Gibbons:1979xm}
\leavevmode\vrule height 2pt depth -1.6pt width 23pt, {\em {Classification of
  Gravitational Instanton Symmetries}}, Commun. Math. Phys., 66 (1979),
  pp.~291--310.

\bibitem{Hatch}
{\sc A.~Hatcher}, {\em Algebraic topology}, Cambridge University Press,
  Cambridge, 2002.

\bibitem{Hauzer-Langer}
{\sc M.~Hauzer and A.~Langer}, {\em Moduli spaces of framed perverse instantons
  on {$\Bbb P^3$}}, Glasg. Math. J., 53 (2011), pp.~51--96.

\bibitem{HKLR}
{\sc N.~J. Hitchin, A.~Karlhede, U.~Lindstr{\"o}m, and M.~Ro{\v{c}}ek}, {\em
  {Hyperkahler metrics and supersymmetry}}, Commun. Math. Phys., 108 (1987),
  p.~535.

\bibitem{itoriddo}
{\sc Y.~Ito and M.~Reid}, {\em The {M}c{K}ay correspondence for finite
  subgroups of {SL}(3,{C})}, in Higher dimensional complex varieties, {E}ditors
  {M}arco {A}ndreatta and Thomas Peternell, De Gruyter, 1994, pp.~221--240.

\bibitem{Joyce-QALE}
{\sc D.~Joyce}, {\em Quasi-{ALE} metrics with holonomy {${\rm SU}(m)$} and
  {${\rm Sp}(m)$}}, Ann. Global Anal. Geom., 19 (2001), pp.~103--132.

\bibitem{Kallosh:1998ph}
{\sc R.~Kallosh and A.~Van~Proeyen}, {\em Conformal symmetry of supergravities
  in {AdS} spaces}, Physical Review D, 60 (1999), p.~026001.
\newblock doi:10.1103/PhysRevD.60.026001 [hep- th/9804099].

\bibitem{witkleb}
{\sc I.~R. Klebanov and E.~Witten}, {\em {Superconformal field theory on
  three-branes at a Calabi-Yau singularity}}, Nucl. Phys., B536 (1998),
  pp.~199--218.

\bibitem{kro1}
{\sc P.~B. Kronheimer}, {\em The construction of {ALE} spaces as
  hyper-{K}\"ahler quotients}, J. Differential Geom., 29 (1989), pp.~665--683.

\bibitem{kro2}
\leavevmode\vrule height 2pt depth -1.6pt width 23pt, {\em A {T}orelli-type
  theorem for gravitational instantons}, J. Differential Geom., 29 (1989),
  pp.~685--697.

\bibitem{Maldacena:1997re}
{\sc J.~Maldacena}, {\em The large-{N} limit of superconformal field theories
  and supergravity}, International Journal of Theoretical Physics, 4 (1999),
  pp.~1113--1133.
\newblock doi:10.1023/A:1026654312961 [hep-th/9711200].

\bibitem{marcovaldo}
{\sc D.~Markushevich}, {\em Resolution of {${\bf C}^3/H_{168}$}}, Math. Ann.,
  308 (1997), pp.~279--289.

\bibitem{mckay}
{\sc J.~McKay}, {\em Graphs, singularities and finite groups}, Proc. Symp. Pure
  Math., 37 (1980), p.~183.

\bibitem{mumford-GIT}
{\sc D.~Mumford, J.~Fogarty, and F.~Kirwan}, {\em Geometric invariant theory},
  vol.~34 of Ergebnisse der Mathematik und ihrer Grenzgebiete (2),
  Springer-Verlag, Berlin, third~ed., 1994.

\bibitem{newstead-GIT}
{\sc P.~E. Newstead}, {\em Geometric invariant theory}, in Moduli spaces and
  vector bundles, vol.~359 of London Math. Soc. Lecture Note Ser., Cambridge
  Univ. Press, Cambridge, 2009, pp.~99--127.

\bibitem{Reid-Kino}
{\sc M.~Reid}, {\em Mc{K}ay correspondence}, in Proc. of Algebraic Geometry
  Symposium, Kinosaki, 1997, pp.~14--41.
\newblock T. {Katsura} Ed.

\bibitem{roanno}
{\sc S.-S. Roan}, {\em Minimal resolutions of {G}orenstein orbifolds in
  dimension three}, Topology, 35 (1996), pp.~489--508.

\bibitem{SardoInfirri:1994is}
{\sc A.~V. Sardo~Infirri}, {\em {Resolutions of orbifold singularities and
  representation moduli of McKay quivers}}, PhD thesis, Kyoto U., RIMS, 1994.

\bibitem{SardoInfirri:1996ga}
\leavevmode\vrule height 2pt depth -1.6pt width 23pt, {\em {Partial resolutions
  of orbifold singularities via moduli spaces of HYM type bundles}}.
\newblock {\tt arXiv:alg-geom/9610004}, 1996.

\bibitem{SardoInfirri:1996gb}
\leavevmode\vrule height 2pt depth -1.6pt width 23pt, {\em {Resolutions of
  orbifold singularities and flows on the McKay quiver}}.
\newblock {\tt arXiv:alg-geom/9610005}, 1996.

\bibitem{smith}
{\sc H.~J.~S. Smith}, {\em On systems of linear indeterminate equations and
  congruences}, Phil. Trans. R. Soc. Lond., 151 (1) (1861), pp.~293--326.
\newblock Reprinted (pp. 367--409) in The Collected Mathematical Papers of
  Henry John Stephen Smith, Vol. I, edited by J. W. L. Glaisher. Oxford:
  Clarendon Press (1894), xcv+603 pp.

\bibitem{gicovo6istica}
{\sc A.~Zhivkov}, {\em Resolution of degree = 6 algebraic equations by genus
  two theta constants}, Journal of Geometry and Symmetry in Physics, 11 (2008),
  pp.~77--93.

\end{thebibliography}

\end{document}